\documentclass[a4paper]{article}
\pdfoutput=1
\usepackage{jhepmod}

\usepackage{multirow}
\usepackage{hyperref}
\usepackage{graphicx}
\usepackage{mathtools}
\usepackage{amsmath,amssymb,amsfonts,amsthm,latexsym,stmaryrd,physics,mathrsfs}
\allowdisplaybreaks[4]
\usepackage{color}
\usepackage{bm}
\usepackage{float}
\usepackage{nicefrac}
\usepackage{microtype} 
\usepackage{booktabs}
\usepackage{verbatim}
\usepackage{setspace}
\usepackage{stfloats}
\usepackage{diagbox}
\usepackage{dsfont}
\usepackage{longtable}
\usepackage{afterpage}

\usepackage{xargs}
\usepackage[colorinlistoftodos,prependcaption,textsize=tiny]{todonotes}

\newcommandx{\stilltodo}[2][1=]{\todo[linecolor=red,backgroundcolor=red!25,bordercolor=red,#1]{#2}}
\newcommandx{\change}[2][1=]{\todo[linecolor=blue,backgroundcolor=blue!25,bordercolor=blue,#1]{#2}}
\newcommandx{\info}[2][1=]{\todo[linecolor=purple,backgroundcolor=purple!25,bordercolor=purple,#1]{#2}}
\newcommandx{\improvement}[2][1=]{\todo[linecolor=purple,backgroundcolor=purple!25,bordercolor=purple,#1]{#2}}
\newcommandx{\modified}[2][1=]{\todo[linecolor=gray,backgroundcolor=gray!25,bordercolor=gray,#1]{#2}}
\newcommandx{\unsure}[2][1=]{\todo[linecolor=cyan,backgroundcolor=cyan!25,bordercolor=cyan,#1]{#2}}
\newcommandx{\thiswillnotshow}[2][1=]{\todo[disable,#1]{#2}}

\def\beq{\begin{equation}}
\def\eeq{\end{equation}}
\newcommand{\bea}{\begin{eqnarray}}
\newcommand{\eea}{\end{eqnarray}}
\def\bi{\begin{itemize}}
\def\ei{\end{itemize}}
\def\ba{\begin{array}}
\def\ea{\end{array}}
\def\bfig{\begin{figure}}
\def\efig{\end{figure}}

\def\C{\mathbb{C}}
\def\R{\mathbb{R}}
\def\Z{\mathbb{Z}}

\def\sgn{\text{sgn}}

\newcommand{\Slc}{\mathrm{SL}(2,\mathbb{C})}

\newcommand{\Su}{\mathrm{SU}(2)}

\def\be{\begin{eqnarray}}
\def\ee{\end{eqnarray}}

\newcommand{\cj}{\mathcal J}
\newcommand{\ck}{\mathcal K}
\newcommand{\cl}{\mathcal L}

\newcommand{\cs}{\mathcal S}

\newcommand{\cz}{\mathcal Z}

\newcommand{\sm}{\mathscr{M}}

\newcommand{\fa}{\mathfrak{a}}

  \newcommand{\Fl}{\mathfrak{L}}

  \newcommand{\Fq}{\mathfrak{Q}}

\renewcommand{\a}{\alpha}

\newcommand{\g}{\gamma}

\newcommand{\sig}{\sigma}

\renewcommand{\l}{\lambda}

\renewcommand{\O}{\Omega}
\renewcommand{\t}{\tau}

\newcommand{\rmd}{\mathrm d}

\newcommand{\lt}{\left}
\newcommand{\rt}{\right}

\newcommand{\act}{\rhd}

\newcommand{\sn}{\mathscr{N}}

\newcommand{\re}{\mathrm{Re}}

\title{Complex critical points in Lorentzian spinfoam quantum gravity: 4-simplex amplitude and effective dynamics on double-$\Delta_3$ complex }

\author[1,2]{Muxin Han}
\author[2]{\ Hongguang Liu}
\author[3,1]{\ Dongxue Qu}
\affiliation[1]{Department of Physics, Florida Atlantic University, 777 Glades Road, Boca Raton, FL 33431-0991, USA}
\affiliation[2]{Department Physik, Institut f\"ur Quantengravitation, Theoretische Physik III, Friedrich-Alexander Universit\"at Erlangen-N\"urnberg, Staudtstr. 7/B2, 91058 Erlangen, Germany}
\affiliation[3]{Perimeter Institute for Theoretical Physics, 31 Caroline St N, Waterloo, ON N2L 2Y5, Canada}

\emailAdd{hanm(AT)fau.edu}
\emailAdd{hongguang.liu(AT)gravity.fau.de}
\emailAdd{dqu(AT)perimeterinstitute.ca}

\abstract{The complex critical points are analyzed in the 4-dimensional Lorentzian Engle-Pereira-Rovelli-Livine (EPRL) spinfoam model in the large-$j$ regime. For the 4-simplex amplitude, taking into account the complex critical point generalizes the large-$j$ asymptotics to the situation with non-Regge boundary data and relates to the twisted geometry. For generic simplicial complexes, we present a general procedure to derive the effective theory of Regge geometries from the spinfoam amplitude in the large-$j$ regime by using the complex critical points. The effective theory is analyzed in detail for the spinfoam amplitude on the double-$\Delta_3$ simplicial complex. We numerically compute the effective action and the solution of the effective equation of motion on the double-$\Delta_3$ complex. The effective theory reproduces the classical Regge gravity when the Barbero-Immirzi parameter $\gamma$ is small. 
}

\begin{document}
\maketitle

\section{Introduction}

The perturbative expansion is widely used in quantum theory to make approximate predictions order by order in certain parameter. The method of perturbative expansion is well-connected to the path integral formulation, whose stationary phase approximation results in the semiclassical expansion in $\hbar$. By the stationary phase approximation, the path integral is approximately computed by the dominant contribution from the \emph{critical point} and neighborhood. The critical point is the solution of the equation of motion, which is obtained from variating the action in the path integral. Given a path integral in terms of real variables, traditionally, the semiclassical expansion only takes into account critical points inside the real integration cycle. However, the recent progress in many research areas demonstrates that the \emph{complex critical point} generically away from the real integration cycle plays a crucial role in the semiclassical expansion of the path integral (see e.g. \cite{Witten:2010zr,Witten:2010cx,Basar:2013eka,Dunne:2016jsr,Cristoforetti:2013qaa,Witten:2021nzp}). The complex critical point is the critical point of the analytically continued path integral, where the integrand is analytically extended to the complexification of the real integration cycle.

The method of stationary phase approximation has been applied extensively to the spinfoam amplitude in Loop Quantum Gravity (LQG) (see e.g. \cite{Conrady:2008mk,Barrett:2009mw,Han:2011re,Bianchi:2009ri,Han:2013hna}). The importance of the complex critical point has been demonstrated in the recent progress in the semiclassical analysis of spinfoam amplitude \cite{Han:2021kll,Asante:2020qpa,Han:2021rjo}. A key result is that the semiclassical curved spacetime geometry can only emerge from the complex critical point of the spinfoam amplitude. Taking into account the complex critical point provides the resolution to the long-standing ``flatness problem'', i.e., the problem of discovering only the flat spacetime geometry in the spinfoam amplitude. This problem turns out to be the confusion from ignoring the complex critical point. 

The present work continues from the earlier work \cite{Han:2021kll} and further study the complex critical points and their implications in spinfoam amplitude. The discussion in this work focuses on the 4-dimensional Lorentzian Engle-Pereira-Rovelli-Livine (EPRL) spinfoam model. Our results demonstrate the impact of the complex critical points mainly from two perspectives: 

\begin{itemize}

\item At the level of one 4-simplex amplitude, taking into account the complex critical point generalizes the large-$j$ asymptotics by Barrett et al \cite{Barrett:2009mw} to the case of non-Regge boundary data. The geometry of the non-Regge boundary data gives the boundary tetrahedra that are glued only with area-matching but without shape-matching, in contrast to the Regge boundary data that requires the shape-matching condition (as well as the orientation matching condition) and determines the Regge boundary geometry. The generalized 4-simplex amplitude asymptotic behavior depends analytically on the boundary data. This analytic dependence is not manifest in the original asymptotic formula in \cite{Barrett:2009mw}. The computation of the generalized asymptotic behavior relies on the numerical method. The discussion in Section \ref{Single4Simplex} provides the general algorithm of computing the complex critical point of the amplitude, and demonstrates the numerical results of the asymptotics for a 1-parameter family of non-Regge boundary data.

\item Based on the application of complex critical points, we develop a formalism to derive the effective theory of Regge geometry from the large-$j$ spinfoam amplitude. As the result, given a simplicial complex $\ck$ with $M$ internal segments, the spinfoam amplitude $A(\ck)$ with Regge boundary data reduces to the integral over the internal line-segment lengths $l_I$, $I=1,\cdots,M$,
\be
A(\ck)\sim\int \prod_{I=1}^M\rmd\mu( l_I)\ e^{\l \cs (\vec{l})}\lt[1+O(1/\l)\rt],\qquad \l\gg 1,\label{effectiveA0}
\ee
within the neighborhood of the integration domain of $A(\ck)$. $\l$ is the scaling parameter of spins $j_f$. $e^{\l \cs (\vec{l})}$ with the effective action $\cs(\vec{l})$ comes from evaluating the analytically continued integrand of $A(\ck)$ at the complex critical point, which depend analytically on $l_I$. The integral in \eqref{effectiveA0} reduced from $A(\ck)$ is over the Regge geometries with the fixed boundary condition. The equation of motion $\partial_{l_I}\cs(\vec{l})=0$ gives the effective dynamics of Regge geometry implied by the spinfoam amplitude. The formalism of deriving the effective theory is discussed in Section \ref{Real critical points and complex critical points}. In Sections \ref{DoubleDelta} and \ref{Solutions of effective dynamics on}, we apply the formalism to the double-$\Delta_3$ simplicial complex, which contains only a single internal segment, i.e., $M=1$. The complex critical points and the effective action $\cs (\vec{l})$ are computed numerically following the general algorithm. The spinfoam amplitude depends on the Barbero-Immirzi parameter $\gamma$. The computations are performed for many different values of the Barbero-Immirzi parameter $\gamma$, ranging from small to large. The resulting $\cs (\vec{l})$ are compared with the Regge action on the double-$\Delta_3$ complex. $\cs (\vec{l})$ is well-approximated by the classical Regge action in the small-$\gamma$ regime, and $\cs (\vec{l})$ provides the correction to the Regge action with increasing $\gamma$. The solutions of the effective dynamics are computed numerically for different values of $\gamma$ and compared to the solution of Regge equation. The solution from $\cs (\vec{l})$ well-approximates the Regge solution for small $\g$ and gives larger correction when increasing $\gamma$. Recovering the classical Regge action and solution from the effective dynamics of spinfoam amplitude gives evidence of the semiclassical consistency of spinfoam quantum gravity.

\end{itemize}

Recovering the classical Regge gravity from the spinfoam amplitude with small $\gamma$ has been argued earlier in \cite{propagator1,Bianchi:2009ri,propagator3,Magliaro:2011zz,Perini:2011uk,Magliaro:2011dz,Han:2013ina}. Our numerical result confirms this property for the spinfoam amplitude on the double-$\Delta_3$ complex.

The numerical computations are performed for different $\gamma$'s ranging from small to large. Fixing the boundary data, the solutions of the effective dynamics give a trajectory in the space of Regge geometries parametrized by $\gamma$. The trajectory approaches the solution of the classical Regge equation for small $\gamma$ as mentioned above. For large $\gamma$, the trajectory stablizes at the Regge geometry that is different from the classical Regge solution. It suggests that the effective theory for large $\gamma$ differs significantly from the Regge gravity. The solutions both at small and large $\gamma$ give non-suppressed contributions to the spinfoam amplitude. In particular, the solutions for large $\gamma$ violate the known bound $|\gamma \delta_h|\lesssim\l^{-1/2}$ \cite{Han:2021kll,Asante:2020qpa,Han:2013hna} ($\delta_h$ is the deficit angle of the Regge geometry), which is valid for non-suppressed contributions to the amplitude with finite and small $\gamma$.

Studying the complex critical points in the spinfoam amplitude closely relates to the recent progress in numerical studies of spinfoam amplitudes \cite{Dona:2022yyn}. Given the complexity of the spinfoam amplitude, the complex critical point and the corresponding contribution to the spinfoam amplitude has to be computed numerically. The numerical analysis of complex critical points connects to the Lefschetz-thimble and Monte-Carlo computation for the spinfoam integral \cite{Han:2020npv}, because every complex critical point associates to an integration cycle known as Lefschetz thimble, and the integral on the Lefschetz thimble collects all contributions associated to the complex critical point. Another related numerical result is the semiclassical expansion of the spinfoam amplitude to the next-to-leading order from the stationary phase approximation \cite{Han:2020fil}. We also would like to mention a few other numerical approaches for spinfoam quantum gravity, including the ``sl2cfoam-next'' code for the non-perturbative computation of the spinfoam amplitude \cite{Gozzini:2021kbt,Frisoni:2022urv,Dona:2022dxs}, the effective spinfoam model \cite{Asante:2020qpa,Asante:2021zzh}, the hybrid algorithm \cite{Asante:2022lnp}, and the spinfoam renormalization \cite{Asante:2022dnj,Bahr:2016hwc}, etc.

This paper is organized as follows: Section \ref{SpinfoamAmplitude} gives a brief review of the integral representation of the EPRL spinfoam amplitude and the definition of the large-$j$ regime. In Section \ref{Real critical points and complex critical points}, we define the real and complex critical points and discuss the general formalism of deriving the effective dynamics of Regge geometry. Section \ref{Single4Simplex} studies the complex critical point of the 4-simplex amplitude and generalizes the large-$j$ asymptotics to include the non-Regge boundary data. Section \ref{Revisit the amplitude} revisits the known results on the spinfoam amplitude on $\Delta_3$ complex as the preparation for analyzing the amplitude on the double-$\Delta_3$ complex. Section \ref{DoubleDelta} discusses the complex critical point in the spinfoam amplitude on the double-$\Delta_3$ complex and computes the effective action. Section \ref{Solutions of effective dynamics on} discusses the numerical solution of the effective dynamics on the double-$\Delta_3$ complex. In Section \ref{Conclusion and Discussion}, we conclude and discuss some outlooks.

\section{Spinfoam amplitude}\label{SpinfoamAmplitude}

A 4-dimensional simplicial complex $\ck$ contains 4-simplices $v$, tetrahedra $e$, triangles $f$, line segments, and points. The internal and boundary triangles are denoted by $h$ and $b$ ($f$ is either $h$ or $b$). The SU(2) spins $j_h,j_b\in\mathbb{N}_0/2$ are assigned to internal and boundary triangles $h,b$. The spins label the quanta of triangle areas. The LQG area spectrum indicates that the quantum area of triangle $f$ is given by $\fa_f=8\pi\g G\hbar\sqrt{j_f(j_f+1)}$ \cite{Rovelli1995,ALarea}. In the large-$j$ regime, which we will focus on, the area spectrum gives $\fa_f\simeq 8\pi \g G\hbar j_f $, or $\fa_f\simeq \g j_f$ when we set the unit such that $8\pi G\hbar=1$.

The Lorentzian EPRL spinfoam amplitude on $\ck$ is given by summing over internal spins $\{j_h\}$:
\be
A(\ck)&=&\sum_{\{j_{h}\}}\prod_h \bm{d}_{j_h}^{|V(f)|+1}\int [\rmd g\rmd\mathbf{z}]\, e^{S
	\left(j_{h}, g_{v e}, \mathbf{z}_{vf};j_b,\xi_{eb}\right)}, \label{amplitude}\\
&& [\rmd g\rmd \mathbf{z}]=\prod_{(v, e)} \mathrm{d} g_{v e} \prod_{(v,f)} \mathrm{d}\O_{\mathbf{z}_{v f}},
\ee
where $\bm{d}_{ j_h}=2j_h+1$. The boundary states are SU(2) coherent states $|j_b,\xi_{eb}\rangle$ where $\xi_{eb}=u_{eb}\act(1,0)^\mathrm{T}$, $u_{eb}\in \Su$. $j_b$ and $\xi_{eb}$ are determined by the area and the 3-normal of the boundary triangle $b$. The summed/integrated variables are $g_{ve}\in\Slc$, $\mathbf{z}_{vf}\in\mathbb{CP}^1$, and $j_h$. $\rmd g_{ve}$ is the Haar measure on $\Slc$,
\be 
\mathrm{d} g=\frac{\mathrm{d} \beta \mathrm{d} \beta^{*} \mathrm{d} \gamma \mathrm{d} \gamma^{*} \mathrm{d} \delta \mathrm{d} \delta^{*}}{|\delta|^{2}}, \quad \forall g=\left(\begin{array}{cc}
	\alpha & \beta \\
	\gamma & \delta
\end{array}\right)\in \Slc,
\ee 
and $\mathrm{d}\O_{\mathbf{z}_{v f}}$ is the scaling invariant measure on $\mathbb{CP}^1$:
\be 
\mathrm{d}\O_{\mathbf{z}_{v f}} 
&=&\frac{i}{2} \frac{\left(z_{0} \mathrm{~d} z_{1}-z_{1} \mathrm{~d} z_{0}\right) \wedge\left(\bar{z}_{0} \mathrm{~d} \bar{z}_{1}-\bar{z}_{1} \mathrm{~d} \bar{z}_{0}\right)}{\left\langle Z_{v e f}, Z_{v e f}\right\rangle\left\langle Z_{v e^{\prime} f}, Z_{v e^{\prime} f}\right\rangle}, \quad \forall \ {\bf z}_{vf}=(z_0,z_1)^{\rm T}, 
\ee 
where $Z_{vef} = g_{ve}^{\dagger} {\bf z}_{vf}$, $\langle\cdot,\cdot\rangle$ is the Hermitian inner product on $\mathbb{C}^2$, and ${\bf z}_{vf}$ is a 2-component spinor for the face $f$. 

The spinfoam action $S$ in Eq.(\ref{amplitude}) is complex and linear to $j_h,j_b$ in an expression of the form \cite{Han:2013gna}, 
\be
S
&=&\sum_{e'}j_hF_{(e',h)}+\sum_{(e,b)}j_bF^{in/out}_{(e,b)}+\sum_{(e',b)}j_bF^{in/out}_{(e',b)},\label{SjFjF} 
\\
F_{(e,b)}^{out}&=&2 \ln\dfrac{\left\langle Z_{v e b},\xi_{e b}\right\rangle}{\left\| Z_{v e b}\right\|}+i\g \ln \left\| Z_{v e b}\right\|^2,\\
F_{(e,b)}^{in}&=& 2 \ln \dfrac{\left\langle\xi_{e b}, Z_{v' e b}\right\rangle}{\left\| Z_{v' e b}\right\|}-i\g \ln \left\| Z_{v' e b}\right\|^2,\\
F_{(e',f)}&=&2 \ln \dfrac{\left\langle Z_{v e' f}, Z_{v^{\prime} e' f}\right\rangle}{\left\| Z_{v e' f}\right\|\left\| Z_{v^{\prime} e' f}\right\|} + i\g \ln\frac{\left\| Z_{v e' f}\right\|^2}{\left\| Z_{v^{\prime} e' f}\right\|^2}.
\ee
Here, $e$ and $e'$ are boundary and internal tetrahedra, respectively. In the dual complex $\ck^*$, the orientation of $\partial f^*$ is outgoing from the vertex dual to $v$ and incoming to another vertex dual to $v'$, and the orientation of the face $f^*$ dual to $f$ induces $\partial f^*$'s orientation. As for the logarithms in the spinfoam action, we fix all the logarithms to be the principal values. The derivation of the spinfoam action $S$ is given in \cite{Han:2013gna}. 

The spinfoam amplitude in the formulation \eqref{amplitude} has the following three types of continuous gauge degrees of freedom, and thus some gauge fixings are needed to remove the redundant degrees of freedom:
\bi
\item Firstly, there is $\Slc$ gauge transformation at each $v$:  
\be 
g_{ve}\mapsto x_v^{-1}g_{ve},\quad\textbf{z}_{vf}\mapsto x_v^{\dagger}\textbf{z}_{vf}, \quad x_v\in\Slc. 
\ee
To remove this gauge degree of freedom, we fix one $g_{ve}$ to be a constant $\Slc$ matrix for each 4-simplex. The amplitude is independent of the choices of constant matrices.
\item Secondly, there is $\Su$ gauge transformation on each internal $e$: 
\be 
g_{v'e}\mapsto g_{v'e}h_e^{-1},\quad g_{ve}\mapsto g_{ve}h_e^{-1}, \quad h_e\in\Su.
\ee 
To fix this $\Su$ gauge freedom, one can parameterize one of two $\Slc$ elements: $g_{ve}$, or $g_{v'e}$ by the upper triangular matrix
\begin{equation}
	k=\left(\begin{matrix}
		\lambda^{-1}&\mu\\0&\lambda
	\end{matrix}\right),\ \lambda\in\R\setminus\{0\},\ \mu\in\mathbb{C} \label{upper}
\end{equation}  
Here, we use the fact that any $g\in\Slc$ can be decomposed as $g=kh$ with $h\in\Su$ and $k$ an upper triangular matrix in Eq.(\ref{upper}).

\item Thirdly, for each $\textbf{z}_{vf}$, there is the scaling gauge freedom:
\be 
\textbf{z}_{vf}\mapsto\lambda_{vf} \textbf{z}_{vf}, \quad \lambda_{vf}\in\mathbb{C}.  
\ee 
Here, we fix the gauge by setting the first component of $\textbf{z}_{vf}$ to 1, i.e. ${\textbf{z}}_{vf}=\left(1, {\a}_{vf}\right)^\mathrm{T}$, where ${\a}_{vf}\in\C$.
\ei

Furthermore, in Eq.(\ref{amplitude}), we assume the summation over internal $j_h\in\mathbb{N}_0/2$ is bounded by $j^{\rm max}$. In some situations, $j^{\rm max}$ is determined by boundary spins $j_b$ via the triangle inequality, otherwise $j^{\rm max}$ are imposed as the cut-off to regularize the infinite sum over spins. To prepare for the stationary phase analysis, we would like to change the summation over $j_h$ in Eq.(\ref{amplitude}) to integrals. The idea is to apply the Poisson summation formula. Firstly, we replace each $\bm{d}_{j_h}$ by a smooth compact support function $\t_{[-\epsilon,j^{\rm max}+\epsilon]}(j_h)$ satisfying
\be
\t_{[-\epsilon,j^{\rm max}+\epsilon]}( j_h)=\bm{d}_{j_h}^{|V(f)|+1},\ \text{for}\  j_h\in[0,j^{\rm max}],\quad \text{and}\quad \t_{[-\epsilon,j^{\rm max}+\epsilon]}(j_h)=0,\ \text{for}\  j_h\not\in[-\epsilon,j^{\rm max}+\epsilon],\nonumber
\ee
for any $0<\epsilon<1/2$. This replacement does not change the value of the amplitude $A(\ck)$ but makes the summand of $\sum_{j_h}$ smooth and compact support in $j_h$. Then, by applying the Poisson summation formula,
\[
\sum_{n\in\Z} f(n)=\sum_{k \in \mathbb{Z}} \int_\R \mathrm{d} n f(n) \,{e}^{2\pi i k n},
\]
the discrete summation over $j_h$ in Eq.(\ref{amplitude}) becomes summing of integrals: 
\be
\!\!\!\!\!\!A(\ck)&=&\sum_{\{k_h\in\mathbb{Z}\}} \int
\prod_h\mathrm{d} j_{h}\prod_h 2\t_{[-\epsilon,j^{\rm max}+\epsilon]}(j_h)\int [\rmd g\rmd \mathbf{z}]\, e^{S^{(k)}},\\ &&S^{(k)}=S+4\pi i \sum_h j_h k_h.\label{integralFormAmp1}
\ee
By the area spectrum, the classical area $\fa_f$ and small $\hbar$ imply the large spin $j_f\gg1$. This motivates understanding the large-$j$ regime as the semiclassical regime of $A(\ck)$. Then, to probe the semiclassical regime, we scale uniformly both the boundary spins $j_b$ and the internal spin cut-off $j^{\rm max}$ by 
\be
j_b\to \l j_b,\qquad j^{\rm max}\to \l j^{\rm max},\quad \l\gg1,
\ee
so $S\rightarrow\lambda S$ as a result from $S$ being linear in $j_b,j_h$. As a consequence, the spinfoam amlitude $A(\ck)$ in the large-$j$ regime is  
\be
\!\!\!\!\!\!A(\ck)&=&\sum_{\{k_h\in\mathbb{Z}\}}\int_\R
\prod_h\mathrm{d} j_{h}\prod_h 2 \l\,\t_{[-\epsilon,\l j^{\rm max}+\epsilon]}(\l j_h)\int [\rmd g\rmd \mathbf{z}]\, e^{\l S^{(k)}},\label{integralFormAmp}\\
&&S^{(k)}=S+4\pi i \sum_h j_h k_h, \label{Stot}
\ee
by the change of integration variables $j_h\to\l j_h$, and $j_h$ is continous.

\section{Complex critical point and effective dynamics}\label{Real critical points and complex critical points}

The integral in \eqref{integralFormAmp} at each $k_h$ can be analyzed with the stationary phase method in the regime $\l\gg1$. By the standard argument of the stationary phase approximation, by fixing the boundary data, the integral with $\l\gg1$ is approximated by the dominant contributions from the solutions of critical equations and neighborhood. In the case of the integrals in \eqref{integralFormAmp}, the critical equations are 
\be
\re(S)&=&\partial_{g_{ve}}S=\partial_{\mathbf{z}_{vf}}S=0,\label{eom1}\\
\partial_{j_h}S&=&4\pi i k_h, \qquad k_h\in\Z.\label{eom2}
\ee
The solutions inside the integration domain are denoted by $\{\mathring{j}_h,\mathring{g}_{ve},\mathring{\bf z}_{vf}\}$. The integration domain is viewed as a real manifold, and the integration variables are real and imaginary parts of the matrix elements in $g_{ve}$ and $\mathbf{z}_{vf}$. We call $\{\mathring{j}_h,\mathring{g}_{ve},\mathring{\bf z}_{vf}\}$ the \textit{real critical point} accordingly.

The existence of the real critical point in \eqref{integralFormAmp} depends on the boundary condition. The real critical point may not exist for the generic boundary condition. We know that $S$ is a complex action with $n$ real variables $x$, and $\partial_x S=0$ gives $n$ complex thus $2n$ real equations, which is over-constrained for $n$ real variables. Consequently, the critical equations \eqref{eom1} and \eqref{eom2} coupled with one more equation $\mathrm{Re}(S)=0$ result in the nonexistence of the general real solution, unless for some special boundary conditions. 

As a solution to this problem of over-constrained equations, the integration variables have to be complexified, and action $S$ has to be analytically continued to the complex variables $z$. We are only interested in the integration domain where the spinfoam action $S$ is analytic. The analytically continued action is denoted by $\cs$. On the space of complex variables, the complex critical equation $\partial_z\cs=0$ is not over-constrained anymore because it gives $n$ complex equations for $n$ complex variables. $\mathrm{Re}(S)=0$ is dropped when we study $\cs$ instead of $S$. In the space of complex variables, the solutions of $\partial_z\cs=0$ are called the \emph{complex critical points}, which play the dominant role for the asymptotics of $A(\ck)$ in the case that the real critical point is absent.

Before discussing the complex critical point, let us firstly review some known results from the critical equations \eqref{eom1} and \eqref{eom2} with the boundary data corresponding to Regge geometry on $\partial\ck$. The real solutions of the part \eqref{eom1} have been well-studied in the literature \cite{Han:2013gna,Han:2011re,Barrett:2009mw,Conrady:2008mk}. We call these solutions the \emph{pseudo-critical points}. As one of the results, the pseudo-critical point satisfying a nondegeneracy condition endows a Regge geometry on $\ck$ with certain 4-simplex orientations. When focusing on the pseudo-critical points endowing the uniform orientations to all 4-simplices, further imposing \eqref{eom2} to them gives the accidental flatness constraint to their corresponding Regge geometries, i.e., every deficit angle $\delta_h$ hinged by the internal triangle $h$ \cite{Bonzom:2009hw,Han:2013hna} satisfies:
\be
\g \delta_h= 4\pi k_h, \qquad k_h\in\Z. \label{flat}
\ee
When $k_h=0$, $\delta_h$ at every internal triangle is zero, and the Regge geometry endowed by the real critical point is flat. Eq.\eqref{flat} is a strong constraint to the allowed geometry from the spinfoams and can be satisfied only for special boundary conditions that admit the flat bulk geometry (mod $4\pi \Z$). The accidental flatness constraint is consistent with the above argument about over-constrained equations, and it has been demonstrated explicitly in the example well-studied in, e.g., \cite{Han:2021kll,Engle:2020ffj}. If one only considers the real critical point for the dominant contribution to $A(\ck)$, Eq.\eqref{flat} would imply that only the flat geometry (mod $4\pi \Z$) exists. This confusion leading to the flatness problem results from ignoring the complex critical point in the stationary phase analysis.

In the following discussion, we show that the large-$\l$ spinfoam amplitude does receive dominant contributions from the complex critical points away from the real integration domain. The complex critical points precisely correspond to the curved Regge geometries emergent from the spinfoam amplitude. Interestingly, the application of complex critical points leads to a derivation of effective dynamics of Regge geometry from the spinfoam amplitude. The emergent curved Regge geometries are constrained by the effective dynamics. We firstly provide a general formalism below, then we apply the formalism to the concrete models with several different $\ck$ in the following sections.

Motivated by relating to the dynamics of Regge geometry, we separate the integral in the amplitude \eqref{integralFormAmp} into two parts. Suppose $\ck$ has $M$ internal segments, the dynamics of Regge geometry should relate to the dynamics of these internal segment-lengths. Motivated by this, we separate $M$ internal areas $j_{h_o}$ ($h_o=1,\cdots,M$) from other $j_{\bar{h}}$ ($\bar{h}=1,\cdots,F-M$), where $j_{h_o}$ relates to the segment-lengths. Here, $F$ is the total number of internal triangles in $\ck$, and $M$ equals the number of the separated internal segments. 
The spinfoam amplitude \eqref{integralFormAmp} then becomes
\be
A(\ck)&=&\sum_{\{k_h\}}\int\prod_{h_o=1}^M \rmd j_{h_o} \cz^{\{k_h\}}_\ck\lt(j_{h_o}\rt),\label{integralFormAmp01}
\ee
where $\cz_\ck^{\{k_h\}}$, called the partial amplitude, is given by 
\be
\cz_\ck^{\{k_h\}}\lt(j_{h_o}\rt)=\int \prod_{\bar{h}}\mathrm{d} j_{\bar h}\, \prod_h\lt(2\l \bm{d}_{\l j_{h}}\rt)\int [\rmd g\rmd \mathbf{z}] e^{\lambda S^{(k)}}.\label{cz0000}
\ee
We can then change variables from the areas $j_{h_o}$ to the internal segment-lengths $\{l_I\}_{I=1}^M$, with $I$ denoting the internal segment. The internal triangles $h_o=1,\cdots,M$ are suitably chosen such that the change of variables is well-defined in the interested region, e.g. a neighborhood of $\{\mathring{j}_{h_o}\}$ of $\{\mathring{j}_h,\mathring{g}_{ve},\mathring{\bf z}_{vf}\}$ corresponding to the flat geometry. Indeed, the chosen $M$ areas $\{j_{h_o}\}$ are related to $M$ segment-lengths $\{l_I\}$ by Heron's formula. Inverting the relation between $\{j_{h_o}\}_{h_o=1}^M$ and $\{l_I\}_{I=1}^M$ defines the local change of variables $(j_{h_o},j_{\bar{h}})\to (l_I,j_{\bar{h}})$ in a neighborhood $K$ of a given Regge geometry in the integration domain of \eqref{integralFormAmp}. This procedure is just changing variables without imposing any restrictions. When focusing on the integrals in the neighborhood $K$, we have $\rmd^{M+N} j_h=\cj_l\rmd^M l_I\,\rmd^{F-M} j_{\bar{h}}$, where $\cj_l=\det(\partial j_{h_o}/\partial l_I)$ is the jacobian obtained by the derivatives of Heron's formula. Therefore, the contribution to $A(\ck)$ from the neighborhood $K$ is expressed as
\be
&&\sum_{\{k_h\}}\int\prod_{I=1}^M \rmd l_I  \cj_l \cz^{\{k_h\}}_\ck\lt(l_I\rt),\label{integralFormAmp0}
\ee

The partial amplitude $\cz_\ck^{\{k_h\}}$ has the external parameters $r\equiv\{l_I,j_b,\xi_{eb}\}$ including not only the boundary data $j_b,\xi_{eb}$ but also internal segment-lengths $l_I$. The above decomposition of $j_h$-integrals closely relates to the earlier proposal \cite{Han:2017isy,Han:2017xwo} (see also \cite{Dittrich:2022yoo} in the context of area Regge calculus). $l_I$ parametrizes a submanifold $\sm_{Regge}$ in the space of $j_h$. The submanifold $\sm_{Regge}$ collects $j_h$'s that can be interpreted as areas determined by the segment lengths $l_I$ (by Heron's formula). Generically the space of $j_h$ is much larger than the space of segment lengths \cite{Barrett:1997tx}. $j_{\bar{h}}$ parametrizes the direction transverse to $\sm_{Regge}$.

\begin{figure}[h]
    \centering
    \includegraphics[scale=0.25]{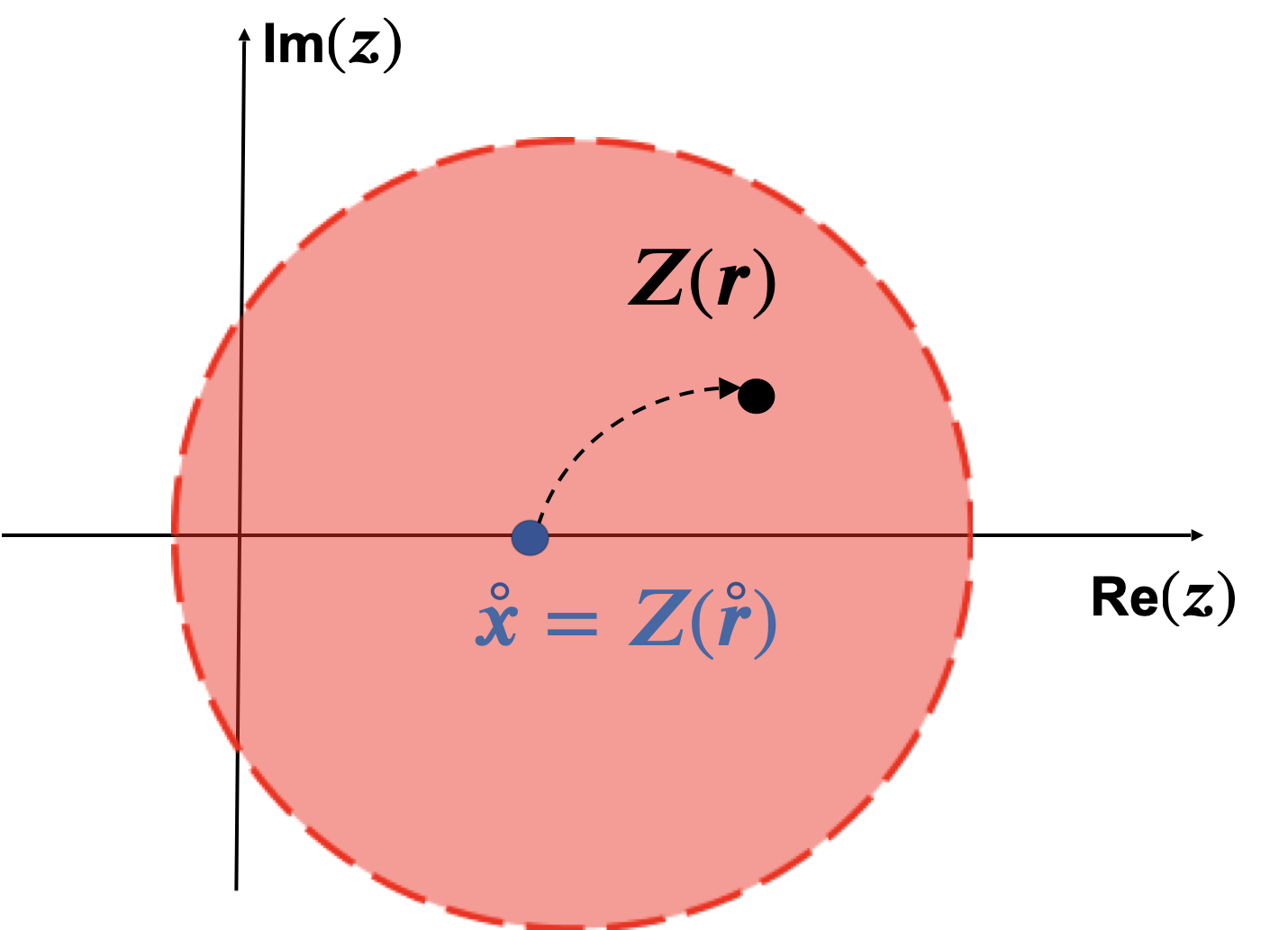}
    \caption[Caption for LOF0]{The real and complex critical points $\mathring{x}$ and $Z(r)$. $\cs(r,z)$ is analytic extended from the real axis to the complex neighborhood illustrated by the red disk.}
    \label{Figure0}
\end{figure}

To study the partial amplitude $\cz_\ck^{\{k_h\}}$, we apply the theory of stationary phase approximation for complex action with parameters \cite{10.1007/BFb0074195,Hormander}. In the following, we only consider the partial amplitude with $k_h=0$, while the situation with other $k_h$ can be studied analogously. We consider the large-$\l$ integral $\int_K e^{\l S(r,x)}\rmd^N x$, and regard $r$ as the external parameters. $S(r,x)$ is an analytic function of $r\in U\subset \R^k,x\in K\subset \R^N$. $U\times K$ is a neighborhood of $(\mathring{r},\mathring{x})$, where $\mathring{x}$ is a real critical point of $S(\mathring{r},x)$. $\mathcal{S}(r,z)$ with $z=x+i y \in \mathbb{C}^{N}$ is the analytic extension of $S(r,x)$ to a complex neighborhood of $\mathring{x}$. The complex critical equation is
\be 
\partial_{z} \mathcal{S}=0,\label{criticaleqn111}
\ee 
whose solution is ${z}=Z(r)$. Here, $Z(r)$ is an analytic function of $r$ in the neighborhood $U$. When $r=\mathring{r}$, $Z(\mathring{r})=\mathring{x}$ reduces to the real critical point. When $r$ deviates away from $\mathring{r}$, $Z(r)\in\C^N$ can move away from the real plane $\R^N$, thus is called the \emph{complex critical point} (see Figure. \ref{Figure0}).  
With this in mind, we have the following large-$\l$ asymptotic expansion for the integral
\be
\int_K e^{\lambda S(r,x)}  \mathrm{d}^N x = \left(\frac{1}{\lambda}\right)^{\frac{N}{2}} \frac{e^{\lambda \mathcal{S}(r,Z(r))}}{\sqrt{\det\left(-\partial^2_{z,z}\mathcal{S}(r,Z(r))/2\pi\right)}} \lt[1+O(1/\l)\rt]
\label{asymptotics0}
\ee
where $\cs(r,Z(r))$ and $\delta^2_{z,z}\mathcal{S}(r,Z(r))$ are the action and Hessian at the complex critical point. In addition, the real part of $\cs$ is zero or negative. More precisely, there exists a constant $C>0$ such that
\be
\operatorname{Re}(\mathcal{S}) \leq-C|\operatorname{Im}(Z)|^{2}.\label{negativeReS}
\ee
See \cite{10.1007/BFb0074195,Hormander} for the proof of this inequality. This inequality indicates that $\operatorname{Re}(\mathcal{S})=0$ resulting in the oscillatory phase in \eqref{asymptotics0} can only happen at the real critical point, where $\operatorname{Im}(Z)=0$ and $r=\mathring{r}$. When $r$ deviates from $\mathring{r}$ with a finite distance, such that $\operatorname{Im}(Z)$ is finite and $\operatorname{Re}(\mathcal{S})$ is negative, \eqref{asymptotics0} is exponentially suppressed when scaling $\lambda$ to large. The asymptotic formula (\ref{asymptotics0}) depends analytically on $r$ and interpolates the two different behaviors smoothly in the parameter space of $r$:
\begin{itemize}
	\item The critical point is not real, then $\operatorname{Re}(\mathcal{S})<0$, which gives the exponentially decaying amplitude.
	\item The critical point is real, then $\operatorname{Re}(\mathcal{S})=0$, and thus $e^{\lambda \cs}$ gives an oscillatory phase.
\end{itemize}
These two distinct behaviors are obtained by fixing $r$ and scaling $\lambda$. But since the asymptotic formula \eqref{asymptotics0} depends on $r$ analytically, we can vary $r$ simultaneously as scaling $\l$. Then we can arrive at the regime where the asymptotic behavior \eqref{asymptotics0} is not suppressed at the complex critical point. Indeed, for any large $\l$, there always exists $r\neq \mathring{r}$ but sufficiently close to $\mathring{r}$,  such that $\operatorname{Im}(Z)$ and $\mathrm{Re}(\cs)$ are small enough, then $e^{\lambda \cs}$ in \eqref{asymptotics0} is not suppressed at the complex critical point. 

The importance of \eqref{asymptotics0} is that the integral can receive a dominant contribution from the complex critical point away from the real plane. These complex critical points indeed give the curved Regge geometries missing in the argument of the flatness problem. The parameter $r$ including both the boundary data and internal segment lengths determines the Regge geometry that is generically curved. Hence the asymptotic formula \eqref{asymptotics0} computes the weight of the Regge geometry contributing to the amplitude and reduces $A(\ck)$ in $K$ to
\be
\left(\frac{1}{\lambda}\right)^{\frac{N}{2}}\!\!\!\int\prod_{I=1}^M \rmd l_I \sn_l\,{e^{\lambda \mathcal{S}(r,Z(r))}} \lt[1+O(1/\l)\rt]\label{pathintegral0}
\ee
at each ${k_h}$. Here, $\sn_l\propto\prod_h\lt(4 j_{h}\rt)\cj_l[\det(-\delta^2_{z,z}\cs/2\pi)]^{-1/2}$ at $Z(r)$, and $r=\{l_I,j_b,\xi_{eb}\}$. Given that $\{l_I\}$ determines the Regge geometry on $\ck$, Eq.\eqref{pathintegral0} is a path integral of Regge geometries with the effective action $\cs$. The integration domain of $l_I$ includes curved geometries. The integral \eqref{pathintegral0} derived from the spinfoam amplitude defines an effective theory of Regge geometries. Indeed, if we focus on the dominant contribution and neglect corrections of $O(1/\l)$, by the stationary phase approximation of \eqref{pathintegral0}, the effective action $\cs$ gives the equation of motion
\be
\frac{\partial\cs}{\partial l_I}=0,\quad I=1,\cdots,M,\label{effeqn0}
\ee 
which determines the effective dynamics of Regge geometry. $\cs$ is generally complex, so \eqref{effeqn0} should be analytically continued to complex $l_I$, and thus the solution is generally not real. As we are going to see in Section \ref{Solutions of effective dynamics on}, we are mainly interested in the regime where the imaginary part of the solution $l_I$ is negligible, then the solution has the interpretation of the Regge geometry.

In the following, we make the above general analysis concrete by considering the examples of spinfoam amplitudes on a single 4-simplex and the double-$\Delta_3$ complex. We also revisit briefly the existing results on $\Delta_3$ complex for the completeness. We compute numerically the complex critical points and $\mathcal{S}$, confirming the contribution of the complex critical points to the spinfoam amplitude. In particular, the double-$\Delta_3$ model corresponding to $M=1$ exhibits the non-trivial effective dynamics of the Regge geometries. The effective dynamics approximates the classical Regge calculus in the small-$\g$ regime.

\section{Four-simplex amplitude} \label{Single4Simplex}

\begin{figure}[h]
    \centering
    \includegraphics[width=1\textwidth]{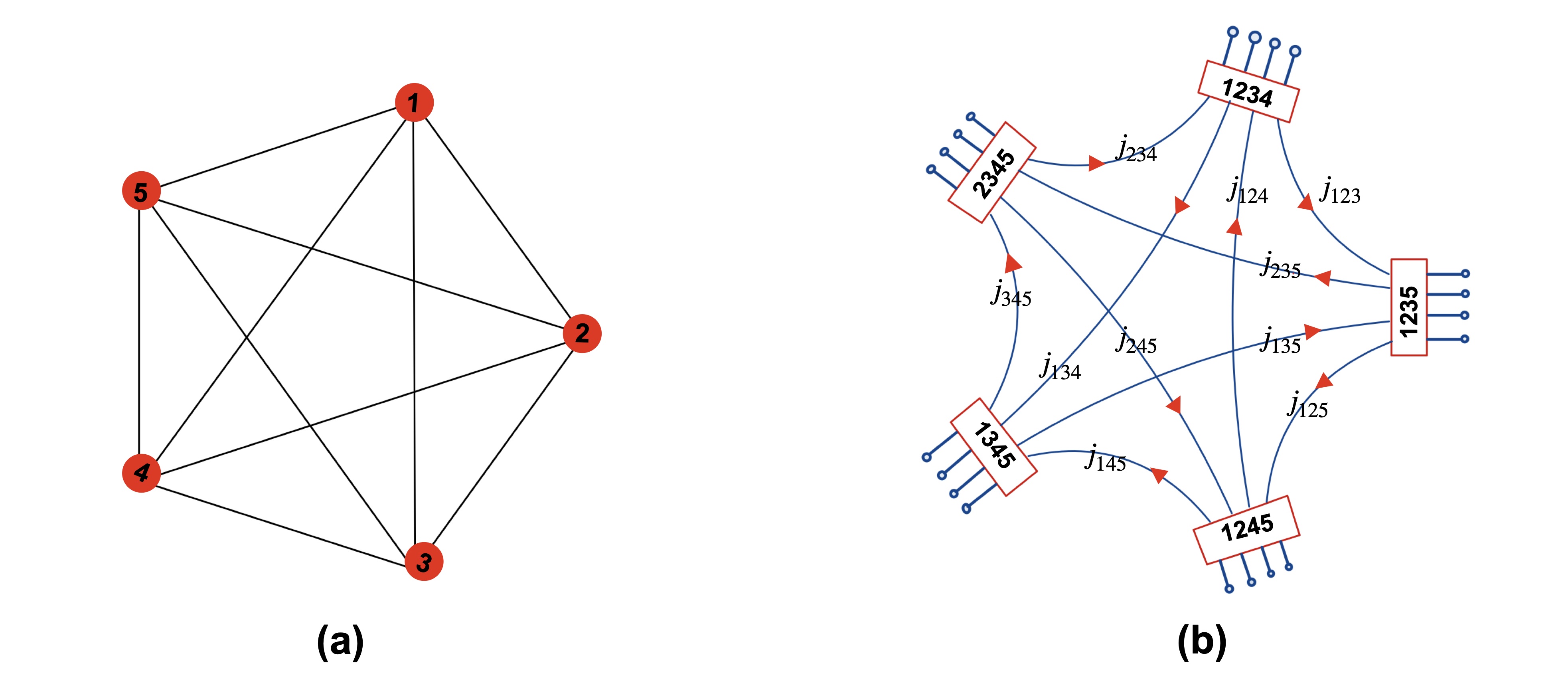}
    \caption[Caption for LOF]{Panel (a) plots the 4-simplex $v=(1,2,3,4,5)$. The boundary comprises five tetrahedra $e_i$ sharing ten faces $f_i$\protect \footnotemark. Panel (b) is the dual complex of the 4-simplex. Five boxes correspond to boundary tetrahedra carrying $g_{ve}\in \Slc$. The strands correspond to triangles carrying spins $j_{f}$. The circles as endpoints of strands carry boundary states $\xi_{ef}$. The arrows represent the orientations of strands. }
    \label{Figure1}
\end{figure}

This section applies the above general procedure to the simplest situation: the 4-simplex amplitude. In this case, there is no internal triangle: $F=M=0$. The external parameter $r$ only contains the boundary data $r=(j_b,\xi_{eb})$. The 4-simplex and its dual diagram are illustrated in Figure \ref{Figure1} (a) and (b). The points of the 4-simplex $v$ are labelled by $(1,2,3,4,5)$. The five tetrahedra on the boundary are labelled by 
\be 
\{e_1,e_2,e_3,e_4,e_5\}=\{(1,2,3,4),(1,2,3,5),(1,2,4,5),(1,3,4,5),(2,3,4,5)\}.\nonumber
\ee
These tetrahedra carry group variable $g_{ve}\in \Slc$. The triangle is shared by the tetrahedra and carries an $\Su$ spin $j_{f}$, e.g., the tetrahedron $e_1=(1,2,3,4)$ and the tetrahedron $e_2=(1,2,3,5)$ share the face $f_1=(1,2,3)$. 
\footnotetext{The shared faces are labelled by $\{f_1,f_2,...,f_{10}\}=\{(1,2,3),(1,2,4),(1,2,5),(1,3,4),(1,3,5),(2,3,4),(2,3,5),(3,4,5)\}$. For convenience, in this section, the notations $e$ and $f$ mean that $e\in\{e_1,...,e_5\}$ and $f\in\{f_1,...,f_{10}\}$.}

\subsection{The amplitude and parametrization of variables}

According to (\ref{amplitude}), the EPRL 4-simplex amplitude with the boundary state has the following expression \cite{Conrady:2008ea,Conrady:2008mk,Barrett:2009gg,Barrett:2009mw,Han:2011re,Han:2011rf}: 
\be 
A_{v}\left(j_{f}, \xi_{e f}\right)=\int \prod_{e} \mathrm{d} g_{ve}\, \delta_{i\sigma_3}\left(g_{ve_1}\right) \int_{\left(\mathbb{C P}^{1}\right)^{10}} e^{S} \prod_{f} \frac{d_{j_{f}}}{\pi} \mathrm{d} \Omega_{{\bf z}_{vf}}. \label{VertexAmp}
\ee 
Here, all triangles are on the boundary, $j_f=j_b$. To fix the $\Slc$ gauge, $g_{ve_1}$ is fixed to be constant matrix $\mathrm{diag}(i,-i)$ (the timelike normal of the reference tetrahedron $e_1$ is past-pointing). The integrand in (\ref{VertexAmp}) is written as an exponential $e^{S}$ with the action
\be 
S=\sum_{f} 2 j_{f} \ln \frac{\left\langle \xi_{ef}, Z_{vef}\right\rangle\left\langle Z_{ve'f}, \xi_{e'f}\right\rangle}{\left\|Z_{vef}\right\|\left\|Z_{ve'f}\right\|}+i \gamma j_{f} \ln \frac{\left\langle Z_{ve'f}, Z_{ve'f}\right\rangle}{\left\langle Z_{vef}, Z_{vef}\right\rangle}. \label{4SimplexAction}
\ee 
The orientations of dual faces follow from Figure \ref{Figure1}(c). To study the large-$j$ behavior of the amplitude, we scale all boundary spins $j_{f}\rightarrow\lambda j_{f}$ by the parameter $\lambda\gg 1$. The scaling of spins results in the scaling of action $S\mapsto\lambda S$, such that the integral \eqref{VertexAmp} can be studied by the stationary phase approximation. In the following, we firstly compute the real critical point $\{\mathring{g}_{ve},\mathring{z}_{vf}\}$, which is the solution of the critical equation (\ref{eom1}) and then describe the algorithm to compute the complex critical point in the neighborhood. 

To obtain the real critical point, we adopt the 4-simplex geometry used in \cite{Han:2020npv,Han:2020fil,Dona:2019dkf} to generate the boundary state. The coordinates of the five vertices $P_a$ in Figure \ref{Figure1}(a) in the Minkowski spacetime are set as
\be \label{P1P5}
\begin{gathered}
	P_{1}=(0,0,0,0), P_{2}=\left(0,0,0,-2 \sqrt{5} / 3^{1 / 4}\right), P_{3}=\left(0,0,-3^{1 / 4} \sqrt{5},-3^{1 / 4} \sqrt{5}\right) \\
	P_{4}=\left(0,-2 \sqrt{10} / 3^{3 / 4},-\sqrt{5} / 3^{3 / 4},-\sqrt{5} / 3^{1 / 4}\right) \\
	P_{5}=\left(-3^{-1 / 4} 10^{-1 / 2},-\sqrt{5 / 2} / 3^{3 / 4},-\sqrt{5} / 3^{3 / 4},-\sqrt{5} / 3^{1 / 4}\right)
\end{gathered}
\ee 
Then, the 4-d normals of the tetrahedra are
\be 
\begin{gathered}
	N_{e_1}=(-1,0,0,0), N_{e_2}=\left(\frac{5}{\sqrt{22}}, \sqrt{\frac{3}{22}}, 0,0\right), N_{e_3}=\left(\frac{5}{\sqrt{22}},-\frac{1}{\sqrt{66}}, \frac{2}{\sqrt{33}}, 0\right) \\
	N_{e_4}=\left(\frac{5}{\sqrt{22}},-\frac{1}{\sqrt{66}},-\frac{1}{\sqrt{33}}, \frac{1}{\sqrt{11}}\right), N_{e_5}=\left(\frac{5}{\sqrt{22}},-\frac{1}{\sqrt{66}},-\frac{1}{\sqrt{33}},-\frac{1}{\sqrt{11}}\right).
\end{gathered}
\ee 
The spinor $\xi_{ef}$ relates to the 3d normals $n_{ef}$ by $n_{ef}=\left\langle\xi_{ef}, \vec{\sigma} \xi_{ef}\right\rangle$ ($\vec{\sigma}$ are Pauli matrices). The Regge boundary data of ten areas $\mathring{j}_{f}$, 3d normals $\mathring{n}_{ef}$ and the corresponding spinors $\mathring{\xi}_{ef}$ of the 4-simplex are listed in Appendix \ref{Appendix4Simplex}.

With the Lorentzian Regge boundary data $\mathring{r}=(\mathring{j}_f,\mathring{\xi}_{ef})$, we solve for the \emph{real critical point} $(\mathring{g}_{ve},\mathring{\bf z}_{vf})$ which satisfies $\re(S)=\partial_{g_{ve}} S=\partial_{z_{vf}}S=0$. The results in the literature \cite{Barrett:2009mw,Han:2011re} show that there are exactly 2 real critical points, which have the interpretations as the geometrical 4-simplex with opposite 4-orientations. The 4-simplex geometrical interpretation of the critical points results in the same geometry as the one given by \eqref{P1P5}. We compute the real critical point following the strategy described in \cite{Han:2021kll, Dona:2019dkf, Han:2021rjo}, where the boundary data and critical points for a single 4-simplex are studied in detail. The data of the real critical point  $(\mathring{g}_{ve},\mathring{\bf z}_{vf})$ is given in Appendix \ref{Appendix4Simplex}. 

By fixing the re-scaling gauge of ${\bf z}_{vf}$, each ${\bf z}_{vf}$ can be parameterized with two real variables $x_{v f}, y_{v f}$:
\be 
{\bf z}_{vf}=\left(1, x_{v f}+\mathrm{i} y_{v f}\right)^{\mathrm{T}}.\label{para0}
\ee 
$g_{ve_{i}}, i= (2,3,4,5)$ are parameterized as 
\be 
\left(\begin{array}{cc}
	1+\left(x^{1}_{ve}+\mathrm{i} y^{1}_{ve}\right) / \sqrt{2} & \left(x^{2}_{ve}+\mathrm{i} y^{2}_{ve}\right) / \sqrt{2} \\
	\left(x^{3}_{ve}+\mathrm{i} y^{3}_{ve}\right) / \sqrt{2} & \frac{1+\left(x^{2}_{ve}+\mathrm{i} y^{2}_{ve}\right)\left(x^{3}_{ve}+\mathrm{i} y^{3}_{ve}\right) / 2}{1+\left(x^{1}_{ve}+\mathrm{i} y^{1}_{ve}\right) / \sqrt{2}}
\end{array}\right), \quad x^{1}_{ve}, y^{1}_{ve}, x^{2}_{ve}, y^{2}_{ve}, x^{3}_{ve}, y^{3}_{ve} \in \mathbb{R}.\label{para1}
\ee 
Therefore, the 4-simplex action is a function in terms of all real variables $x=(x_{vf},y_{vf},x^{1}_{ve}, y^{1}_{ve}, x^{2}_{ve}, y^{2}_{ve}, x^{3}_{ve}, y^{3}_{ve})$ for all $f$ in $\{f_1,...f_{10}\}$ and $e$ in $\{e_2,..e_5 \}$. The real critical point $\mathring{\bf z}_{vf}$ is in the form $\mathring{\bf z}_{vf}=\left(1, \mathring{\a}_{vf}\right)^T$, where $\mathring{\a}_{vf} =\mathring{x}_{vf}+ \mathrm{i}  \mathring{y}_{vf} \in\C$. It is convenient to set one of the critical points at the origin $\mathring{x}=\{0,0,...,0\}$ by modifying (\ref{para0}) and (\ref{para1}) to 
\be 
{\bf z}_{vf}&=&\left(1, \mathring{\alpha}_{vf}+x_{v f}+\mathrm{i} y_{v f}\right)^{\mathrm{T}},\nonumber\\
g_{ve}&=&\mathring{g}_{ve}\left(\begin{array}{cc}
	1+\left(x^{1}_{ve}+\mathrm{i} y^{1}_{ve}\right) / \sqrt{2} & \left(x^{2}_{ve}+\mathrm{i} y^{2}_{ve}\right) / \sqrt{2} \\
	\left(x^{3}_{ve}+\mathrm{i} y^{3}_{ve}\right) / \sqrt{2} & \frac{1+\left(x^{2}_{ve}+\mathrm{i} y^{2}_{ve}\right)\left(x^{3}_{ve}+\mathrm{i} y^{3}_{ve}\right) / 2}{1+\left(x^{1}_{ve}+\mathrm{i} y^{1}_{ve}\right) / \sqrt{2}}
\end{array}\right). \label{para2}
\ee
With the parameterization  in (\ref{para2}), the measures $\rmd g_{ve}$ and $\mathrm{d}\O_{\mathbf{z}_{v f}}$ are
\be 
\begin{gathered}
	\rmd g_{ve}=\frac{1}{128 \pi^{4}} \frac{\rmd x^{1}_{ve} \rmd x^{2}_{ve} \rmd x^{3}_{ve} \rmd y^{1}_{ve} \rmd y^{2}_{ve} \rmd y^{3}_{ve}}{\left|1+\frac{x^{1}_{ve}+\mathrm{i} y^{1}_{ve}}{\sqrt{2}}\right|^{2}}, \\
	\mathrm{d}\O_{\mathbf{z}_{v f}} =\frac{\mathrm{d} x_{v f} \mathrm{~d} y_{v f}}{\left\langle Z_{v e f}, Z_{v e f}\right\rangle\left\langle Z_{v e^{\prime} f}, Z_{v e^{\prime} f}\right\rangle} .
\end{gathered}
\ee 
As a result, the 4-simplex amplitude is in the form
\be 
A_v=\int \mathrm{d}^{44}x \,\mu(x) \,e^{\lambda S(r, x)},
\ee 
where $r=(j_f,\xi_{ef})$ are boundary data. The integral is 44 real-dimensional. In the following, we focus on a neighborhood $K$ of $\mathring{x}$. We have defined the local coordinates $x\in\mathbb{R}^{44}$ covering $K$.

\subsection{Deviating from the shape-matching}\label{Deviating from the shape-matching}


The amplitude $A_v$ has the real critical points with the non-degenerate Regge boundary data $\mathring{r}$. However, the real critical point disappears when the boundary data deviates away from $\mathring{r}$. Considering a neighborhood $U$ of $\mathring{r}$ in the space of boundary data, such that any $r\in U$ (different from $\mathring{r}$) does not correspond to any Regge geometry or vector geometry\footnote{In the Lorentzian EPRL spinfoam amplitude, the critical points corresponding to the non-degenerate Regge geometry are isolated critical points.}. If we fix $r\in U$ and scale the spins with a large $\lambda$, there are two possible behaviors for the amplitude \cite{Barrett:2009gg,Barrett:2009mw}
\begin{itemize}
	\item For $r=\mathring{r}$, the amplitude has two critical points whose geometrical interpretations have opposite orientations. $S$ evaluated at critical points gives the Regge action of the 4-simplex with opposite sign. Therefore, the asymptotic amplitude of the 4-simplex gives two oscillatory phases 
	\be 
	A_{v} \simeq \lambda^{-12}\left(N_{+} e^{i \lambda S_{\text {Regge }}}+N_{-} e^{-i \lambda S_{\text {Regge }}}\right).  \label{eom0}
	\ee

	\item For $r\neq \mathring{r}$, it leads to no solutions to (\ref{eom1}) and the exponentially suppressed amplitude. 
\end{itemize}
To interpolate smoothly between the oscillatory phases and the exponential suppression in the asymptotics \eqref{eom0}, the discussion in section \ref{Real critical points and complex critical points} suggests making $r$ vary and introducing the complex critical points. 

The boundary data $\mathring{r}=\{\mathring{j}_{f},\mathring{\xi}_{ef}\}$ of the Lorentzian Regge geometry satisfies the shape-matching condition, i.e., five geometrical tetrahedra determined by $\mathring{r}$ on the boundary are glued with the triangles matching in shapes. Consider the 4-simplex action $S(r,x)$ in the neighborhood $K\times U$ of $(\mathring{r},\mathring{x})$. We define $z\in\C^{44}$ as the complexification of $x$, and $\cs(r,z)$ extends holomorphically $S(r,x)$ to a complex neighborhood of $\mathring{x}$. To avoid confusion, we note that the integration variables $x$ are complexified, while the boundary data $r=(j_f,\xi_{ef})$ is real.

Next, we let $r=\mathring{r}+\delta r$ vary, such that the shape-matching condition violates. We describe below a parametrization of the tetrahedron shapes. A tetrahedron in $\mathbb{R}^3$ is determined by $4$ points $\{\tilde{P}_a,\tilde{P}_b,\tilde{P}_c,\tilde{P}_d \}$ up to a $\mathbb{R}^3 \rtimes \text{O}(3)$ symmetry. We gauge fix the $\mathbb{R}^3 \rtimes \text{O}(3)$ symmetry by choosing $\tilde{P}_a$ at the origin, $\tilde{P}_b$ along the $z$ axis, and $\tilde{P}_c$ within the $(y,z)$-plane. The last point $\tilde{P}_d$ is not constrained. Given the tetrahedron's segment lengths, the coordinates of the points are fixed in $\mathbb{R}^3$ by the above gauge fixing. For example, for the tetrahedron $e_2=\{1,2,3,5\}$, $\mathring{r}$ implies that the coordinates of the points in $\mathbb{R}^3$ are given by
\be
&&\tilde{P}_1=(0,0,0),\quad \tilde{P}_2=(0,0,-3.40),\quad \tilde{P}_3=(0,-2.94, -1.70),\nonumber\\
&&\tilde{P}_5=(-0.651, -0.981, -1.70). \label{3dvertices}
\ee 
All other four tetrahedra can be described similarly, and the coordinates of the points in $\mathbb{R}^3$ are determined by $\mathring{r}$. The 3d face-normals $\vec{n}$ implied by the coordinates match with the data in Table \ref{4Simplex3Dnormals} up to a simultaneous SO(3) rotation. The spinors $\xi$ associating with each face are given by 
\be 
\xi=\frac{1}{\sqrt{2}}\lt(\sqrt{1+w},\frac{x+\mathrm{i} y}{\sqrt{1+w}}\rt)^{\mathrm{T}},  \quad \text{if }\,\vec{n}=(x,y,w)^{\mathrm{T}}. \label{newxi}
\ee 

When we deform the boundary data, we keep the areas $j_f=\mathring{j}_f$ unchanged, while $\xi_{ef}$ are deformed, such that the boundary data $r$ is deformed to violate the shape-matching condition. We move the vertices $\tilde{P}_a\in\R^3$ to deform the tetrahedron shapes. For example, the vertices in (\ref{3dvertices}) are moved to new positions
\be
	 &&\tilde{P}_1=(0,0,0),\quad  \tilde{P}_2=(0,0,-3.40+\delta w^{(2)}_{2}),\quad \tilde{P}_3=(0,-2.94+\delta y^{(2)}_3, -1.70+\delta w^{(2)}_3),\nonumber\\
	 &&\tilde{P}_5=(-0.651+\delta x^{(2)}_5, -0.981+\delta y^{(2)}_5, -1.70+\delta w^{(2)}_5). \label{newcoord}
\ee 
In the notations $\delta x_i^{(a)},\delta y_i^{(a)}$,$\delta w_i^{(a)}$, $a=1,\cdots,5$ labels the tetrahedron, and $i=1,\cdots,5$ labels the variables associated to the vertex $\tilde{P}_i$. There are 30 variables $\delta x_i^{(a)},\delta y_i^{(a)}$,$\delta w_i^{(a)}$ in total. We keep the face areas unchanged. Then in each tetrahedron, Heron's formula gives 4 constraint equations, each corresponding to a face area. For example, in the tetrahedron $e_{2}=\{1,2,3,5\}$, the equations are 
\be
\left\{
\begin{array}{c}
	A_{123}(\delta w^{(2)}_2, \delta y^{(2)}_3,\delta w^{(2)}_3)=5\\
	A_{125}(\delta w^{(2)}_2, \delta x^{(2)}_5, \delta y^{(2)}_5, \delta w^{(2)}_5)=2\\
	A_{135}(\delta y^{(2)}_3,\delta w^{(2)}_3, \delta x^{(2)}_5,\delta y^{(2)}_5, \delta w^{(2)}_5)=2\\
	A_{235}(\delta w^{(2)}_2, \delta y^{(2)}_3,\delta w^{(2)}_3, \delta x^{(2)}_5,\delta y^{(2)}_5, \delta w^{(2)}_5)=2. \label{areasfunc}
\end{array}
\right. 
\ee 
At least in a neighborhood of the deformation, $\delta w^{(2)}_2, \delta y^{(2)}_3,\delta w^{(2)}_3, \delta x^{(2)}_5$ can be solved in terms of $\delta y^{(2)}_5, \delta w^{(2)}_5$ from \eqref{areasfunc}. The shape of the tetrahedron is parameterized by 2 variables $\delta y^{(2)}_5, \delta w^{(2)}_5$. This way of parametrization is convenient in our computation. However, it is different from the known strategy, such as the Kapovich-Millson phase space \cite{10.4310/jdg/1214459218} or using dihedral angles \cite{Rovelli:2006fw}. For each tetrahedron, we adopt the same strategy. We have in total ten variables $\mathrm{B}\equiv (\delta y^{(1)}_4,\delta w^{(1)}_4, \delta y^{(2)}_5,\delta w^{(2)}_5,\delta y^{(3)}_5, \delta w^{(3)}_5,\delta y^{(4)}_5, \delta w^{(4)}_5, \delta w^{(5)}_5,\delta w^{(5)}_5)$ to parameterize the deformation of five tetrahedra. The spinors $\xi_{ef}$ of each face can be expressed in terms of $\mathrm{B}$ according to (\ref{newxi}). At this point, the boundary data is $r({\rm B})=(j_f,\xi_{ef}(\mathrm{B}))$. We insert $r({\rm B})$ into the action $S(r({\rm B}),x)$ in (\ref{4SimplexAction}), whose analytical extension is $\mathcal{S}(r({\rm B}),z)$. Then, the complex critical equations are $F({\rm B},z)={\partial}_z \cs(r({\rm B}),z)=0$, from which we solve for the complex critical point $z({\rm B})$.

The asymptotics of the 4-simplex amplitude with the boundary data violating the shape-matching condition is given by \eqref{asymptotics0}. Here, the complex critical point $z({\rm B})$ inserting into the analytic continued action gives $\cs(r({\rm B}),z({\rm B}))$. In contrast to the Regge action obtained from spinfoam asymptotics in \cite{Barrett:2009mw}, $\cs(r({\rm B}),z({\rm B}))$ is an action of the twisted geometry\cite{Freidel:2010aq}. \footnote{The condition for shape matching differs from the shape matching condition discussed in \cite{Freidel:2010aq}. In their work, Freidel et al. \cite{Freidel:2010aq} introduced an additional angle variable as a degree of freedom in twisted geometry, which is canonically conjugate to the area variable. While these two conditions share an intuitive similarity, they are not precisely identical.} Indeed, $\cs(r({\rm B}),z({\rm B}))$ depends on the degrees of freedom of semiclassical tetrahedra, which are not constrained by the shape-matching condition. These degrees of freedom are beyond the Regge geometry and belong to the twisted geometry of the boundary.

\begin{figure}[h]
	\centering
	\includegraphics[scale=0.3]{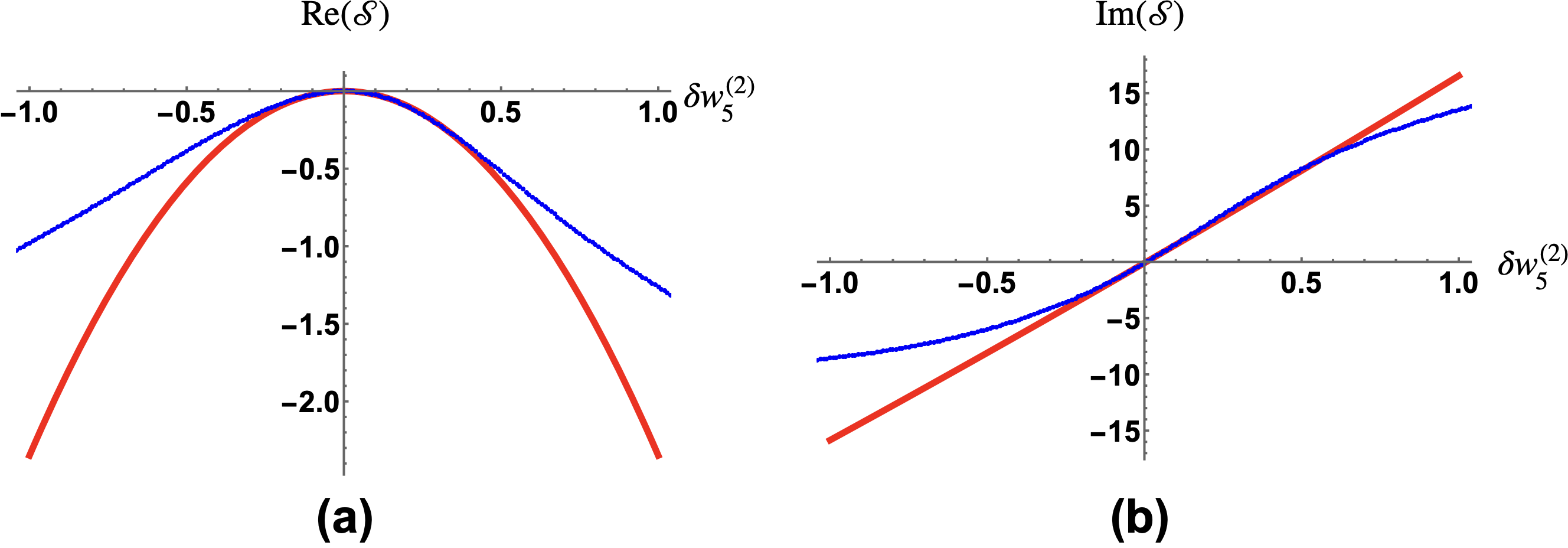}
	\caption{In both panels, the blue curves are the numerical results with the Newton-Raphson method, and the red curves are the results from the linear solution. Panel (a) is the real part of the analytically continued action $\mathcal{S}$ at the complex critical points varying with $\delta w^{(2)}_5$. Panel (b) is the imaginary part of $\mathcal{S}$ at the complex critical points varying with $\delta w^{(2)}_5$. The range of $\delta w^{(2)}_5$ is [-1,1].}
	\label{Figure2}
\end{figure}

To solve the complex critical point, we can linearize \eqref{areasfunc} and obtain the linear solution $(\delta w^{(2)}_2, \delta y^{(2)}_3,\delta w^{(2)}_3, \delta x^{(2)}_5)$ in terms of $\delta y^{(2)}_5, \delta w^{(2)}_5$. We can also linearize the complex critical equation at ${\rm B}=(0,\cdots,0)$, and then solve for the complex critical point $z=z^{({\rm lin})}({\rm B})$. The solution $z^{({\rm lin})}({\rm B})$ is a linear function of the perturbations ${\rm B}$. The coefficients in the linear function can be computed numerically. Inserting this linear solution into the action, we obtain $\mathcal{S}(r(\mathrm{B}),z^{({\rm lin})}({\rm B}))$ as a function of $\mathrm{B}$ and expand it to the second order: 
\be
\mathcal{S}(r(\mathrm{B}),z^{({\rm lin})}({\rm B}))=\Fq_{ij}{\rm B}^i{\rm B}^j+\Fl_j{\rm B}^j+\cs_0
\ee
where the coefficients $\Fq_{ij},\Fl_j$ can be computed numerically. $\cs_0$ is the spinfoam action evaluated at the real critical point with ${\rm B}=(0,\cdots,0)$. In Figure \ref{Figure2}, we let $\mathrm{B}=(0,0,0,\delta w^{(2)}_5,0,0,0,0,0,0)$, the red curves in (a) and (b) are the real part and imaginary part of $\mathcal{S}(r(\mathrm{B}),z^{({\rm lin})}({\rm B}))$ with $\delta w^{(2)}_5$ varying from -1 to 1. 

The linear solution may have a large error when components in ${\rm B}$ are large. We apply the Newton-Raphson method to numerically search for the solution, which is more accurate than the linear solution. To compare with the linear solution in Figure \ref{Figure2}, we still only focus on the deformation of $e_2=\{1,2,3,5\}$ and set $\delta y_5^{(2)}=0$. We outline the procedure in the following. 

For any given $\delta w^{(2)}_5$, we can numerically solve equations (\ref{areasfunc}) for $(\delta w^{(a)}_2, \delta y^{(a)}_3,\delta w^{(a)}_3, \delta x^{(a)}_5)$. There are multiple solutions. We select the solution that is within a neighborhood at $(0,0,0,0)$, by requiring $|\delta w_2^2+ \delta y_3^2+\delta w_3^2+ \delta x_5^2|\leq 4|\delta w^2_5|$. The coordinates in (\ref{newcoord}) given by the solution result in the 3d face normal vectors $\vec{n}$ and spinors $\xi$, which are the boundary data $r$ violating the shape-matching condition.


We apply the Newton-Raphson method to search for the complex critical point satisfying $\partial_{z} \cs=0$. An outline of the procedure in the Newton-Raphson method is given in Appendix \ref{The Newton-Raphson method}. In Figure \ref{Figure2}, the blue curves in (a) and (b) are the real part and imaginary part of the analytically continued action at the complex critical points. This numerical result (blue curves) and the result from the linear solution (red curves) are close when the deformation is small. However, the linear solution is less accurate when the deformation is large. 

Figure \ref{Figure2} demonstrates the smooth interpolation between the oscillatory and exponential suppression behaviors mentioned at the beginning of this subsection. In addition to scaling large $\l$, we need to consider the smooth deformation ${\rm B}$. For any given $\lambda$, there exists sufficiently small deformation ${\rm B}$ beyond the shape-matching, such that $\mathrm{Re}(\cs)$ is small, and thus the amplitude is not suppressed.

\section{Revisit the $\Delta_3$ amplitude}\label{Revisit the amplitude}

In this section, we revisit briefly the existing result on the spinfoam amplitude on the $\Delta_3$ complex, for the completeness and preparing the discussion of the double-$\Delta_3$ complex in the next section. The $\Delta_3$ complex contains a single internal face $F=1$ but has no internal segment $M=0$. There is an internal $j_h$ that is an integrated variable in the amplitude $A(\Delta_3)$ in \eqref{integralFormAmp}.

\begin{figure}[h]
	\centering
	\includegraphics[width=1\textwidth]{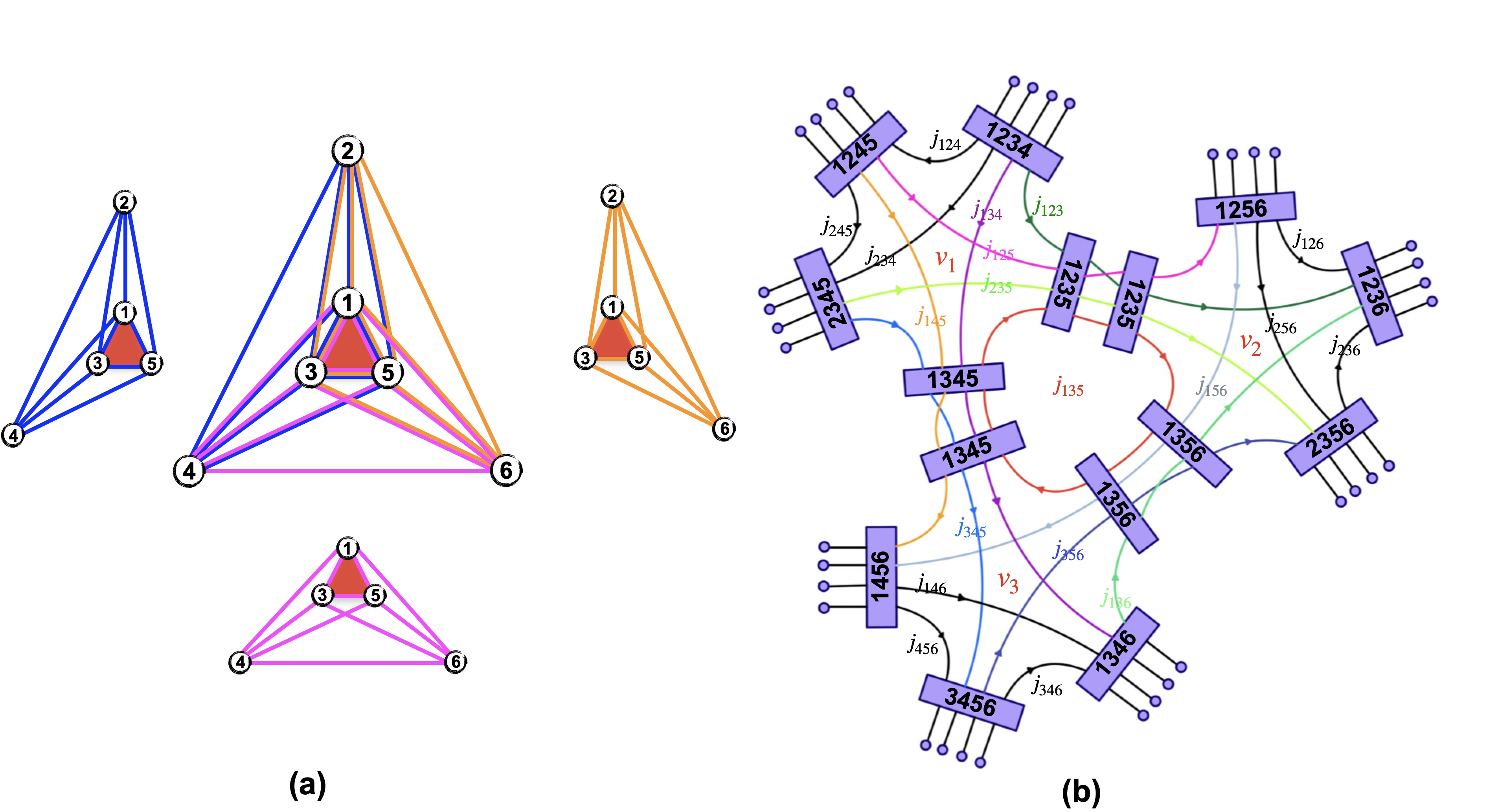}
	\caption[Caption for LOF]{Panel (a) illustrates the simplicial complex $\Delta_3$ made by three 4-simplices $\{v_1,v_2,v_3\}$ and 12 tetrahedra $e_i$ sharing nineteen faces $f_i$. There are eighteen boundary faces and one internal face. Panel (b) is the dual cable diagram of the $\Delta_3$ spinfoam amplitude: The boxes correspond to tetrahedra carrying $g_{ve} \in \Slc$. The strands stand for triangles carrying spins $j_{f}$. The strand with the same color belonging to a different dual vertex corresponds to the triangle shared by the different 4-simplices. The circles as the endpoints of the strands carry boundary states $|j_b,\xi_{eb}\rangle$. The arrows represent orientations. This figure is adapted from \cite{Dona:2020yao}.}	\label{Figure3}
\end{figure}

The $\Delta_3$ complex and its dual cable diagram are represented in Figure \ref{Figure3}. All tetrahedra and triangles are spacelike. The Regge geometry on $\Delta_3$ is completely fixed by the Regge boundary data $\{j_b,\xi_{eb}\}$ that is determined by the boundary segment lengths. In this section, we only focus on the Regge boundary data, in contrast to the discussion of 4-simplex amplitude in the previous section. The generalization to non-Regge boundary data should be straightforward. In terms of the notations in Section \ref{Real critical points and complex critical points}, we have $r=\{j_b,\xi_{eb}\}$ as the boundary data. $\mathring{r}=\{\mathring{j}_b,\mathring{\xi}_{eb}\}$ fixes the flat geometry $\mathbf{g}(\mathring{r})$ with deficit angle $\delta_h=0$. $\mathring{x}=\{\mathring{j}_h,\mathring{g}_{ve},\mathring{\bf z}_{vf}\}$ is the real critical point associated to $\mathring{r}$. The data $\mathring{r}$, $\mathbf{g}(\mathring{r})$, and $\mathring{x}$ are computed numerically in \cite{Han:2021kll}.

According to the general spinfoam amplitude (\ref{integralFormAmp}) and the spinfoam action (\ref{Stot}), the $\Delta_3$ amplitude $A(\Delta_3)$ can be written as
\be
\begin{aligned}
	A\left(\Delta_{3}\right) &=\sum_{k_{h} \in \mathbb{Z}} 2 \lambda \int \mathrm{d} j_{h} d_{\lambda j_{h}} \int [\mathrm{d} g \mathrm{d} \mathbf{z}] e^{\lambda S^{(k)}}, \\
	S^{(k)} &=S+4 \pi i \sum_{h} j_{h} k_{h}. \label{delta3Amp}
\end{aligned} 
\ee  
For each ${k_h}$ in \eqref{delta3Amp},  the \textit{real critical point} $\{\mathring{j}_h,\mathring{g}_{ve},\mathring{\bf z}_{vf}\}$ happens only when the boundary data satisfies the accidental flatness constraint \eqref{flat}. 

Given the boundary data $\mathring{r}$ corresponding to $\delta_h=0$, we consider its neighborhood $U$ in the space of the non-degenerate Regge boundary data, such that any boundary data $r\in U$ satisfies $|\g \delta_h|< 4\pi$. For large $\l$, the sectors with $k_h\neq 0$ do not give dominant contribution to $A(\Delta_3)$ as far as $r\in U$. If we arbitrarily fix the boundary data $r\in U$ and scale $\l$ large, the amplitude has two asymptotic behaviors analogs to the discussion at the beginning of Section \ref{Deviating from the shape-matching}
\begin{itemize}
	\item For the boundary data that corresponds to a flat Regge geometry, there is a real critical point, and the amplitude gives an oscillatory phase. 
	\item For the boundary data corresponding to a curved Regge geometry, there are no real critical points, and the amplitude is exponentially suppressed. 
\end{itemize}

\begin{figure}[h]
	\centering
	\includegraphics[width=1\textwidth]{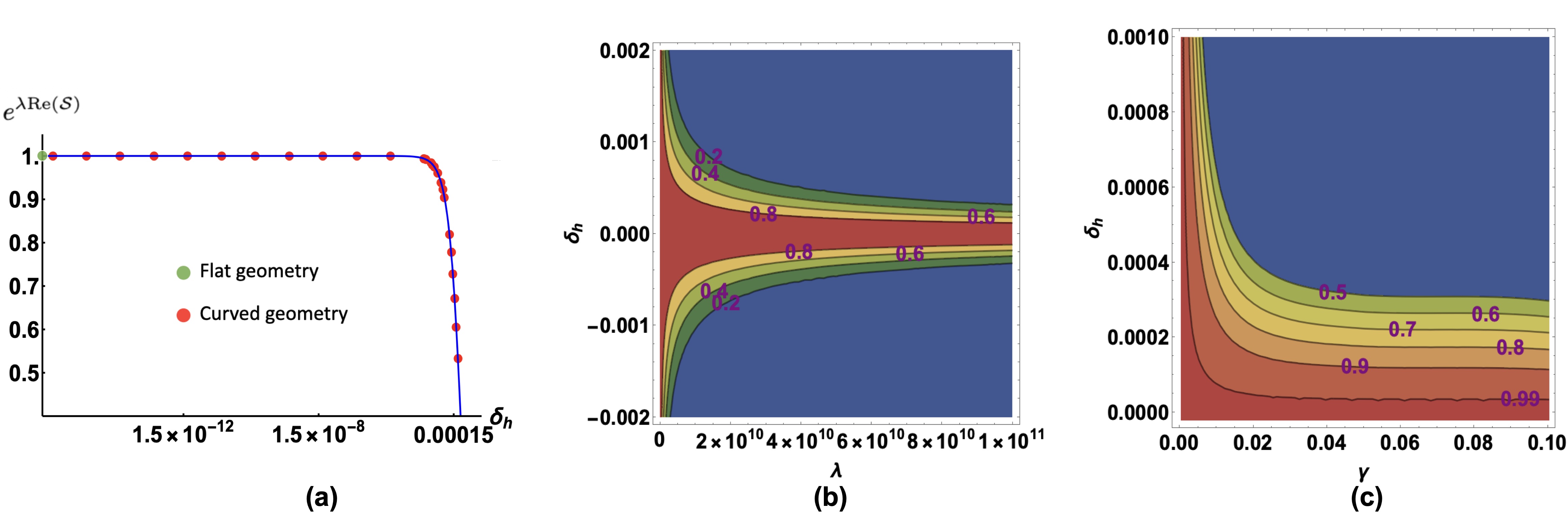}
	\caption{Panel (a) plots $e^{\lambda\re(\cs)}$ versus the deficit angle $\delta_{h}$ at $\lambda=10^{11}$ and $\gamma=0.1$ in $A(\Delta_3)$. The panels (b) and (c) are the contour plots of $e^{\lambda\re(\cs)}$ as functions of $(\lambda,\delta_h)$ at $\gamma=0.1$ and of $(\g,\delta_{h})$ at $\lambda=5\times10^{10}$ in $A(\Delta_3)$. They demonstrate the (non-blue) regime of curved geometries where the spinfoam amplitude is not suppressed. These figures first appeared in \cite{Han:2021kll}.
	}
	\label{Figure4}
\end{figure}

However, this way of presenting the asymptotic behavior leads to confusion about the flatness problem. From the discussion in Section \ref{Real critical points and complex critical points}, it is clear that there is a smooth interpolation between the oscillatory phase and the exponential suppression behaviors, since the boundary data varies smoothly. The interpolation is obtained by applying the method of the complex critical point. The formal discussion of the complex critical point and the asymptotic behavior of this model have been given in \cite{Han:2021kll}. Figure \ref{Figure4}(a) plots $e^{\l\mathrm{Re}(\cs)}$ in the asymptotic formula \eqref{asymptotics0} versus $\delta_h$ determined by the boundary data and demonstrates the smooth interpolation between the above two asymptotic behaviors. Letting the boundary data vary at the same time as scaling $\l$, we find the boundary data for the curved geometries with small nonzero $\delta_h$ for any $\lambda$, such that the amplitude $A(\Delta_3)$ is not suppressed, shown in Figure \ref{Figure4}(b). The range of $\delta_h$ for non-suppressed $A(\Delta_3)$ is nonvanishing as far as $\l$ is finite. The range of $\delta_h$ is enlarged when $\g$ is small, shown in Figure \ref{Figure4}(c). $\delta_h$ that leads to non-suppressed $e^{\lambda\Re[\mathcal{S}(Z(r))]}$ satisfies the bound
\be
 |\gamma \delta_h| \lesssim \lambda^{-1 / 2}. \label{bound}
\ee 
The above result provides evidence for the emergence of curved geometries from the spinfoam amplitude. The bound \eqref{bound} is consistent with the earlier proposal \cite{Han:2013hna} and the result in the effective spinfoam model \cite{Asante:2020qpa,Asante:2020iwm,Asante:2021zzh}. So far, the bound \eqref{bound} has only been confirmed in the regime of small or finite $\gamma$ as we are going to see in Section \ref{Solutions of effective dynamics on}, in the large-$\gamma $ regime, geometries are violating the bound \eqref{bound} but still giving a non-suppressed contribution to the spinfoam amplitude.



\section{Double-$\Delta_3$ amplitude and effective action}\label{DoubleDelta}

\subsection{Some setups}

\begin{figure}[h]
    \centering
    \includegraphics[scale=0.19]{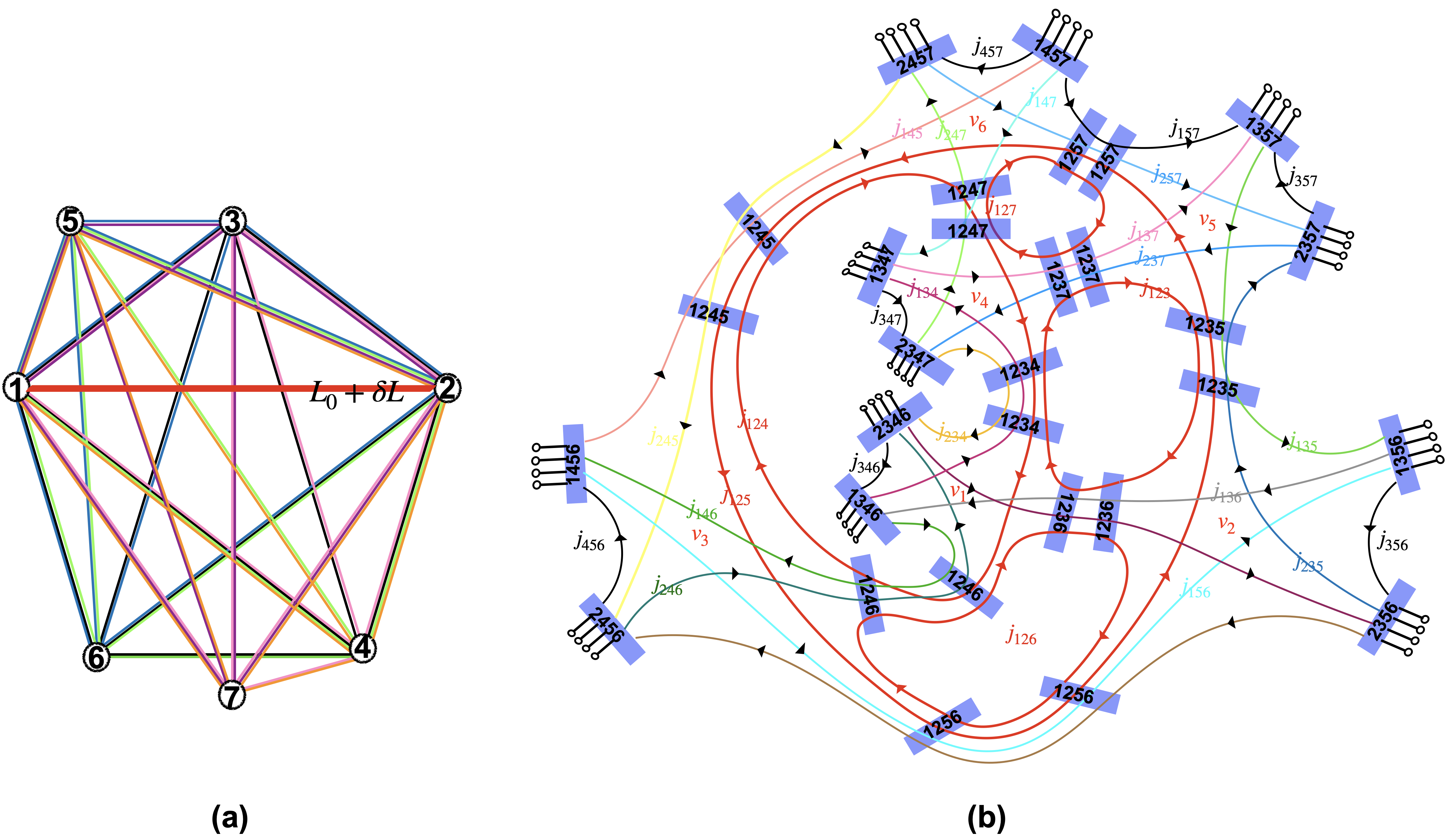}
    \caption{A complex made of six simplices sharing the bulk edge $(1,2)$ with length $l_{12}$ (the red line in panel (a)). In panel (a), the boundary edges are colored black, blue, violet and cyan. The bulk edge is colored red. Panel (b) is the dual complex of the triangulation. The internal faces carrying $j_{123},j_{124},j_{125},j_{126},j_{127}$ are bounded by red loops, and other faces are boundary faces.}
    \label{Figure5}
\end{figure}

The $\Delta_3$ complex does not have any internal segment, and the boundary data determines the Regge geometry completely. $A(\Delta_3)$ does not give the $l_I$-integral as in \eqref{pathintegral0} by $M=0$, so the effective dynamics of Regge geometry is trivial. In this section, we study the spinfoam amplitude on the ``double-$\Delta_{3}$'' complex (see Figure \ref{Figure5}(a)), which is denoted by $\Delta_3^2$. The double-$\Delta_{3}$ complex contains a single internal segment, so $M=1$, and thus $A(\Delta_3^2)$ gives \eqref{pathintegral0} as 1-dimensional integral. So the double-$\Delta_3$ complex admits non-trivial effective dynamics of the Regge geometry. Note that the same complex is also considered in the context of the effective spinfoam model \cite{Asante:2020iwm}. 

The double-$\Delta_3$ complex glues a pair of $\Delta_{3}$ complex around the internal segment $(1,2)$. The complex has seven points $P_1..., P_7$. The 4-simplices are given by
\be 
\{v_1,\cdots,v_6\}=\{(1,2,3,4,6),(1,2,3,5,6),(1,2,4,5,6),(1,2,3,4,7),(1,2,3,5,7),(1,2,4,5,7)\}. \nonumber
\ee 
The tetrahedra are labelled by $\{e_1,\cdots,e_{21}\}$\protect \footnotemark.
There are twelve boundary tetrahedra and nine internal tetrahedra among them. $j_h=\{j_{123},j_{124},j_{125},j_{126},j_{127}\}$ are carried by $5$ internal triangles, whose dual faces are bounded by red loops shown in the dual diagram Figure \ref{Figure5} (b). Since there is only one internal segment $(1,2)$ and all other segments are on the boundary,  the boundary data and the length $l_{12}$ of the internal segment determine the Regge geometry $\textbf{g}(r)$ on $\Delta_3^2$. 
\footnotetext{The tetrahedra are $\{e_1,\cdots,e_{21}\}=\{\{1,2,3,4\},\{1,2,3,6\},\{1,2,4,6\},\{1,3,4,6\},\{2,3,4,6\},\{1,2,3,5\},\{1,2,5,6\},\\\{1,3,5,6\},\{2,3,5,6\},\{1,2,4,5\},\{1,4,5,6\},\{2,4,5,6\},\{1,2,3,7\},\{1,2,4,7\},\{1,3,4,7\},\{2,3,4,7\},\{1,2,5,7\},\{1,3,5,7\},\\\{2,3,5,7\},\{1,4,5,7\},\{2,4,5,7\}\}$.} 
Following the procedure described in \eqref{integralFormAmp0} and \eqref{cz0000}, we pick up the internal spin $j_{123}$ and express the spinfoam amplitude as
\be
\begin{gathered}
	A\left(\Delta_{3}^{2}\right) =\int \mathrm{d} j_{123} \,\mathcal{Z}\left(j_{123};j_b,\xi_{eb}\right), \\
	\mathcal{Z}\left(j_{123};j_b,\xi_{eb}\right) =\sum_{\left\{k_{h}\right\}} \int \prod_{\bar{h}=1}^{4} \mathrm{~d} j_{\bar{h}} \prod_{h=1}^{5} 2 \l\,\t_{[-\epsilon,\l j^{\rm max}+\epsilon]}(\l j_h) \int \mathrm{d} \mu(g, \mathbf{z})\, e^{\lambda S^{(k)}},
\end{gathered} \label{Zj123}
\ee 
where $j_{\bar{h}}=\{j_{124},j_{125},j_{126},j_{127}\}$. The external data of $\mathcal{Z}$ is $r_l=\{j_{123}(l_{12});j_b,\xi_{eb}\}$ including both the boundary data and $j_{123}(l_{12})$. Identifying $\g j_f$ to be the area of $f$ (in Planck unit), the Heron's formula
\be
\g j_{123}(l_{12})=\frac{1}{4} \sqrt{4 l_{12}^{2} l_{13}^{2}-\left(l_{12}^{2}+l_{13}^{2}-l_{23}^{2}\right)^{2}}
\ee 
relates $j_{123}$ to the internal segment length $l_{12}$ and boundary segment lengths $l_{13}, l_{23}$. We consider the Regge boundary data that determines all the boundary segment lengths. We can always make a local change of the real variable $j_{123}\rightarrow l_{12}$ within a neighborhood $K$ of a given Regge geometry, where the correspondence $j_{123}\leftrightarrow l_{12}$ is 1-to-1.

In the following discussion, we only focus on the case with $k_h=0$. The Regge geometries under consideration are of small deficit angles. The following describes the procedure to compute the complex critical points $Z(r_l)$ of $\mathcal{Z}$. 

We embed the double-$\Delta_3$ complex in $(\R^4,\eta_{IJ})$ and determines a flat Regge geometry with all tetrahedra spacelike. We assign the following coordinates to the points, 
\be
\begin{gathered}
P_{1}=(0,0,0,0),\quad P_2=\left(-0.0680, -0.220, -0.532, -1.33\right), \quad P_{3}=\left(0, 0, 0, -3.40\right), \nonumber\\
P_{4}=\left(-0.240, -0.694, -0.981, -1.70\right), \quad P_{5}=\left(0, 0, -2.94, -1.70\right), \quad P_{6}=\left(0, -2.77, -0.981, -1.70\right),\nonumber\\ P_{7}=\left(-2.47, -3.89, -1.36, -1.91\right).\label{P7}
\end{gathered}
\ee
From the coordinates, we can compute the length of the segments of the triangulation by using 
\be 
l_{ij}=\sqrt{\eta_{IJ} (P_i-P_j)^I (P_i-P_j)^J}.
\ee 
with $\eta_{IJ}=\text{Diag}(\{-1,1,1,1\})$ the Minkowski metric.
The segment lengths are shown in Table \ref{EdgeLength}. 
\begin{table}[h]
	\centering \caption{Each cell of the table is the segment length for vertice $P_i$ and $P_j$.}\label{EdgeLength}
	\scalebox{0.8}{
		\begin{tabular}{|c|c|c|c|c|c|c|c|} 
			\hline
			\diagbox{$i$}{$l_{ij}$}{$j$}&1&2&3&4&5&6&7\\
			\hline
			1&\diagbox{}{}&1.45&3.40&2.07&3.40&3.40&3.81\\
			\hline
			2&1.45& \diagbox{}{}& 2.14& 0.729& 2.45& 2.62& 2.96\\
			\hline
			3&3.40& 2.14& \diagbox{}{}& 2.07& 3.40& 3.40& 3.62\\
			\hline
			4&2.07& 0.729& 2.07& \diagbox{}{}& 2.07& 2.07& 2.34\\
			\hline
			5&3.40& 2.45& 3.40& 2.07&\diagbox{}{} & 3.40& 3.41\\
			\hline
			6& 3.40& 2.62& 3.40& 2.07& 3.40&\diagbox{}{}& \diagbox{}{}\\
			\hline
			7& 3.81& 2.96& 3.62& 2.34& 3.41& \diagbox{}{}& \diagbox{}{}\\
			\hline
		\end{tabular}
	}
\end{table}
The triangles within a 4-simplex are classified into two categories \cite{Barrett:2009mw}: The triangle corresponds to the \emph{thin wedge} if the inner product between the timelike normals of the two adjacent tetrahedra is positive, otherwise the triangle corresponds to the \emph{thick wedge}. The dihedral angle $\theta_{v,e_i,e_j}$ are given by:
\be
&\text{thin wedge:}\qquad N_{ve_i}\cdot N_{ve_j}&=\cosh\theta_{v,e_i,e_j},\nonumber\\
&\text{thick wedge:}\qquad N_{ve_i}\cdot N_{ve_j}&=-\cosh\theta_{v,e_i,e_j},\label{dihedral}
\ee
where the inner product is the Minkowski inner product defined by $\eta$. Then we check the deficit angles $\delta_{h_i}$ associated to the shared triangles $h_i$ 
\be
\begin{gathered}
	0=\delta_{h_1}=\theta_{v_1,e_1,e_2}+\theta_{v_2,e_2,e_6}+\theta_{v_4,e_1,e_{13}}+\theta_{v_5,e_6,e_{13}}\approx 0.514+0.464-0.575-0.404,\\
	0=\delta_{h_2}=\theta_{v_1,e_1,e_3}+\theta_{v_3,e_3,e_{10}}+\theta_{v_4,e_1,e_{15}}+\theta_{v_6,e_{10},e_{15}}\approx 1.08 -1.02-1.30+1.24,\\
	0=\delta_{h_3}=\theta_{v_2,e_6,e_7}+\theta_{v_3,e_7,e_{10}}+\theta_{v_5,e_6,e_{17}}+\theta_{v_6,e_{10},e_{17}}\approx -0.360-0.481+0.414+ 0.426,\\
	0=\delta_{h_4}=\theta_{v_1,e_2,e_3}+\theta_{v_2,e_2,e_{7}}+\theta_{v_3,e_7,e_{10}}\approx-0.723-0.208+0.931,\\
	0=\delta_{h_5}=\theta_{v_4,e_1,e_{15}}+\theta_{v_5,e_{13},e_{17}}+\theta_{v_6,e_{15},e_{17}}\approx-0.903+ 1.20 -0.301,								
\end{gathered}
\ee
which implies the Regge geometry is flat. The data of the flat geometry determines the external data $\mathring{r}_l$ for the partial amplitude $\cz$, which has the real critical points $(\mathring{j}_{\bar{h}},\mathring{g}_{ve},\,\mathring{\textbf{z}}_{vf})$ corresponding to this flat Regge geometry and endowing the consistent 4-orientations to all 4-simplices. The boundary data of the flat geometry and the real critical point can be found in Appendix \ref{flatdata}, and Mathematica code can be found in \cite{liu_spinfoam.org} and \cite{toappear.org}. In this case, given the boundary data, the flat Regge geometry is the solution of the classical Regge equation of motion, and it is also the solution $(\mathring{j}_{\bar{h}},\mathring{g}_{ve},\,\mathring{\textbf{z}}_{vf})$ to the critical equations from the spinfoam amplitude.

\begin{figure}[h]
    \centering
    \includegraphics[scale=0.09]{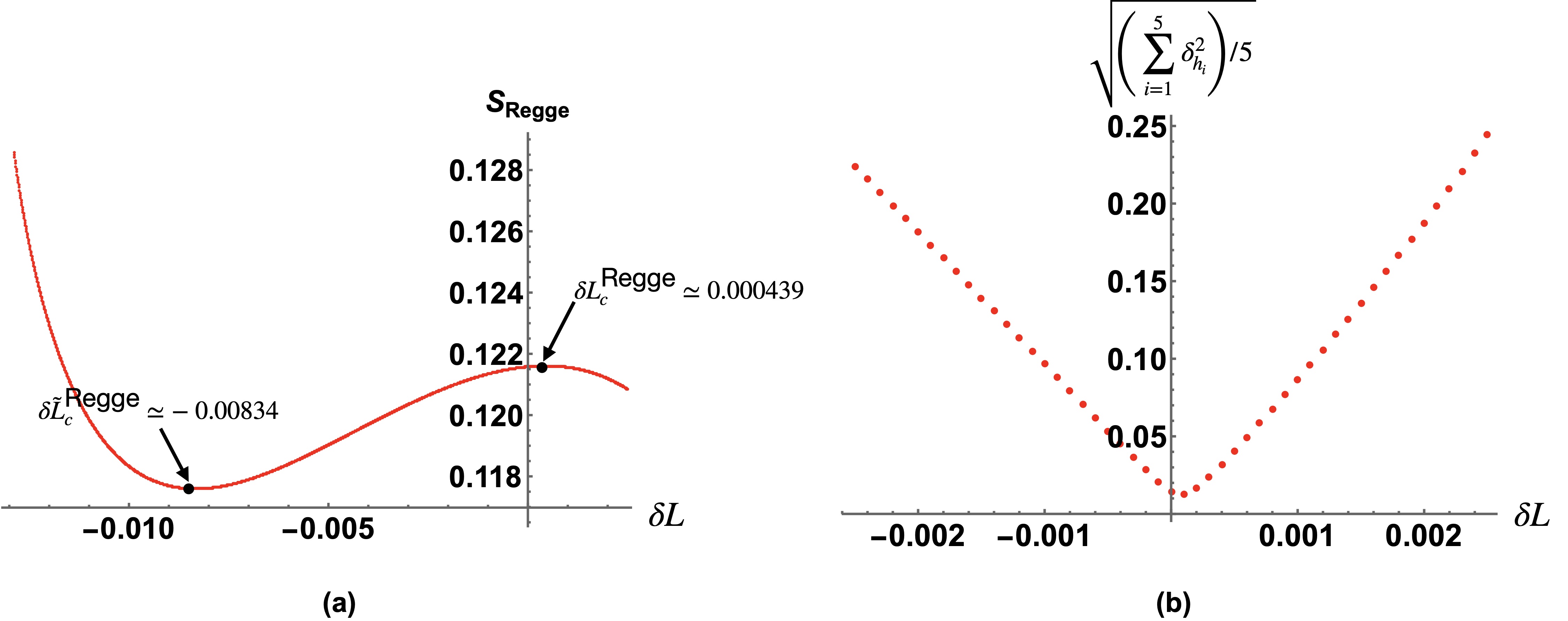}
    \caption{Panel (a) is the Regge action varying with $\delta L$ when we deform the boundary segment length $l_{35}\rightarrow l_{35}+10^{-3}$ from the boundary data of the flat geometry. In this case, the Regge solutions are given by $\delta L_c^{\rm Regge}\simeq 0.000439$ and $\delta \widetilde{L}_c^{\rm Regge}\simeq -0.00834$. Panel (b) is $\sqrt{(\sum_{i=1}^5\delta_{h_i}^2)/5}$ versus $\delta L$ with the deformed boundary data. All geometries in the range of $\delta L$ are not flat. The minimum of $\sqrt{(\sum_{i=1}^5\delta_{h_i}^2)/5}$ is $0.013$.}
    \label{FigureRegge}
\end{figure}

We are going to compare the classical Regge dynamics and the spinfoam effective dynamics for curved geometries. This comparison is based on the numerical computations. In concrete, we deform the boundary segment length $l_{35}\rightarrow l_{35}+10^{-3}$ but keep the other boundary segment lengths unchanged. The boundary data does not admit any flat geometry on $\Delta_3^2$ (see Figure \ref{FigureRegge}(b))\footnote{If the boundary data admitted a flat Regge geometry on the complex, the flat geometry would be a solution to the Regge equation. However, the solution of the Regge equation is a curved geometry with the given boundary data, contradicting the assumption of admitting the flat geometry.}. With this deformation, a classical Regge solution (i.e. the solution to the classical Regge equation $\delta S_{\rm Regge}=0$) gives the deficit angles 
\be\label{nonzerodelta}
 \begin{gathered}
 	\delta_{h_1}=0.0118, \quad \delta_{h_2}=0.0661,\quad \delta_{h_3}=-0.0215,\\
 	 \delta_{h_4}=-0.0236, \quad \delta_{h_5}=-0.0252,	
 \end{gathered}
\ee 
which implies that the classical Regge dynamics gives curved geometry. We fix the boundary data and vary the internal segment length $l_{12}=L_0+\delta L$ where $L_0=1.45$ is the length $l_{12}$ in the flat geometry. The change of $l_{12}$ is denoted by $\delta L$ with $\delta L\in [-0.0129,0.00251]$ \footnote{The range used here is restricted by the existence of curved Regge geometry with all tetrahedra spacelike.}. The classical Regge action $S_{\rm Regge}$ as a function of $\delta L$ is plotted in Figure \ref{FigureRegge}(a). The above solution leading to \eqref{nonzerodelta} is close to the origin $\delta L=0$ and is denoted by $\delta L_c^{\rm Regge}$. There exists another Regge solution in $\delta L<0$ and far from $\delta L=0$ as shown in Figure \ref{FigureRegge}(a). We denote this solution by $\delta\widetilde{L}_c^{\rm Regge}$. 

Likely, the solution $\delta\widetilde{L}_c^{\rm Regge}$ is a discretization artifact because when smoothly deforming the boundary data $l_{35}$ back to the one for the flat geometry, $\delta L_c^{\rm Regge}$ reduces back to the flat solution. In contrast, $\delta\widetilde{L}_c^{\rm Regge}$ still reduces to a curved Regge geometry. Some boundary data also exist such that the second solution $\delta\widetilde{L}_c^{\rm Regge}$ disappears. Nevertheless, we will take into account both solutions $\delta{L}_c^{\rm Regge}$ and $\delta\widetilde{L}_c^{\rm Regge}$ in discussing the effective dynamics in Section \ref{Solutions of effective dynamics on}.

The boundary data $(j_b,\xi_{ef})$ and the corresponding pseudo-critical points $(j_h^0,g^0_{ve},\bold{z}^0_{vf})$ for the curved geometry with the boundary segment length $l_{35}\rightarrow l_{35}+10^{-3}$ and the internal edge $l_{12}=L_0+\delta L_{c}^{\text{Regge}}$ are listed in Appendix \ref{curveddata}.

Notice that the geometrical areas in the boundary data relate to $j_b$ by $\fa_b=\gamma j_b$, and the area $\fa_b$ relates to the lengths $l_{ij}$ by Heron's formula. The following discussion involves fixing the geometrical area $\fa_b$ and performing computations at different Barbero-Immirzi parameter $\g$, so this leads to different $j_b$ at different $\gamma$. Fixing the geometrical area instead of fixing $j_b$ is useful when we compare with the Regge action $S_{\rm Regge}$, since $S_{\rm Regge}$ only depends on the geometrical boundary data.

\subsection{Numerical computing the effective action}

Given the boundary condition $(j_b,\xi_{eb})$ corresponds to the above Regge boundary data with the deformed $l_{35}$, and given any $l_{12}$ and $j_{123}(l_{12})$ taking value inside a neighborhood of the value for the flat geometry, we find the \emph{pseudo-critical point} $({j}^0_{\bar h},{g}^0_{ve},{\textbf{z}}^0_{vf})$ close to the real critical point inside the real integration domain. The pseudo-critical point only satisfies $\re(S)=\partial_{g_{ve}} S=\partial_{\mathbf{z}_{vf}}S=0$ but does not necessarily satisfy $\partial_{j_{\bar h}}S=0$. The {pseudo-critical point} $({j}^0_{\bar h},{g}^0_{ve},{\textbf{z}}^0_{vf})$ is the critical point of the spinfoam amplitude with fixed $j_h,j_b$ \cite{Han:2011re}, and endows the Regge geometry $\mathbf{g}(r)$ and consistent 4-simplex orientations to $\Delta_3^2$ complex\footnote{Since the correspondence between $j_{123}$ and $l_{12}$ is not 1-to-1 globally, it might be possible to have multiple pseudo-critical points corresponding to different Regge geometries with the same value of $j_{123}$. However, in our numerical analysis, the other $l_{12}$ from the same $j_{123}$ does not satisfy the triangle inequality. Therefore all pseudo-critical points correspond to the same Regge geometry but with different 4-simplex orientations, although we only focus on a fixed orientation. }. It reduces to the real critical point $(\mathring{j}_{\bar h},\mathring{g}_{ve},\mathring{\textbf{z}}_{vf})$ when $r_l=\mathring{r}_l$ corresponds to the flat geometry on $\Delta_3^2$. As the deformation of segment length $l_{35}$ is small, this curved geometry is close to the flat geometry, so $({j}^0_{\bar h},{g}^0_{ve},{\textbf{z}}^0_{vf})$ is close to $(\mathring{j}_{\bar h},\mathring{g}_{ve},\mathring{\textbf{z}}_{vf})$ in the integration domain. The data for the pseudo-critical point is listed in Appendix \ref{curveddata}.

In this computation, we still adopt the similar parametrizations of variables as in (\ref{para0}), (\ref{para1}), and (\ref{para2}), but with the pseudo-critical points as the origin. The parametrizations of the group element $g_{v_1 e_2}, g_{v_2 e_7}$, $g_{v_3 e_3}$,  $g_{v_4 e_{13}}, g_{v_5 e_{17}}$, $g_{v_6 e_{15}}$, $g_{v_1 e_1}, g_{v_2 e_6}$, and $g_{v_3 e_{10}}$ are upper-triangular matrices due to the SU(2) gauge fixing at $9$ internal tetrahedra
\be
g_{ve}=g^{0}_{ve}\left(\begin{array}{cc}
	1+\frac{x_{ve}^{1}}{\sqrt{2}} & \frac{x_{ve}^{2}+i y_{ve}^{2}}{\sqrt{2}} \\
	0 & *
\end{array}\right), 
\ee 
where the entry $*$ is determined by $\det(g_{ve})=1$. The internal spin $j_{\bar{h}}$ is parametrized as 
\be
j_{\bar{h}}=j^{0}_{\bar{h}}+ \mathrm{j}_{\bar{h}},\quad \mathrm{j}_{\bar{h}} \in \mathbb{R}.
\ee 
As a result,  for $k_h=0$, the spinfoam amplitude $A(\Delta_3^2)$ and $\mathcal{Z}(j_{123})$ in (\ref{Zj123}) can be written in the form of 
\be
\begin{gathered}
	A(\Delta_3^2)=\int \mathrm{d}l_{12}\left|\frac{\partial j_{123}}{\partial l_{12}}\right|\mathcal{Z}(j_{123}(l_{12});j_b,\xi_{eb}),\\
	\mathcal{Z}(j_{123}(l_{12});j_b,\xi_{eb})\sim \int \mathrm{d}^{241}x\,\mu(x) e^{\lambda S(r_l,x)},\qquad r_l=(j_{123}(l_{12}),j_b,\xi_{eb})
\end{gathered}
\ee 
where $x\equiv (x^{1}_{ve}, y^{1}_{ve}, x^{2}_{ve}, y^{2}_{ve}, x^{3}_{ve}, y^{3}_{ve},x_{vf},y_{vf},\mathrm{j}_{\bar{h}})$. The parametrizations of $(l_{12},x)$ define the coordinate chart covering the neighborhood $K$ enclosing $\tilde{x}_0= (j_{123},x_0)= ({j}^0_h,{g}^0_{ve},{\textbf{z}}^0_{vf})$, and $\mathring{\tilde{x}} = (\mathring{j}_{123},\mathring{x})=(\mathring{j}_h,\mathring{g}_{ve},\mathring{\textbf{z}}_{vf})$. 
This neighbourhood is large enough since the parametrizations are valid generically. The pseudo-critical point is ${x}_0=(0,0,...,0)$, which contains 241 zero components. Here we use ``$\sim$'' instead of ``$=$'' because (1) we only consider $k_h=0$ but ignore other $k_h$ terms\footnote{The integrals in the neighborhood $K$ with $k_h\neq 0$ give exponentially suppressed contributions. }, (2) we only focus on the contribution from the neighborhood $K$ enclosing a single pseudo-critical point\footnote{there may exist other pseudo-critical points outside $K$ in $\cz$, e.g. the ones corresponding to different orientations of 4-simplices. But our discuss only focuses on the critical points inside $K$.}. In our discussion, we only consider the effective dynamics within a sector of Regge geometries with the fixed 4d orientation. 

We compute the complex critical point of $\mathcal{Z}$ for any given external data $r_l$: Here, both $S(r,x)$ and $\mu(x)$ are analytic in the neighborhood $K$ of $x_0$. 
$S(r,x)$ can be analytically continued to a holomorphic function $\cs(r_l,z)$, and $z\in\C^{241}$ is in a complex neighborhood of $x_0$. The analytic continuation is obtained by simply extending $x\in \R^{241}$ to $z\in \C^{241}$. The formal discussion of the analytic continuation of the spinfoam action is given in \cite{Han:2021rjo}. We fix the boundary data to be the one resulting in \eqref{nonzerodelta} and vary the length $l_{12}=L_0+\delta L$, where $L_0=1.45$ (the value of $l_{12}$ in Table \ref{EdgeLength}) and the change of $l_{12}$, $\delta L\in [-0.0129,0.00251]$. For any given $\delta L$, combining the boundary data, we repeat the steps above (from the beginning of this subsection) to reconstruct the Regge geometry and the corresponding pseudo-critical point. Taking the pseudo-critical point as the starting point, we apply the Newton-Raphson method by repeating the steps in \eqref{step1} - \eqref{step6} to numerically compute the complex critical point $Z(r_l)$ for a sequence of $\delta L$. By evaluating $\cs$ at the complex critical point and apply the asymptotic formula \eqref{asymptotics0}, we obtain the following asymptotic behavior of $\mathcal{Z}$ and $A(\Delta_3^2)$ for the dominant contribution from the integral on $K$
\be\label{AZinK}
\begin{gathered}
\mathcal{Z}\left(j_{123}(l_{12});j_b,\xi_{eb}\right) \sim \lt(\frac{1}{\l}\rt)^{\frac{241}{2}}\sn_l\, e^{\lambda \mathcal{S}\left(r_{l}, Z\left(r_{l}\right)\right)}\lt[1+O(1/\l)\rt], \\
A\left(\Delta_{3}^{2}\right) \sim \lt(\frac{1}{\l}\rt)^{\frac{241}{2}} \int \mathrm{d} l_{12}\left|\frac{\partial j_{123}}{\partial l_{12}}\right|\sn_l\,  e^{\lambda \mathcal{S}\left(r_{l}, Z\left(r_{l}\right)\right)}\lt[1+O(1/\l)\rt] ,
\end{gathered}
\ee 
where $\sn_l=\mu(Z(r_l))\det(-\partial_{z,z}^2\cs(r_l,Z(r_l))/2\pi)^{-1/2}$. Effectively, $A\left(\Delta_{3}^{2}\right) $ gives a path integral of Regge geometry on $\Delta_3^2$. $\mathcal{S}\left(r_{l}, Z\left(r_{l}\right)\right)$ is the effective action for the Regge geometry in the large-$\l$ regime of the spinfoam amplitude. The stationary phase approximation of the $l_{12}$-integral in \eqref{AZinK} relates to the variation of $\mathcal{S}\left(r_{l}, Z\left(r_{l}\right)\right)$ with respect to $l_{12}$. The effective equation of motion
\be
\partial_{l_{12}}\mathcal{S}\left(r_{l}, Z\left(r_{l}\right)\right)=0
\ee
determines the effective dynamics of Regge geometry.

\subsection{Comparing to Regge action}

\begin{figure}[t]
	\centering
	\includegraphics[scale=0.1]{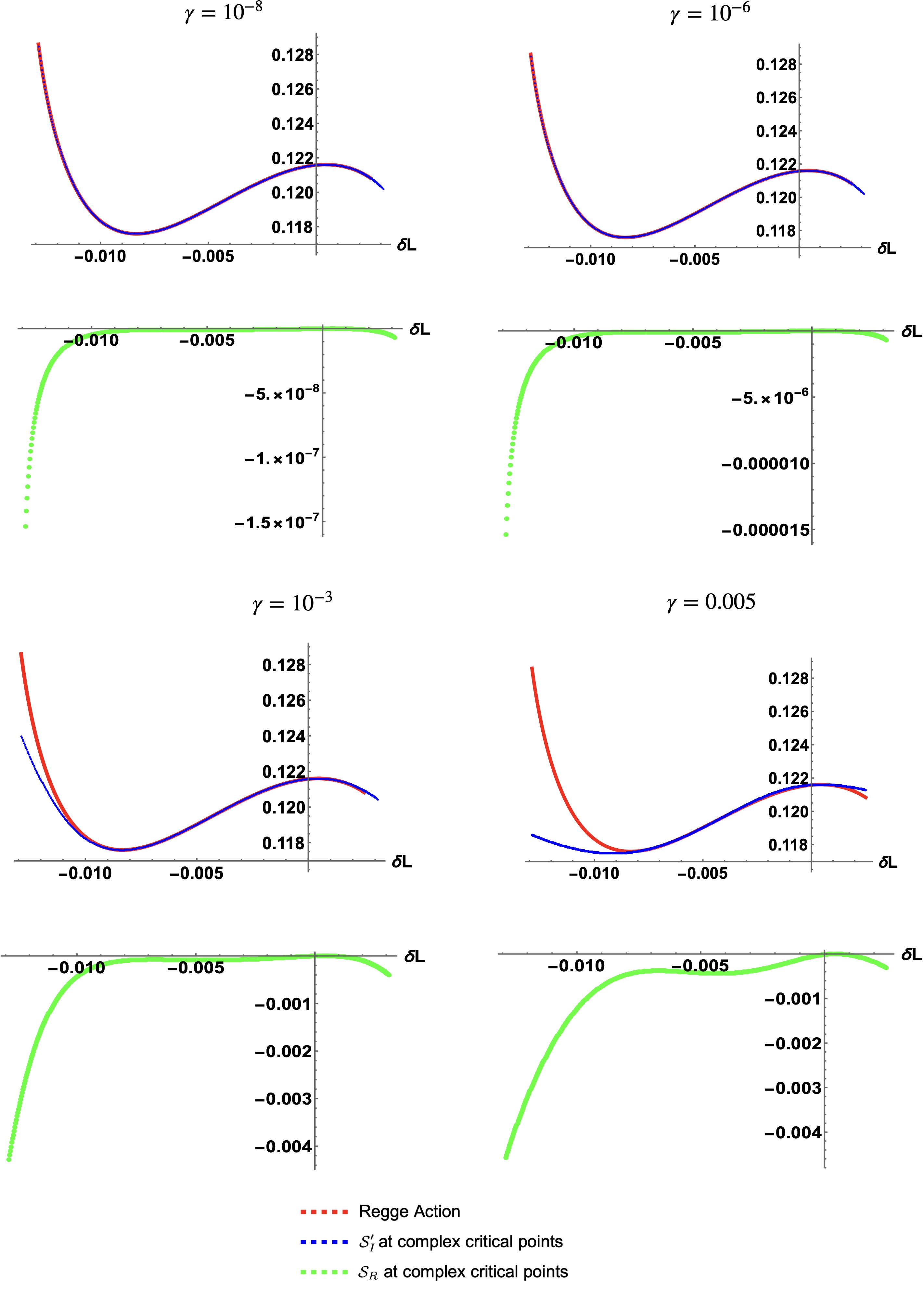}
	\caption{The red curves plots the Regge action as a function of $\delta L$. In comparison to the Regge action, the blue curves plots $\cs'_I$ of the analytic continued spinfoam action at complex critical points. The green curve plots the real part $\cs_R$ of the analytic continued spinfoam action at complex critical points.}
	\label{Figure6}
\end{figure}

It is interesting to compare the effective action $\mathcal{S}\left(r_{l}, Z\left(r_{l}\right)\right)$ to the classical Regge action $S_{\text{Regge}}$ since both actions define the dynamics of Regge geometry. The definition of Regge action $S_{\text{Regge}}(l_{12})$ is reviewed in Appendix \ref{AppendixA2}. In order to compare, we compute and plot the real and imaginary parts $\cs_R$ and $\cs_I$ of $\mathcal{S}\left(r_{l}, Z\left(r_{l}\right)\rt)$ respectively,
\be 
\mathcal{S}\left(r_{l}, Z\left(r_{l}\right)\rt) =\mathcal{S}_{R}(\g, \delta L)+\mathrm{i}  \cs_{I}(\g, \delta L),
\ee 
We view both $\cs_R$ and $\cs_I$ as functions of two variables $\gamma$ and $\delta L$, and we compute the numerical values of $\cs_R$ and $\cs_I$ with samples of $\gamma\in [10^{-9},10^{6}]$ and $\delta L\in [-0.0129,0.00251]$. 

It is known that the spinfoam action contains an overall phase, which needs to be subtracted to compare to the Regge action. We denote the overall phase by $\phi(\gamma)$. This overall phase can be computed numerically by inserting the pseudo-critical point $({j}^0_{\bar h},{g}^0_{ve},{\textbf{z}}^0_{vf})$ in the spinfoam action $S$ and subtracting the Regge action at the corresponding geometry. Generally, we have
\be
\phi(\g)=\a/\gamma
\ee
where the coefficient $\a$ depends on the boundary data. In terms of the spinfoam variables, the overall phase comes from the $\g$-independent terms in $S$ and is linear to the boundary spins $\phi\sim j_b$, but here we fix the boundary area and let $\g$ vary, then $\phi \sim \fa_b/\g$. The numerical value of $\a$ is $\a=0.003993$ resulting from our setup of the boundary data. In general, the overall phase in the spinfoam action can be cancelled by the phase choice of boundary $\xi_{eb}$. To remove the overall phase from $\cs_I$, we define $\cs_I'$ by
\be
\cs_{I}(\g,\delta L)=-\cs_I'(\g,\delta L)+\phi(\g).
\ee
$\cs'_{I}$ as a function of $\delta L$ is compared to the classical Regge action for different values of $\g$ in Figure \ref{Figure6}. The minus sign in front of $\cs_I'$ relates to the 4-simplex orientation in the real and pseudo-critical points. 
As indicated by Figure \ref{Figure6}, $\cs_{I}'$ well-approximates the Regge action for small $\gamma$ with negligible corrections. When increasing $\g$, $\cs'_{I}$ gives nontrivial corrections to the Regge action. 

For any given $\gamma$, the real part $\cs_R$ is always negative, and $|\cs_R|$ is larger for larger $|\delta L|$, so $e^{\l \cs}$ is smaller for larger $|\delta L|$. However, if we fix $\delta L$ and vary $\gamma$, $|\cs_R|$ is smaller so $e^{\l \cs}$ is less suppressed for any $\l$, when $\gamma$ is smaller. In other words, the smaller $\g$ opens a larger range of $\delta L$, in which $|\cs_R|$ is small and $e^{\l \cs}$ is not suppressed for a given $\l$. In this range of $\delta L$, the numerical result indicates that $\cs\left(r_{l}, Z\left(r_{l}\right)\rt)$ well-approximates the Regge action. The similar situation has appeared in the $\Delta_3$ amplitude, where the amplitude with smaller $\gamma$ admits a wider range of curved geometries (see Figure \ref{Figure4}(c)).

\section{Solutions of effective dynamics on double-$\Delta_3$}\label{Solutions of effective dynamics on}

\subsection{Spinfoam complex critical point and the Regge solution $\delta L_c^{\rm Regge}$}

\begin{figure}[t]
	\centering
	\includegraphics[scale=0.115]{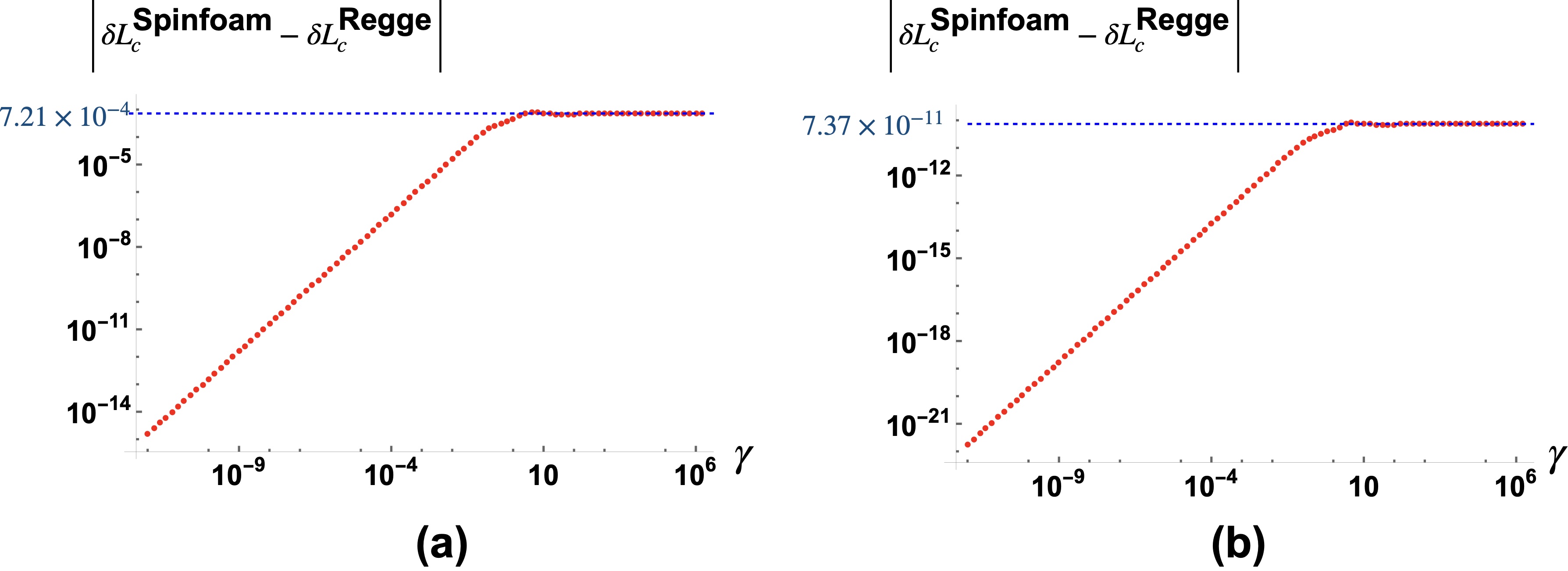}
	\caption{Panels (a) and (b) are log-log plots of the distances \eqref{distanceLc} between the spinfoam and Regge solutions in a neighbourhood of $\delta \cl=0$ as a function of  $\gamma$. The boundary data has the boundary segment length $l_{35}$ deformed from the flat geometry by $l_{35}\rightarrow l_{35}+10^{-3}$ for (a) and $l_{35}\rightarrow l_{35}+10^{-10}$ for (b). }
	\label{Figure7} 
\end{figure}

\begin{figure}[t]
	\centering
	\includegraphics[scale=0.21]{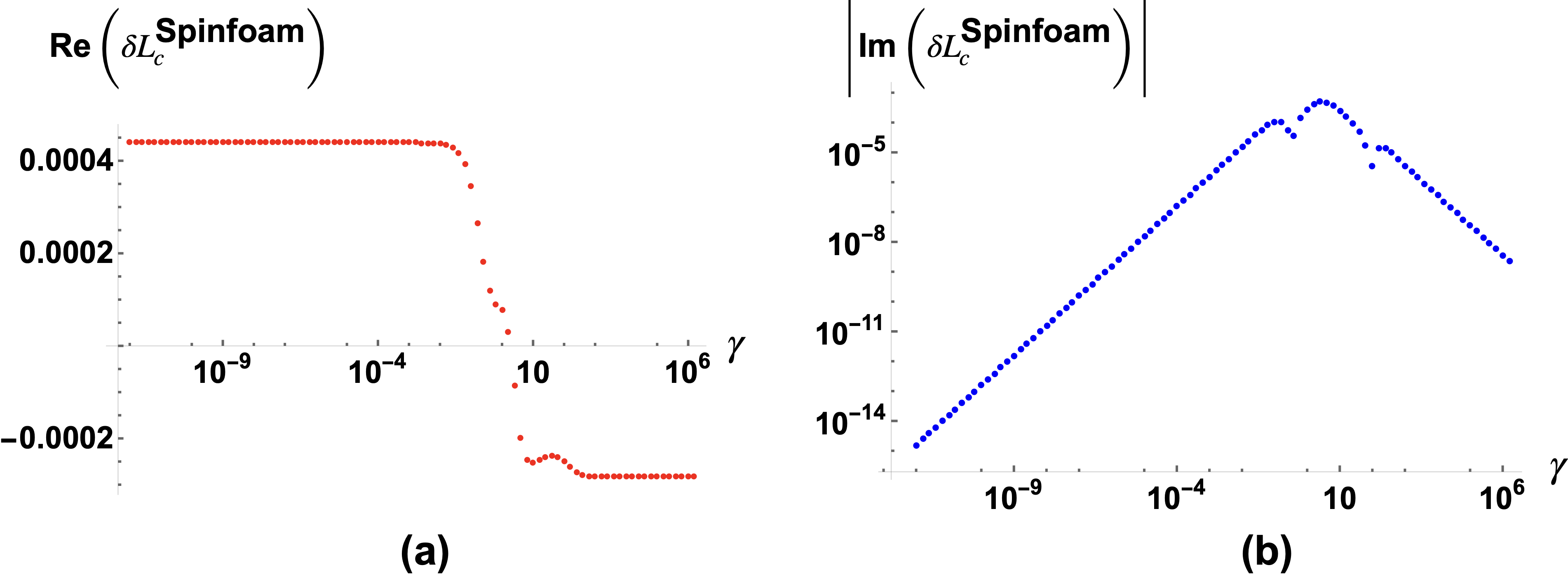}
	\caption{Panels (a) show the real part of the spinfoam solution $\delta L_c^\text{Spinfoam}$ v.s. log-scaled $\gamma$ value with the boundary data deformed from the flat geometry by $l_{35}\rightarrow l_{35}+10^{-3}$. Panels (b) is the log-log plot of the absolute value of the imaginary parts of the spinfoam solution $\delta L_c^\text{Spinfoam}$ as a function of $\gamma$.}\label{ImdlvsGamma} 
\end{figure}

\begin{figure}[t]
	\centering
	\includegraphics[scale=0.14]{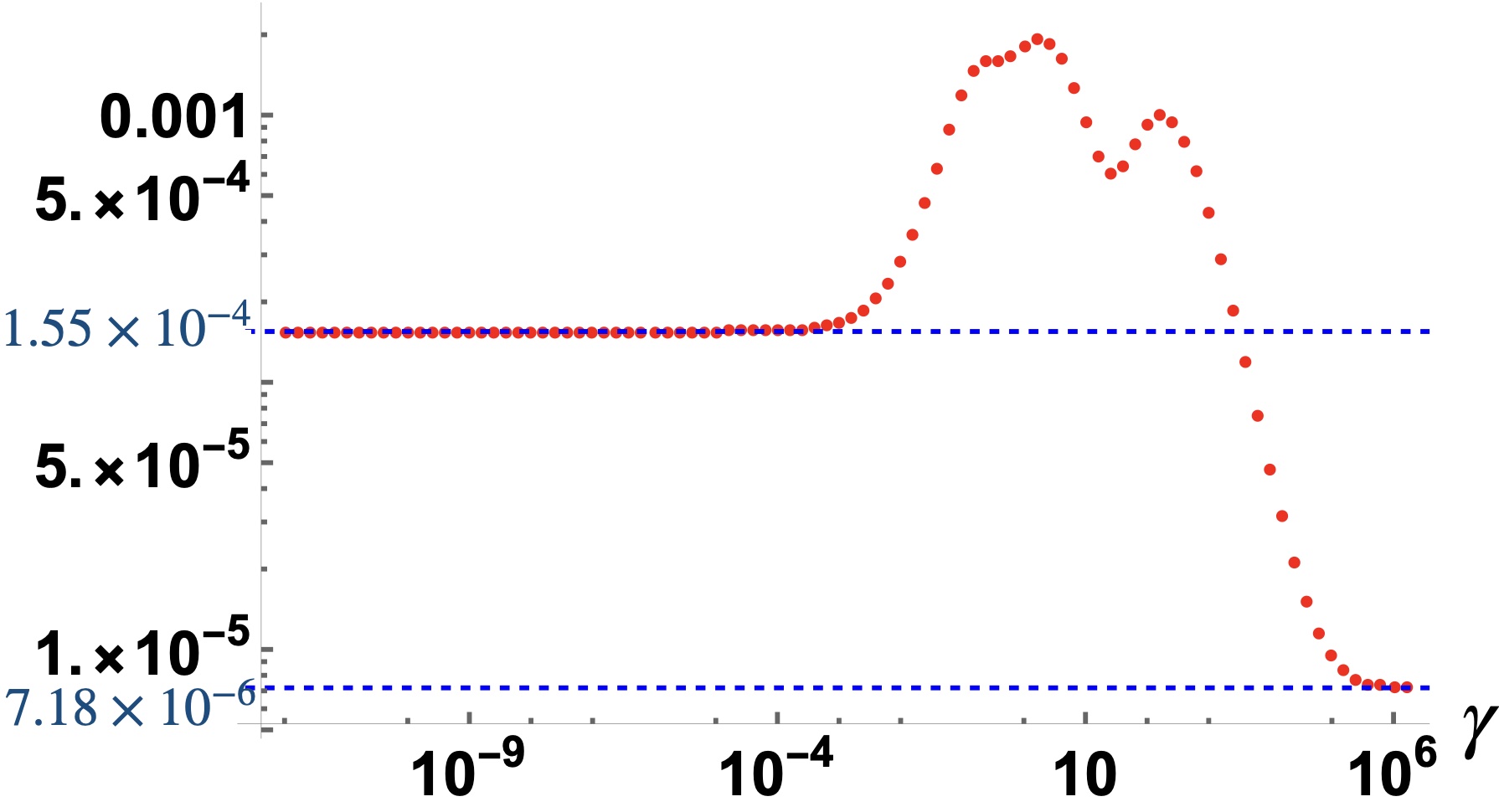}
	\caption{The log-log plot of the average of the absolute value of the imaginary part of the complex critical point v.s. $\gamma$.}\label{AverageZ} 
\end{figure}

\begin{figure}[t]
	\centering
	\includegraphics[scale=0.11]{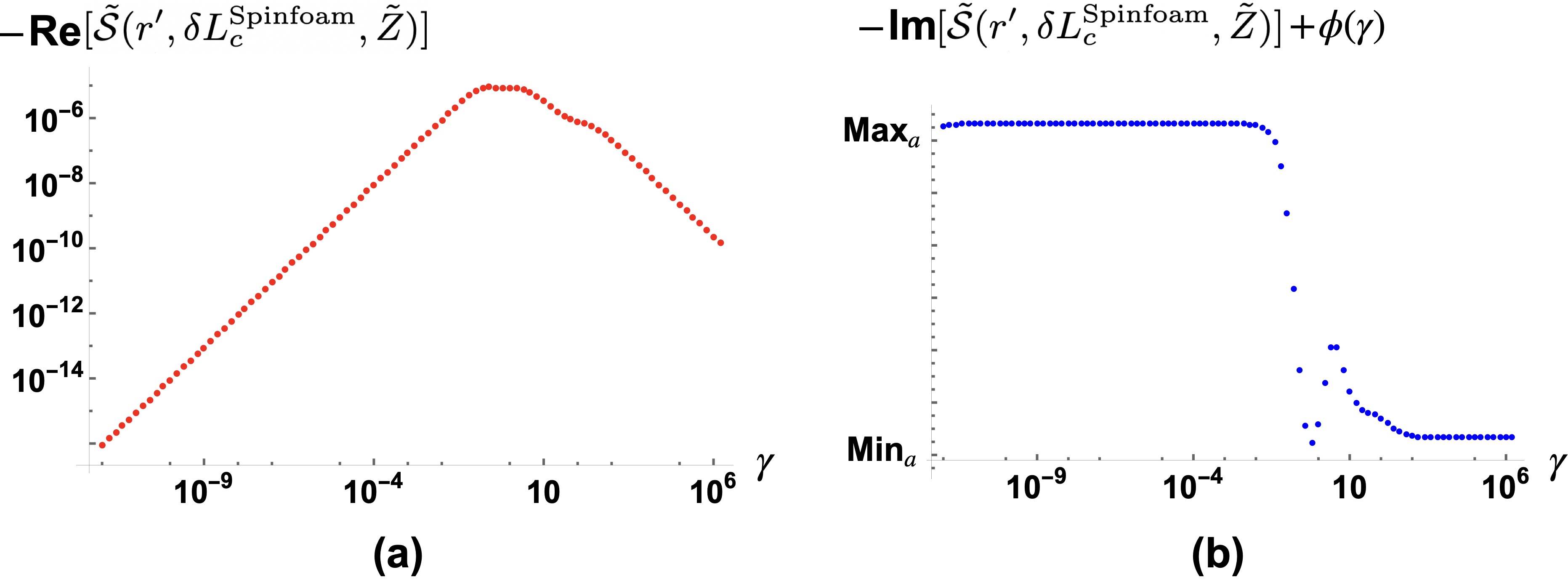}
	\caption{Panels (a) are the log-log plot of the negative real parts of $\tilde{\cs}(r',\delta\mathcal{L},z)$ at the complex critical points $z=\tilde{Z}(r',\delta\mathcal{L})$ as a function of $\gamma$ with the boundary data deformed from the flat geometry by $l_{35}\rightarrow l_{35}+10^{-3}$. Panels (b) show the imaginary parts of $\tilde{\cs}(r',\delta\mathcal{L},z)$ at the complex critical points $z=\tilde{Z}(r',\delta\mathcal{L})$ v.s. log-scaled $\gamma$. We subtract the overall phase $\phi(\g)$ from $\mathrm{Im}[\tilde{\cs}(r',\delta L^{\text{Spinfoam}}_c,\tilde{Z})]$ and add a minus sign in plotting (b). In Panel (b), the overall phase $\phi(\gamma)\simeq {0.003993}{\gamma}^{-1}$, and the maximum and minimum of the plot range are $\text{Max}_a \simeq 0.121606$ and $\text{Min}_a \simeq 0.121596$.}
	\label{StildevsGamma} 
\end{figure}

The above discussion compares the effective action $\cs(r_l,Z(r_l))$ to the classical Regge action. It is also interesting to compare the solution of the effective equation $\partial_{l_{12}} \cs(r_l,Z(r_l))=0$ to the solution of the Regge equation. By the above computation, the real and imaginary parts of $\cs(r_l,Z(r_l))$ are obtained as the numerical function. Numerically solving the effective equation involves finding the possible complex roots of numerical derivatives of the complex $\cs(r_l,Z(r_l))$, which requires an estimation of $\cs(r_l,Z(r_l))$ on the complex $\delta L$ plane and may give a relatively large numerical error. In the following, we introduce an alternative strategy, which computes the solution of the effective equation more efficiently. 

Instead of introducing the partial amplitude $\cz$, we consider the full spinfoam amplitude, which can be written as the following integral for the same contribution as in \eqref{AZinK}
\be
A(\Delta_3^2)\sim\int \mathrm{d}\delta L \mathrm{d}^{241}x\, \mu(\delta L,x) e^{\lambda \tilde{S}(r',\delta L,x)}.
\ee 
Here the external parameter $r'$ is just the boundary data $r'=(j_b,\xi_{eb})$. $\tilde{S}(r',\delta L,x)$ is the spinfoam action $S$ with $j_{123}=j_{123}(l_{12})$ and $l_{12}=L_0+\delta L$.

Recall that $\delta L_c^{\rm Regge}$ is a solution of the classical Regge equation. The Regge geometry with $\delta L_c^{\rm Regge}$ corresponds to a pseudo-critical point of $\tilde{S}(r',\delta L,x)$. Both $\tilde{S}(r',\delta L,x)$ and $\mu(\delta L,x)$ are analytic in the neighbourhood of this pseudo-critical point. Therefore, $\tilde{S}(r',\delta L,x)$ and $\mu(\delta L,x)$ can be analytic continued to the holomorphic functions $\tilde{\mathcal{S}}(r',\delta\mathcal{L},z)$ and $\mu(\delta \cl,z)$, where $(\delta\mathcal{L},z)\in \mathbb{C}^{242}$ is in a complex neighborhood of the pseudo-critical point. We fix the boundary data $r'$ to be the same as the one used in Figure \ref{FigureRegge}. Since $r'$ is a small deformation from the boundary data of the flat geometry, the neighbourhood covers the real critical point corresponding to the flat geometry and the boundary data before the deformation. 



\begin{figure}[h]
\begin{center}
	\includegraphics[scale=0.22]{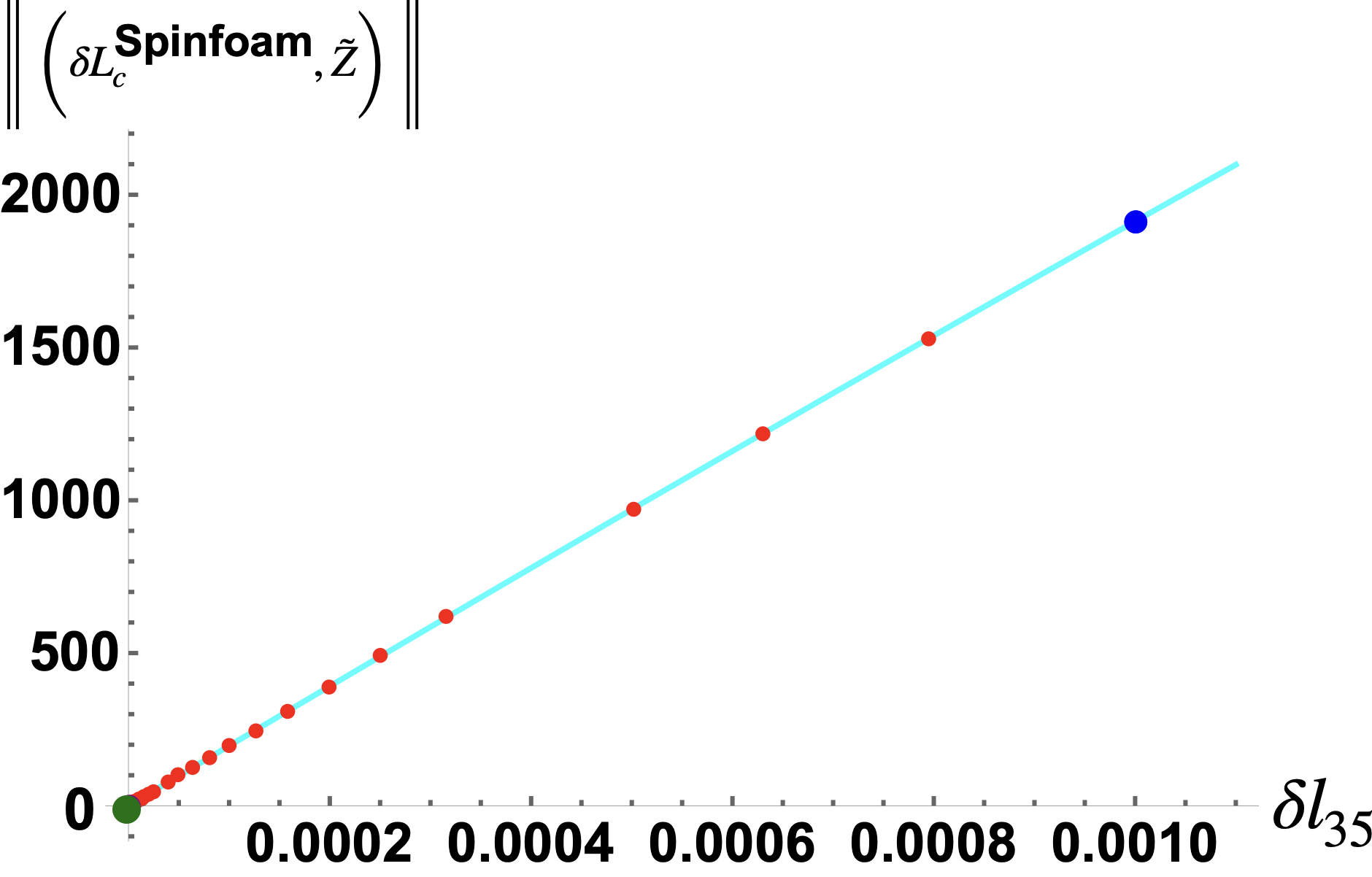}
 \end{center}
\caption{The red points are the list-plot of the norm of the complex critical point $(\delta L_c^{\text{Spinfoam}},\tilde{Z})$ v.s. the deformation of the boundary segment length $\delta l_{35}$. For any complex critical points $(\delta L_c^{\text{Spinfoam}},\tilde{Z})=(\delta L_c^{\text{Spinfoam}},z_1,z_2,\cdots,z_{241})$, the norm is defined as $\lVert(\delta L_c^{\text{Spinfoam}},\tilde{Z})\rVert=\sqrt{\abs{\delta L_c^{\text{Spinfoam}}}^2+\abs{z_1}^2+\abs{z_2}^2+\cdots+\abs{z_{241}}^2}$. Here, the boundary segment length $l_{35}$ is deformed from the flat geometry by $l_{35}\rightarrow l_{35}+\delta l_{35}$ at $\gamma=10^{-6}$, $\delta l_{35}\in [0,10^{-3}]$. The blue point is the complex critical point as $\delta l_{35}=10^{-3}$, and the green point is the real critical point at the origin $(0,0)$ corresponding to the flat geometry. The cyan curve represents the fitted function $\|(\delta L_c^{\text{Spinfoam}},\tilde{Z})\|\simeq 1.97\times 10^6\,\delta l_{35}-5.49\times 10^{7}\,(\delta l_{35})^2$.}
    \label{normZ} 
\end{figure}

For each $\gamma$, we would like to numerically compute the complex critical points $(\delta\cl,z)=(\delta L_c^{\text{Spinfoam}},\tilde{Z})(r')$ as the solution to the following equations,
\be
\partial_z\tilde{\mathcal{S}}(r',\delta\mathcal{L},z)&=&0,\label{newcomplexcritical0}\\
\partial_{\delta\cl}\tilde{\cs}(r',\delta\mathcal{L},z)&=&0\label{newcomplexcritical}.
\ee
Since we fix the boundary data $r'$ and vary $\gamma$, the complex critical points give a continuous trajectory parametrized by $\g$ in the complex space of $(\delta\cl,z)$. In the numerical computation, we sample a sequence of $\gamma \in [10^{-9},10^{6}]$ and compute the complex critical point for each $\gamma$ by the Newton-Raphson method, following the steps in \eqref{step1} - \eqref{step6}. For any $\gamma$, the recursion of the Newton-Raphson method can be initialized at the pseudo-critical point and give the convergent result within the desired tolerance. Moreover, all resulting complex critical points depend smoothly on the boundary data $\delta l_{35}$ and reduces to the real critical point when $\delta l_{35}\to 0$ (see Figure \ref{normZ} for an example).



The solution $\delta \cl$ from \eqref{newcomplexcritical0} and \eqref{newcomplexcritical} is the same as the solution of $\partial_{\delta L}\cs(r_l,Z(r_l))=0$. Indeed, 
\be
0=\partial_{\delta L}\cs(r_l,Z(r_l))&=&\frac{\partial\cs(r_l,Z(r_l))}{\partial{r_l}}\Big|_{Z(r_l)}\cdot \frac{\partial r_l}{\partial \delta L}+\frac{\partial\cs(r_l,Z(r_l))}{\partial Z(r_l)}\Big|_{r_l}\cdot \frac{\partial Z(r_l)}{\partial \delta L}\nonumber\\
&=&\frac{\partial\cs(r_l,Z(r_l))}{\partial{r_l}}\Big|_{Z(r_l)}\cdot \frac{\partial r_l}{\partial \delta L}=\lt[\partial_{\delta L}\cs(r_l,z)\rt]_{z=Z(r_l)},\label{qidiansan}
\ee
where we have used ${\partial\cs(r_l,Z(r_l))}/{\partial Z(r_l)}|_{r_l}=0$. $Z(r_l)$ depends on $\delta L$. $z=Z(r_l)$ is the solution of \eqref{newcomplexcritical0}, when analytic continuing $\delta L\to\delta\cl$. The result $\lt[\partial_{\delta L}\cs(r_l,z)\rt]_{z=Z(r_l)}=0$ from \eqref{qidiansan}, followed by analytic continuing $\delta L\to\delta\cl$, is equivalent to \eqref{newcomplexcritical} with the solution of \eqref{newcomplexcritical0} inserted.

\begin{figure}[h]
	\centering
	\includegraphics[scale=0.09]{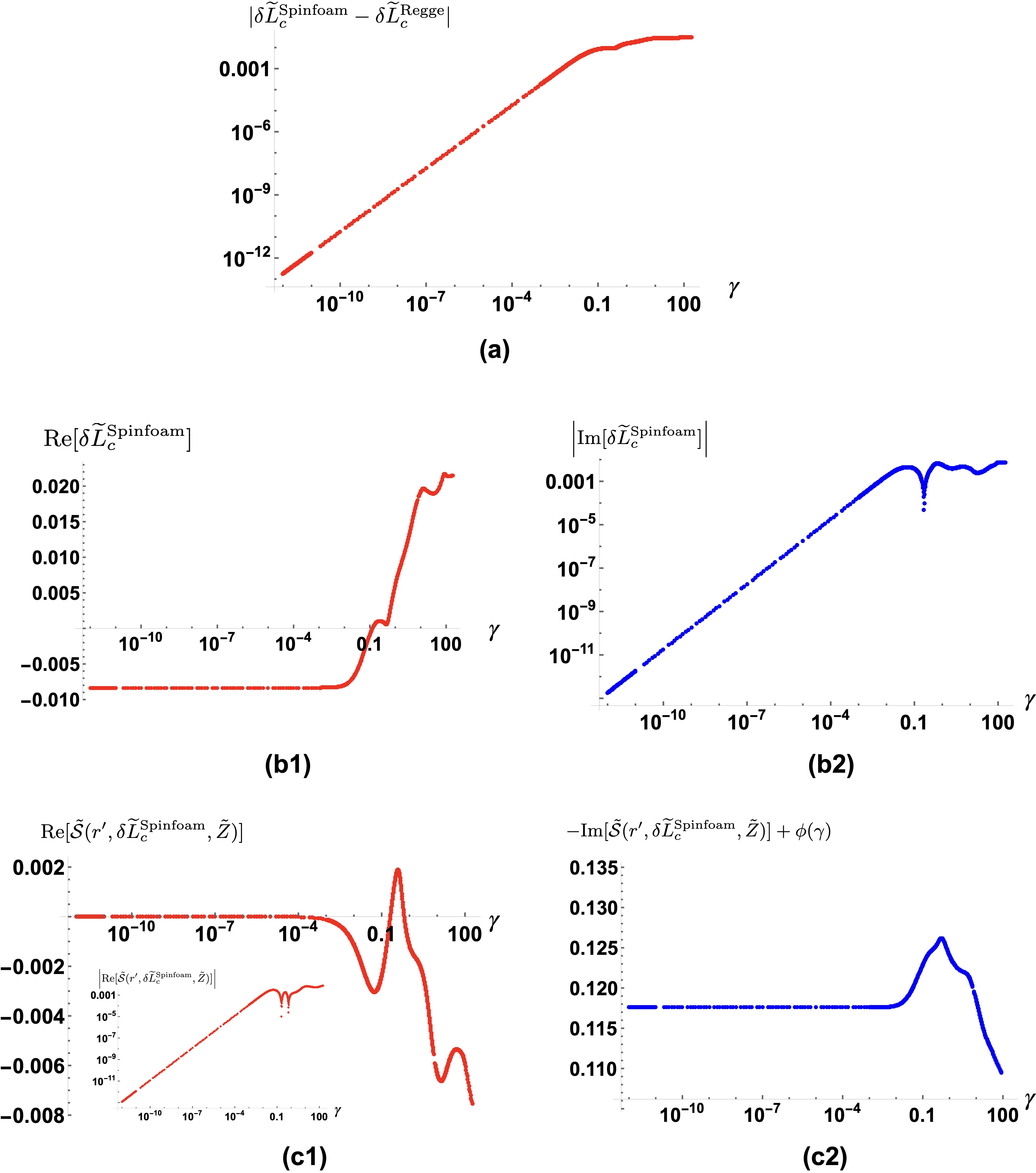}
	\caption{Panel (a) is the log-log plot of the distance between the spinfoam solution and the Regge solution in a neighborhood of $\delta \widetilde{L}=\delta \widetilde{L}_c^{\text{Regge}}$ as a function of $\gamma$. Panel (b1) shows the real of the spinfoam solution $\delta \widetilde{L}_c^{\text{Spinfoam}}$ v.s. $\gamma$. Panel (b2) is the log-log plot of the imaginary parts of the spinfoam solution $\delta \widetilde{L}_c^{\text{Spinfoam}}$ v.s. $\gamma$. Panel (c1) is the real parts of $\tilde{\cs}(r',\delta{\widetilde{\mathcal{L}}},z)$ at the complex critical points v.s. $\gamma$, and the small figure in (c1) is the log-log plot. Panel (c2) plots the imaginary parts of $\tilde{\cs}(r',\delta{\widetilde{\mathcal{L}}},z)$ at the complex critical points v.s. $\gamma$. }
	\label{soln2} 
\end{figure}

\begin{figure}[h]
	\centering
	\includegraphics[scale=0.12]{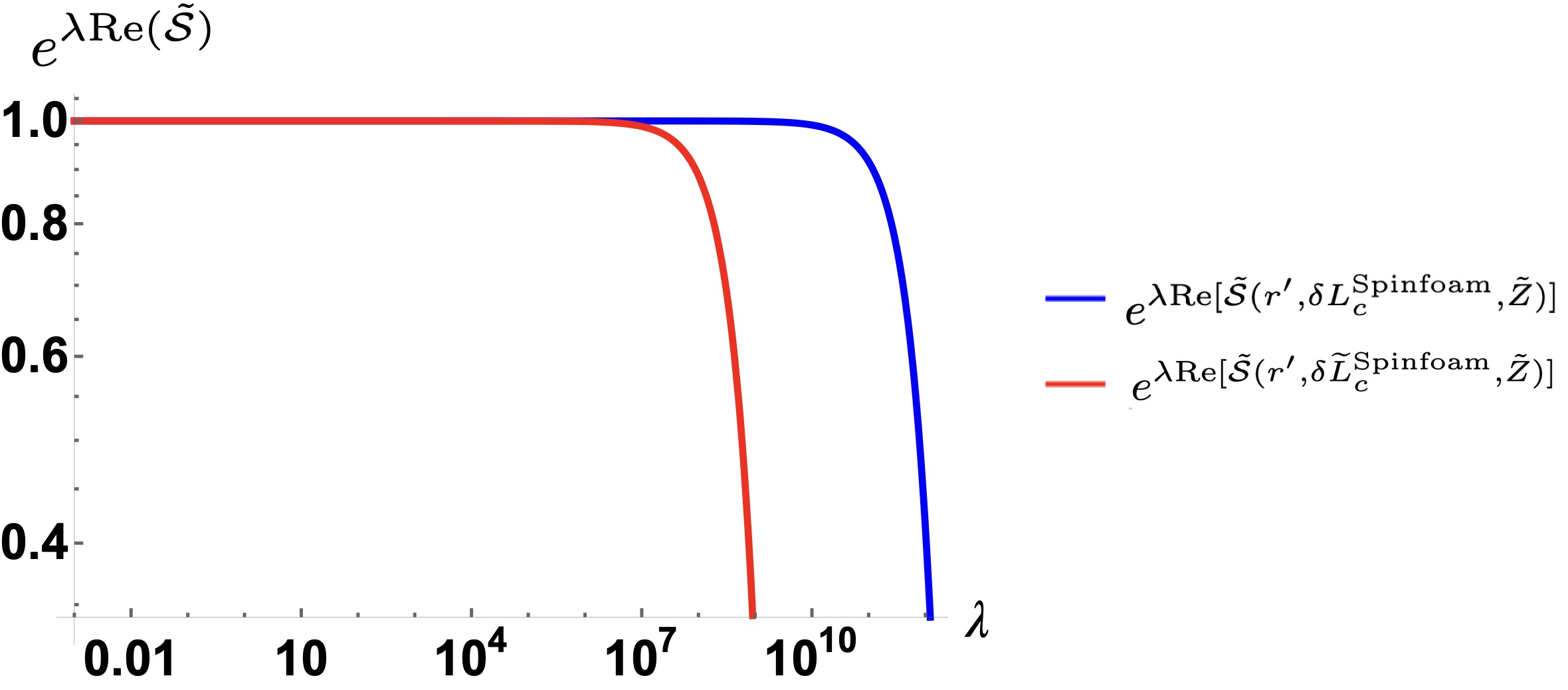}
	\caption{Figure is the log-log plot of $e^{\lambda\re[\tilde{\cs}(r',\delta L_c^{\text{Spinfoam}},\tilde{Z})]}$ (blue curve) and $e^{\lambda\re[ \tilde{\cs}(r',\delta\widetilde{L}_c^{\text{Spinfoam}},\tilde{Z})]}$ (red curve) as a function of $\lambda$ at $\gamma=10^{-8}$.}
	\label{lambda} 
\end{figure}

The complex critical point gives $\delta\cl\equiv \delta L_c^{\text{Spinfoam}}(\g)$ as a trajectory parametrized by $\gamma$ in a complex neighborhood at $\delta\cl =0$. This solution is compared to the Regge solution $\delta L_c^{\rm Regge}\simeq 0.000439$ (recall Figure \ref{FigureRegge}(a)). This solution $\delta L_c^{\text{Spinfoam}}(\g)$ is complex generically, although it is close to the real axis, especially for small $\g$. Figure \ref{Figure7} (a) demonstrates the distance (in the complex plane) between the spinfoam solution $\delta L_c^{\text{Spinfoam}}(\g)$ and the classical Regge solution $\delta L_c^{\text{Regge}}$:
\be
\left|\delta L_c^{\text{Spinfoam}}(\gamma)-\delta L_c^{\text{Regge}}\right|.\label{distanceLc}
\ee  
This distance is small in the small-$\gamma$ regime. So the classical Regge dynamics is reproduced by the spinfoam effective dynamics for small $\g$. This result is consistent with comparing the actions in Figure \ref{Figure6}. This result is also consistent with some earlier arguments in \cite{Magliaro:2011zz,Perini:2011uk,Magliaro:2011dz,Han:2013ina} about the semiclassical approximation of spinfoams with small $\gamma$.

The distance \eqref{distanceLc} becomes larger when increasing $\gamma$. It indicates that the spinfoam amplitude with larger $\gamma$ gives larger correction to the classical Regge solution. Therefore the effective theory in the large-$\gamma$ regime has more significant difference from the Regge gravity. Furthermore, the distance \eqref{distanceLc} stabilizes in the large-$\gamma$ regimes, as shown in Figure \ref{Figure7}(a). The distance value where it stablizes becomes smaller when the boundary data is closer to the one for the flat geometry, by comparing Figure \ref{Figure7}(a) and (b). The small and large $\gamma$ regimes might be viewed as two phases of the spinfoam amplitude. The effective dynamics is closer to the Regge dynamics for small $\gamma$ but more different from the Regge dynamics for large $\gamma$.

The critical point $(\delta L_c^{\text{Spinfoam}},\tilde{Z})(r')$ is generally complex for every $\gamma$ (see Figure \ref{AverageZ}). Figure \ref{StildevsGamma}(a) and (b) plot the analytic continued action $\tilde{\cs}(r',\delta\cl,z)$ (with the overall phase $\phi(\g)$ removed) evaluated at the complex critical points for a large number of samples of $\gamma$. The real part $\mathrm{Re}(\tilde{\cs})$ is close to zero for both the small-$\g$ and large-$\gamma$ regimes, so $e^{\lambda \tilde{\cs}}$ in the asymptotic formula \eqref{asymptotics0} is not suppressed for large $\l$ for both the small and large $\gamma$. The non-suppressed $e^{\lambda \tilde{\cs}}$ for small $\gamma$ has been anticipated since it can be predicted by the bound \eqref{bound}. But the non-suppressed $e^{\lambda \tilde{\cs}}$ with large $\lambda$ in the large-$\gamma$ regime violates the bound \eqref{bound}. This result suggests that the bound \eqref{bound} is not universal but only valid for the small or finite $\gamma$.

Figures \ref{Figure7}(b) plots $\left|\delta L_c^{\text{Spinfoam}}-\delta L_c^{\text{Regge}}\right|$ for the different boundary data, which deform the boundary data of the flat geometry by $l_{35}\rightarrow l_{35}+10^{-10}$. This boundary data is closer to the boundary data for the flat geometry. The results are qualitatively similar to the results from the previous boundary data, although the maximum of $\left|\delta L_c^{\text{Spinfoam}}-\delta L_c^{\text{Regge}}\right|$ become smaller comparing to the results from the previous boundary data. Changing the boundary data seems not to shift the location in the $\gamma$-space, where the small-$\g$ phase (where \eqref{distanceLc} is small) transits to the large-$\gamma$ phase (where \eqref{distanceLc} is stablizes), as suggested by comparing Figures \ref{Figure7} (a) and (b).

\subsection{Complex critical point and the other Regge solution $\delta \widetilde{L}_c^{\rm Regge}$}

Recall Figure \ref{FigureRegge}(a) that there is another classical Regge solution $\delta L=\delta \widetilde{L}_c^{\rm Regge}$ with the boundary condition under consideration. This solution corresponds to a different pseudo-critical point, which we use as the starting point of initializing the recursion in the Newton-Raphson method. Following the same procedure discussed above, we obtain a new trajectory of complex critical points parameterized by $\gamma$. The complex critical point gives $\delta\cl =\delta \widetilde{L}^{\rm Spinfoam}_c(\g)$, which is generically complex. Figure \ref{soln2} plots the distance $|\delta \widetilde{L}^{\rm Spinfoam}_c(\g)-\delta \widetilde{L}^{\rm Regge}_c|$, the real and imaginary part of the $\delta \widetilde{L}^{\rm Spinfoam}_c(\g)$, and the real and imaginary part of the action $\tilde{\cs}$ evaluated at the complex critical points. For small $\gamma$, $\delta \widetilde{L}^{\rm Spinfoam}_c(\g)$ is approximately real and close to the classical Regge solution $\delta \widetilde{L}_c^{\rm Regge}$. Increasing $\gamma$ results in that $\delta \widetilde{L}^{\rm Spinfoam}_c(\g)$ makes larger corrections to $\delta \widetilde{L}_c^{\rm Regge}$. 

Both the complex critical point here, denoted by $(\delta \widetilde{L}_c^{\rm Spinfoam},\tilde{Z})(r')$, and $(\delta {L}_c^{\rm Spinfoam},\tilde{Z})(r')$ discussed in the last subsection give contributions to $A(\Delta_3^2)$. When we compare their contributions. $e^{\l \cs}$ is suppressed faster at the critical point here than at the one in the last subsection (see Figure \ref{lambda}) for fixed $\g<0.1$. This relates to the fact that $\delta \widetilde{L}_c^{\rm Regge}$ gives larger deficit angles. Therefore the complex critical point here contributes to the amplitude much less than the one in the last subsection for generic small $\g$ and large $\l$. Recall that $\delta \widetilde{L}_c^{\rm Regge}$ likely relates to the discretization artifact. The result suggests that the spinfoam amplitude should suppress the contribution from the discretization artifact, in favor of a good continuum limit.

The complex critical points used in Figure \ref{soln2} are likely beyond the stationary phase approximation (for complex action) described above and below \eqref{criticaleqn111}, because these complex critical points do not analytically relate to the real critical point $(\mathring{j}_h,\mathring{g}_{ve},\mathring{z}_{vf})$ for the flat geometry. It relates to the existence of complex critical points with $\mathrm{Re}(\tilde{\cs})>0$ in Figure \ref{soln2}(c1) violating \eqref{negativeReS}. Indeed, when we continuously deform the boundary data $r'$ by the deformation by $l_{35}\rightarrow l_{35}+\delta l_{35}$ from the boundary data of flat geometry to the one that does not admit flat geometry, the solution of \eqref{newcomplexcritical0} and \eqref{newcomplexcritical} deforms analytically from the real critical point to the previous complex critical point $(\delta L_c^{\rm Spinfoam},\tilde{Z}) (r')$ (see Figure \ref{normZ}, and the similar property holds for the complex critical points in Section \ref{DoubleDelta}), but not to any of the complex critical points used in Figure \ref{soln2}.

The complex critical point used in Figure \ref{soln2} has to be studied by the fully-fledged Picard-Lefschetz theory (see, e.g. \cite{Lefschetz1975,Alexandru:2020wrj,Han:2020npv}). Consequently, given that the spinfoam amplitude is defined on the real integration cycle where $\mathrm{Re}(S)\leq 0$, the complex critical point with $\mathrm{Re}(\tilde{\cs})>0$ does not contribute to the asymptotics of the amplitude, because the steepest-ascent flow associated to this critical point turns out to have no intersection with the real integration cycle. Therefore, the contributions from the complex critical points in Figure \ref{lambda} are vanishing or suppressed for finite or larger $\gamma$, where $\mathrm{Re}(S)> 0$ or $e^{\l \mathrm{Re}(S)}$ is suppressed.

\section{Conclusion and Outlook}\label{Conclusion and Discussion}

Our above analysis demonstrates the importance of complex critical points in understanding the asymptotic behaviour of the spinfoam amplitude in the large-$j$ regime. In the case of the 4-simplex amplitude, taking into account the complex critical point generalizes the asymptotics to non-Regge boundary data and relates to the twisted geometry. In the case of the simplicial complex, the complex critical point plays an important role in deriving the effective dynamics from the spinfoam amplitude. The effective dynamics closely relate to the Regge gravity in the small $\gamma$ regime, as demonstrated by the numerical computation for the amplitude on the double-$\Delta_3$ complex.

In this paper, we examine the semiclassical behavior of the spinfoam amplitude within the regime of large-$j$. The semiclassical limit characterizes a scenario where the spinfoam amplitude exhibits behavior akin to classical gravity. This limit relates to the region where the values of Planck's constant are small, leading to the emergence of classical behavior. On the other hand, the continuous limit relates to the situation in which a discrete system approaches a continuous or smooth description. This typically involves taking a large number of discrete elements or degrees of freedom and allowing them to become infinitely numerous, resulting in a continuous and infinitely divisible system. It may relate to the situation that the triangulations underlining spinfoams are refined such that the geometries are made by refined Planckian size cells. Note that it is actually possible to relate certain refinement and small-$j$ spinfoam amplitudes to some semiclassical behaviors, as shown in \cite{Han:2019emp}. Generally speaking, while these two limits are related, they are not interchangeable. For spinfoams, both limits are relevant and may be taken simultaneously. It is indeed possible, as shown in \cite{Han:2018fmu,Han:2017xwo}, where the semiclassical gravity on the continuum is recovered in certain regime with both the large-$j$ and refinement. 

Our work provides a general procedure to derive the effective theory in the large-$j$ regime. From the perspective of semiclassical analysis, our numerical computation should be generalized to triangulations larger than double-$\Delta_3$, which has more internal segments. One should check if the Regge gravity still can be reproduced by the large-$j$ effective dynamics on larger triangulations.

The effective dynamics in LQG has been primarily investigated in the context of symmetry-reduced models, such as Loop Quantum Cosmology (LQG) and black holes, see, e.g. \cite{Ashtekar:2006wn,Ashtekar:2018lag}. The effective dynamics is useful in deriving the singularity resolution. Our result shows that the spinfoam amplitude also results in certain effective dynamics. However, this effective dynamics is in terms of the discrete Regge geometry, in contrast to the effective dynamics in terms of smooth fields in LQC and black holes. A research in progress is to understand if the effective dynamics from the spinfoam amplitude can relate to LQC and black holes. If the relation exists, it might provide a new approach toward embedding LQC and black hole models in the full theory of LQG.  

It is also interesting to investigate the behavior of the effective dynamics under the lattice refinement for spinfoam amplitudes. The Regge geometries approach to the continuum limit under the refinement, so we expect that the effective dynamics of Regge geometries from spinfoams should reduce to certain effective dynamics of the smooth geometry.

\section*{Acknowledgments}

The authors acknowledge the helpful communications with Bianca Dittrich, Carlo Rovelli, and Simone Speziale. M.H. receives support from the National Science Foundation through grants PHY-1912278 and PHY-2207763, and the sponsorship provided by the Alexander von Humboldt Foundation during his visit at FAU Erlangen-N\"urnberg. In addition, M.H. acknowledges IQG at FAU Erlangen-N\"urnberg, IGC at Penn State University, Perimeter Institute for Theoretical Physics, and University of Western Ontario for the hospitality during his visits. Research at Perimeter Institute is supported in part by the Government of Canada through the Department of Innovation, Science and Economic Development and by the Province of Ontario through the Ministry of Colleges and Universities.

\begin{appendix}
\section{Boundary data for single 4-simplex}\label{Appendix4Simplex}
In Section \ref{4Simplex3Dnormals}, we introduce the real critical points of the 4-simplex, which corresponds to the Regge geometry. We construct the Regge boundary geometry, Table \ref{4SimplexAreas}, \ref{4Simplex3Dnormals} and \ref{4SimplexXi} record areas $\mathring{\fa}_f=\gamma \mathring{j}_f$, 3d normals $\mathring{n}_{ef}$ and the corresponding spinors $\mathring{\xi}_{ef}$ of the single 4-simplex. 
\begin{table}[h]
	\centering\caption{Each cell shows the area of the face shared by line number tetrahedra and column number tetrahedra.}
	\label{4SimplexAreas}
	\scalebox{0.8}
	{
		\begin{tabular}{|c|c|c|c|c|c|}
			\hline
			\diagbox{$e$}{$\mathring{\fa}_{f}$}{$e'$}&$e'_1$&$e'_2$&$e'_3$&$e'_4$&$e'_5$\\
			\hline
			$e_1$&\diagbox{}{} &5&\diagbox{}{}&5& \diagbox{}{}\\
			\hline
			$e_2$&\diagbox{}{}&\diagbox{}{}&2 & \diagbox{}{}& 2 \\
			\hline
			$e_3$&5&\diagbox{}{}&\diagbox{}{}&2 & \diagbox{}{}\\
			\hline
			$e_4$&\diagbox{}{}&2&\diagbox{}{}&\diagbox{}{}&2\\
			\hline
			$e_5$&5&\diagbox{}{}&2&\diagbox{}{}&\diagbox{}{}\\
			\hline	
		\end{tabular}
	}
\end{table}

\begin{table}[h]
	\centering \caption{Each cell shows the 3d normal vectors of the face shared by line number tetrahedra and column number tetrahedra.}
	\label{4Simplex3Dnormals}
	\scalebox{0.72}
	{
		\begin{tabular}{|c|c|c|c|c|c|}
			\hline
			\diagbox{$e$}{$\mathring{n}_{ef}$}{$e'$}&$e'_1$&$e'_2$&$e'_3$&$e'_4$&$e'_5$\\
			\hline
			$e_1$&\diagbox{}{}& (1.00, 0, 0)& (-0.333, -0.943, 0)& (-0.333, 0.471, -0.816)& (-0.333, 0.471, 0.816)\\
			\hline
			$e_2$&(0.938, 0, -0.346)& \diagbox{}{}& (-0.782, -0.553, 0.289)& (-0.948, 0.276, -0.160)&(-0.616, 0.276, 0.738) \\
			\hline
			$e_3$&(-0.313, -0.884, -0.346)&(0.782, 0.553, 0.289)&\diagbox{}{}&(0.0553, 0.986, -0.160)&(-0.0553, 0.673, 0.738)\\
			\hline
			$e_4$&(-0.244, 0.345, -0.907)&(0.739, -0.215, 0.639)&(-0.0431, -0.768, 0.639)&\diagbox{}{}&(-0.0862, 0.122, 0.989)\\
			\hline
			$e_5$&(-0.436, 0.617, 0.655)&(0.859, -0.385, -0.338)&(0.0771, -0.938, -0.338)&(0.154, -0.218, -0.964)&\diagbox{}{}\\
			\hline	
		\end{tabular}
	}
\end{table}

\begin{table}[h]
	\centering\caption{Each cell shows a spinor $\xi_{ef}$ corresponding to a 3-normal to the face.}
	\label{4SimplexXi}
	\scalebox{0.68}
	{
		\begin{tabular}{|c|c|c|c|c|c|}
			\hline
			\diagbox{$e$}{$\mathring{\xi}_{ef}$}{$e'$}&$e'_1$&$e'_2$&$e'_3$&$e'_4$&$e'_5$\\
			\hline
			$e_1$&\diagbox{}{}& (0.707, -0.707)& (0.707, -0.236 - 0.667i)&(0.953, 0.175 - 0.247i)&(0.953, -0.175 + 0.247i)\\
			\hline
			$e_2$&(0.820, -0.572)&\diagbox{}{}&(0.803, -0.487 - 0.344i)&(0.762,0.622 - 0.181i)&(0.932, -0.330 + 0.148i) \\
			\hline
			$e_3$&(0.572, -0.273 - 0.774i)& (0.596, -0.655 - 0.463i)&\diagbox{}{}&(0.648, 0.043 + 0.761i)&(0.362, 0.076 - 0.929i)\\
			\hline
			$e_4$&(0.976, 0.125 - 0.177 i)&(0.905, 0.408 - 0.119 i)&(0.425, 0.051 + 0.904i)&\diagbox{}{}&(0.997, -0.0432 + 0.0611i)\\
			\hline
			$e_5$&(0.910, -0.240 + 0.339 i)&(0.818, -0.525 + 0.236i)&(0.576, 0.067 - 0.815 i)&(0.991, -0.0778 + 0.1100)&\diagbox{}{}\\
			\hline	
		\end{tabular}
	}
\end{table}
Table \ref{4Simplexg} and \ref{4Simplexz} record the values of the real critical point  $\mathring{g}_{ve}$ and $\mathring{\bold{z}}_{vf}$ for the 4-simplex with the boundary data $(\mathring{j}_f,\mathring{\xi}_{ef})$. 
\begin{table}[h]
	\centering \caption{Each cell of the table is the critical point of $\mathring{g}_{ve}$.}
	\label{4Simplexg}
	\scalebox{0.68}
	{
		\begin{tabular}{|c|c|c|c|c|c|}
			\hline
			$e$&$e_1$&$e_2$&$e_3$&$e_4$&$e_5$\\
			\hline
			$\mathring{g}_{ve}$ & $\left(\begin{array}{ll}
				0 & -\mathrm{i} \\
				-\mathrm{i} & 0
			\end{array}\right)$ & $\left(\begin{array}{ll}
				0 & -1.03\mathrm{i}\\-0.969\mathrm{i}& -0.358\mathrm{i}
			\end{array}\right)$ &  $\left(\begin{array}{ll}
				0& -1.03\mathrm{i}\\ -0.969\mathrm{i}& 0.337 + 0.119\mathrm{i}
			\end{array}\right)$ & $\left(\begin{array}{ll}
				0&-1.17\mathrm{i}\\-0.855\mathrm{i}& -0.149 + 0.105\mathrm{i}
			\end{array}\right)$&  $\left(\begin{array}{ll}
				0&-0.874\mathrm{i}\\-1.14\mathrm{i}& -0.199 + 0.141\mathrm{i}
			\end{array}\right)$\\
			\hline	
		\end{tabular}
	}
\end{table}

\begin{table}[h]
	\centering\caption{Each cell shows the critical points of $\mathring{\bold{z}}_{vf}$}
	\label{4Simplexz}
	\scalebox{0.7}
	{
		\begin{tabular}{|c|c|c|c|c|c|}
			\hline
			\diagbox{$e$}{$\mathring{\bold{z}}_{vf}$}{$e'$}&$e_1$&$e_2$&$e_3$&$e_4$&$e_5$\\
			\hline
			$e_1$&\diagbox{}{} &(1,-1)&\diagbox{}{}&$(1.00, 1.82 + 2.57 \mathrm{i})$& \diagbox{}{}\\
			\hline
			$e_2$&\diagbox{}{}&\diagbox{}{}&$(1.00, -0.915 + 0.402 \mathrm{i})$ & \diagbox{}{}& $(1.00, -1.41 - 0.31\mathrm{i})$ \\
			\hline
			$e_3$&$(1.00, -0.333 + 0.943\mathrm{i})$&\diagbox{}{}&\diagbox{}{}&$(1.00, 0.086 - 0.690\mathrm{i})$ & \diagbox{}{}\\
			\hline
			$e_4$&\diagbox{}{}&$(1.00, 1.86 + 0.99 \mathrm{i})$&\diagbox{}{}&\diagbox{}{}&$(1.00, 5.72 + 8.08 \mathrm{i})$\\
			\hline
			$e_5$&$(1.00, -1.82 - 2.57 \mathrm{i})$&\diagbox{}{}&$(1.00, 0.071 + 0.470\mathrm{i})$&\diagbox{}{}&\diagbox{}{}\\
			\hline	
		\end{tabular}
	}
\end{table}

All the Regge boundary data $\mathring{r}=(\mathring{j}_{f},\mathring{\xi}_{ef})$ and the data of the real critical point $(\mathring{g}_{ve}, \mathring{\bold{z}}_{vf})$ for the 4-simplex amplitude can be found in the Mathematica notebook \cite{qudx.org}. 

\section{The Newton-Raphson method}\label{The Newton-Raphson method}
The Newton-Raphson method for the single-variable equation $f(x)=0$ is initialized with a starting point $x_0$, and then one iterate
\be
x_{n+1}=x_{n}-\frac{f\left(x_{n}\right)}{f^{\prime}\left(x_{n}\right)}, 
\ee
to approach the solution with higher accuracy. In single 4-simplex case as an example, the equations of motion is 44 dimensions, we denote by
\be
 F\left(\left[\begin{array}{c}
 	z_1 \\
 	\vdots\\
 	z_{44}
 \end{array}\right]\right)=\left[\begin{array}{c}
 	f_{1}(z_1,...,z_{44}) \\
 	\vdots\\
 	f_{44}(z_1,...,z_{44}).
 \end{array}\right]\label{step1}
\ee 
The derivative of this system is the 44$\times$44 Jacobian given by:
\be
J(z_1,...,z_{44})=\left[\begin{array}{ccc}
	\frac{\partial f_{1}}{\partial  z_1} & ... &\frac{\partial f_{1}}{\partial z_{44}} \\
	\vdots& \vdots & \vdots\\
	\frac{\partial f_{44}}{\partial z_1} & ... & \frac{\partial f_{44}}{\partial z_{44}}
\end{array}\right]
\ee 
We define the function $G$ by
\be
G(z)=z-J(z)^{-1} F(z).
\ee 
The functional Newton-Raphson method for nonlinear systems is the iteration procedure that evolves from the initial $z^{(0)}$, which in our case is the real critical point $\mathring{x}$, and generates
\be
z^{(k)}=G\left(z^{(k-1)}\right)=z^{(k-1)}-J\left(z^{(k-1)}\right)^{-1} F\left(z^{(k-1)}\right) ,\qquad k\geq 1.
\ee 
We can write this as 
\be
\left[\begin{array}{c}
	z_{1}^{(k)} \\
	\vdots\\
	z_{44}^{(k)}
\end{array}\right]=\left[\begin{array}{c}
	z_{1}^{(k-1)} \\
	\vdots\\
	z_{44}^{(k-1)}
\end{array}\right]+\left[\begin{array}{c}
	\Delta z_{1}^{(k-1)} \\
	\vdots\\
	\Delta z_{44}^{(k-1)}
\end{array}\right] ,
\ee 
where 
\be
\left[\begin{array}{c}
	\Delta z_{1}^{(k-1)} \\
	\vdots\\
	\Delta z_{44}^{(k-1)}
\end{array}\right]=-J\left(z^{(k-1)}\right)^{-1} F\left(z^{(k-1)}\right).
\ee 
We set the desired tolerance 
$\epsilon=10^{-100}$, and we stop after $n$ iterations when 
\be
\sqrt{\left|(\Delta z_{1}^{(n-1)} )^2+\cdots+(\Delta z_{44}^{(n-1)})^2\right|}<\epsilon\label{step6}
\ee 
The resulting $z^{(n)}$ is the approximated solution within the tolerance. We evaluate the analytic continued 4-simplex action $\mathcal{S}$ at $z^{(n)}$ and apply it to the asymptotic formula \eqref{asymptotics0}.

\section{Boundary data for the $\Delta_3^2$ complex}
\subsection{Boundary data and the real critical point for the flat $\Delta_3^2$ complex} \label{flatdata}
We construct the flat geometry with the segment lengths in Table \ref{EdgeLength}. The corresponding boundary data for flat geometry is shown in Table \ref{flatBDdata1}, \ref{flatBDdata2}, \ref{flatBDdata3}, \ref{flatBDdata4}, \ref{flatBDdata5} and \ref{flatBDdata6}. Here, the area $\fa_f$ and the spins $j_f$ satisfy $\fa_f=\gamma j_f$.   
\begin{table}[h] 
	\centering \caption{Boundary data $(\mathring{\fa}_b,\mathring{\xi}_{eb})$ for the 4-simplex $v_1=\{1,2,3,4,6\}$}\label{flatBDdata1}
	\scalebox{0.6}{
		\begin{tabular}{|c|c|c|c|c|c|} 
			\hline
			\diagbox{\small{$e$}}{$\mathring{\xi}_{eb}$}{\small{$e'$}}&$e'_1$&$e'_2$&$e'_3$&$e'_4$&$e'_5$\\
			\hline
			$e_1$&\diagbox{}{}&\diagbox{}{}&\diagbox{}{}&\diagbox{}{}&(-0.41 + 0.73i, -0.15 - 0.52i)\\
			\hline
			$e_2$&\diagbox{}{}&\diagbox{}{}&\diagbox{}{}&(-0.61 + 0.22i, -0.76i)&\diagbox{}{}\\
			\hline
			$e_3$&\diagbox{}{}&\diagbox{}{}&\diagbox{}{}&\diagbox{}{}&(-0.078 - 0.033i, 0.04 - 1.0i)\\
			\hline
			$e_4$&(0.60, -0.66 - 0.46i)&\diagbox{}{}&(0.76, -0.04 - 0.65i)&\diagbox{}{}&\diagbox{}{}\\
			\hline
			$e_5$&\diagbox{}{}&(0.43, -0.18 - 0.88i)&\diagbox{}{}&(0.95, -0.03 + 0.31i)&\diagbox{}{}\\
			\hline
		\end{tabular}	
		
		\begin{tabular}{|c|c|c|c|c|c|}
			\hline
			\diagbox{\small{e}}{$\mathring{\fa}_b$}{\small{e'}}&$e'_1$&$e'_2$&$e'_3$&$e'_4$&$e'_5$\\
			\hline
			$e_1$&\diagbox{}{}&\diagbox{}{}&\diagbox{}{}&\diagbox{}{}&0.75\\
			\hline
			$e_2$&\diagbox{}{}&\diagbox{}{}&\diagbox{}{}&5&\diagbox{}{}\\
			\hline
			$e_3$&\diagbox{}{}&\diagbox{}{}&\diagbox{}{}&\diagbox{}{}&0.55\\
			\hline
			$e_4$&2&\diagbox{}{}&2&\diagbox{}{}&\diagbox{}{}\\
			\hline
                $e_5$&\diagbox{}{}&2.8&\diagbox{}{}&2.0&\diagbox{}{}\\
                \hline
		\end{tabular}	
	}
\end{table}

\begin{table}[h]
	\centering \caption{Boundary data $(\mathring{\fa}_b,\mathring{\xi}_{eb})$ for the 4-simplex $v_2=\{1,2,3,5,6\}$}\label{flatBDdata2}
	\scalebox{0.58}
	{
		\begin{tabular}{|c|c|c|c|c|c|} 
			\hline
			\diagbox{\small{$e$}}{$\mathring{\xi}_{eb}$}{\small{$e'$}}&$e'_2$&$e'_6$&$e'_7$&$e'_8$&$e'_9$\\
			\hline
			$e_2$&\diagbox{}{}&\diagbox{}{}&\diagbox{}{}&\diagbox{}{}&(-0.72 + 0.13 i, 0.02 - 0.68 i)\\
			\hline
			$e_6$&\diagbox{}{}&\diagbox{}{}&\diagbox{}{}&(0.81 i, -0.59i)&\diagbox{}{}\\
			\hline
			$e_7$&\diagbox{}{}&\diagbox{}{}&\diagbox{}{}&(-0.27 - 0.19i, -0.94i)&\diagbox{}{}\\
			\hline
			$e_8$&(0.71, -0.24 - 0.67 i)&\diagbox{}{}&\diagbox{}{}&\diagbox{}{}&(0.95, -0.17 + 0.25 i)\\
			\hline
			$e_9$&\diagbox{}{}&(0.74, -0.67 + 0.05i)&(1.0, 0.048 - 0.068i)&\diagbox{}{}&\diagbox{}{}\\
			\hline
		\end{tabular}
		
		\begin{tabular}{|c|c|c|c|c|c|}
			\hline
			\diagbox{\small{$e$}}{$\mathring{\fa}_b$}{\small{$e'$}}&$e'_2$&$e'_6$&$e'_7$&$e'_8$&$e'_9$\\
			\hline
			$e_2$&\diagbox{}{}&\diagbox{}{}&\diagbox{}{}&\diagbox{}{}&2.8\\
			\hline
			$e_6$&\diagbox{}{}&\diagbox{}{}&\diagbox{}{}&5&\diagbox{}{}\\
			\hline
			$e_7$&\diagbox{}{}&\diagbox{}{}&\diagbox{}{}&5&\diagbox{}{}\\
			\hline
			$e_8$&5&\diagbox{}{}&\diagbox{}{}&\diagbox{}{}&5\\
			\hline
                $e_9$&\diagbox{}{}&2.6&3.2&\diagbox{}{}&\diagbox{}{}\\
                \hline
		\end{tabular}
	}
\end{table}

\begin{table}[h]
	\centering\caption{Boundary data $(\mathring{\fa}_b,\mathring{\xi}_{eb})$ for the 4-simplex $v_3=\{1,2,4,5,6\}$}\label{flatBDdata3}
	\scalebox{0.57}
	{
		\begin{tabular}{|c|c|c|c|c|c|}
			\hline
			\diagbox{\small{$e$}}{$\mathring{\xi}_{eb}$}{\small{$e'$}}&$e'_3$&$e'_7$&$e'_{10}$&$e'_{11}$&$e'_{12}$\\
			\hline
			$e_3$&\diagbox{}{}&\diagbox{}{}&\diagbox{}{}&(-0.22 - 0.03 i, 0.07 - 0.97 i)&\diagbox{}{}\\
			\hline
			$e_7$&\diagbox{}{}&\diagbox{}{}&\diagbox{}{}&\diagbox{}{}&(-0.10 - 0.073i, -0.99i)\\
			\hline
			$e_{10}$&\diagbox{}{}&\diagbox{}{}&\diagbox{}{}&\diagbox{}{}&(0.18 + 0.98 i, 0.065 - 0.11 i)\\
			\hline
			$e_{11}$&\diagbox{}{}&(0.98, 0.12 - 0.18i)&(0.43, -0.87 + 0.25i)&\diagbox{}{}&\diagbox{}{}\\
			\hline
			$e_{12}$&(0.99, -0.01 - 0.17i)&\diagbox{}{}&\diagbox{}{}&(1.0, -0.018 + 0.025 i)&\diagbox{}{}\\
			\hline
		\end{tabular}
		
		\begin{tabular}{|c|c|c|c|c|c|}
			\hline
			\diagbox{\small{$e$}}{$\mathring{\fa}_b$}{\small{$e'$}}&$e'_3$&$e'_7$&$e'_{10}$&$e'_{11}$&$e'_{12}$\\
			\hline
			$e_3$&\diagbox{}{}&\diagbox{}{}&\diagbox{}{}&2&\diagbox{}{}\\
			\hline
			$e_7$&\diagbox{}{}&\diagbox{}{}&\diagbox{}{}&\diagbox{}{}&3.2\\
			\hline
			$e_{10}$&\diagbox{}{}&\diagbox{}{}&\diagbox{}{}&\diagbox{}{}&0.69\\
			\hline
			$e_{11}$&\diagbox{}{}&5&2&\diagbox{}{}&\diagbox{}{}\\
			\hline
			$e_{12}$&0.55&\diagbox{}{}&\diagbox{}{}&2&\diagbox{}{}\\
			\hline
		\end{tabular}
	}
\end{table}

\begin{table}[h]
	\centering\caption{Boundary data $(\mathring{\fa}_b,\mathring{\xi}_{eb})$ for the 4-simplex $v_4=\{1,2,3,4,7\}$}\label{flatBDdata4}
	\scalebox{0.56}
	{
		\begin{tabular}{|c|c|c|c|c|c|}
			\hline
			\diagbox{\small{$e$}}{$\mathring{\xi}_{eb}$}{\small{$e'$}}&$e'_1$&$e'_{13}$&$e'_{14}$&$e'_{15}$&$e'_{16}$\\
			\hline
			$e_1$&\diagbox{}{}&\diagbox{}{}&\diagbox{}{}&(-0.33 + 0.75 i, -0.11 - 0.56 i)&\diagbox{}{}\\
			\hline
			$e_{13}$&\diagbox{}{}&\diagbox{}{}&\diagbox{}{}&\diagbox{}{}&(-0.52 + 0.71 i, -0.35 - 0.32 i)\\
			\hline
			$e_{14}$&\diagbox{}{}&\diagbox{}{}&\diagbox{}{}&(-0.59 + 0.71 i, -0.18 - 0.35 i)&\diagbox{}{}\\
			\hline
			$e_{15}$&\diagbox{}{}&(0.90, -0.14 - 0.41 i)&\diagbox{}{}&\diagbox{}{}&(0.63, 0.33 + 0.71 i)\\
			\hline
			$e_{16}$&(0.94, -0.25 - 0.22 i)&\diagbox{}{}&(0.94, 0.28 - 0.18i)&\diagbox{}{}&\diagbox{}{}\\
			\hline
		\end{tabular}
		
		\begin{tabular}{|c|c|c|c|c|c|}
			\hline
			\diagbox{\small{$e$}}{$\mathring{\fa}_b$}{\small{$e'$}}&$e'_1$&$e'_{13}$&$e'_{14}$&$e'_{15}$&$e'_{16}$\\
			\hline
			$e_1$&\diagbox{}{}&\diagbox{}{}&\diagbox{}{}&2&\diagbox{}{}\\
			\hline
			$e_{13}$&\diagbox{}{}&\diagbox{}{}&\diagbox{}{}&\diagbox{}{}&3.2\\
			\hline
			$e_{14}$&\diagbox{}{}&\diagbox{}{}& 2.1&\diagbox{}{}&\diagbox{}{}\\
			\hline
			$e_{15}$&\diagbox{}{}&5.6&\diagbox{}{}&\diagbox{}{}&2.3\\
			\hline
			$e_{16}$&0.75&\diagbox{}{}&0.5&\diagbox{}{}&\diagbox{}{}\\
			\hline
		\end{tabular}
	}
\end{table}

\begin{table}[h]
	\centering \caption{Boundary data $(\mathring{\fa}_b,\mathring{\xi}_{eb})$ for the 4-simplex $v_5=\{1,2,3,5,7\}$}\label{flatBDdata5}
	\scalebox{0.6}
	{
		\begin{tabular}{|c|c|c|c|c|c|} 
			\hline
			\diagbox{\small{$e$}}{$\mathring{\xi}_{eb}$}{\small{$e'$}}&$e'_{6}$&$e'_{13}$&$e'_{17}$&$e'_{18}$&$e'_{19}$\\
			\hline
			$e_6$&\diagbox{}{}&\diagbox{}{}&\diagbox{}{}&\diagbox{}{}&(0.04 + 0.77 i, 0.01 - 0.63 i)\\
			\hline
			$e_{13}$&\diagbox{}{}&\diagbox{}{}&\diagbox{}{}&(-0.48 + 0.71 i, -0.31 - 0.41 i)&\diagbox{}{}\\
			\hline
			$e_{17}$&\diagbox{}{}&\diagbox{}{}&\diagbox{}{}&(-0.19 + 0.17 i, -0.18 - 0.95 i)&(-0.05 + 0.25 i, -0.06 - 0.97 i)\\
			\hline
			$e_{18}$&(0.90, -0.43)&\diagbox{}{}&\diagbox{}{}&\diagbox{}{}&\diagbox{}{}\\
			\hline
			$e_{19}$&\diagbox{}{}&(0.71, -0.26 - 0.65 i)&\diagbox{}{}&(0.95, 0.19 + 0.25 i)&\diagbox{}{}\\
			\hline
		\end{tabular}
		
		\begin{tabular}{|c|c|c|c|c|c|}
			\hline
			\diagbox{\small{$e$}}{$\mathring{\fa}_b$}{\small{$e'$}}&$e'_{6}$&$e'_{13}$&$e'_{17}$&$e'_{18}$&$e'_{19}$\\
			\hline
			$e_6$&\diagbox{}{}&\diagbox{}{}&\diagbox{}{}&\diagbox{}{}&2.6\\
			\hline
			$e_{13}$&\diagbox{}{}&\diagbox{}{}&\diagbox{}{}&5.6&\diagbox{}{}\\
			\hline
			$e_{17}$&\diagbox{}{}&\diagbox{}{}&\diagbox{}{}&5.4&3.5\\
			\hline
			$e_{18}$&5&\diagbox{}{}&\diagbox{}{}&\diagbox{}{}&\diagbox{}{}\\
			\hline
                $e_{19}$&\diagbox{}{}&3.2&\diagbox{}{}&5.2&\diagbox{}{}\\
                \hline
		\end{tabular}
	}
\end{table}

\begin{table}[H]
	\centering\caption{Boundary data $(\mathring{\fa}_b,\mathring{\xi}_{eb})$ for the 4-simplex $v_6=\{1,2,4,5,7\}$}\label{flatBDdata6}
	\scalebox{0.55}
	{
		\begin{tabular}{|c|c|c|c|c|c|}
			\hline
			\diagbox{\small{$e$}}{$\mathring{\xi}_{eb}$}{\small{$e'$}}&$e'_{10}$&$e'_{14}$&$e'_{17}$&$e'_{20}$&$e'_{21}$\\
			\hline
			$e_{10}$&\diagbox{}{}&\diagbox{}{}&\diagbox{}{}&(0.20 + 0.91 i, 0.07 - 0.35 i)&\diagbox{}{}\\
			\hline
			$e_{14}$&\diagbox{}{}&\diagbox{}{}&\diagbox{}{}&\diagbox{}{}&(-0.55 + 0.68 i, -0.16 - 0.46 i)\\
			\hline
			$e_{17}$&\diagbox{}{}&\diagbox{}{}&\diagbox{}{}&\diagbox{}{}&\diagbox{}{}\\
			\hline
			$e_{20}$&\diagbox{}{}&(0.76, 0.22 - 0.61 i)&(0.74, 0.57 - 0.36 i)&\diagbox{}{}&(0.85, 0.52 - 0.1 i)\\
			\hline
			$e_{21}$&(0.95, -0.31 + 0.07 i)&\diagbox{}{}&(0.39, 0.89 - 0.23 i)&\diagbox{}{}&\diagbox{}{}\\
			\hline
		\end{tabular}
		
		\begin{tabular}{|c|c|c|c|c|c|}
			\hline
			\diagbox{\small{$e$}}{$\mathring{\fa}_b$}{\small{$e'$}}&$e'_{10}$&$e'_{14}$&$e'_{17}$&$e'_{20}$&$e'_{21}$\\
			\hline
			$e_{10}$&\diagbox{}{}&\diagbox{}{}&\diagbox{}{}&2&\diagbox{}{}\\
			\hline
			$e_{14}$&\diagbox{}{}&\diagbox{}{}&\diagbox{}{}&\diagbox{}{}&0.5\\
			\hline
			$e_{17}$&\diagbox{}{}&\diagbox{}{}&\diagbox{}{}&\diagbox{}{}&\diagbox{}{}\\
			\hline
			$e_{20}$&\diagbox{}{}&2.1&5.4&\diagbox{}{}&2.4\\
			\hline
			$e_{21}$&0.69&\diagbox{}{}&3.5&\diagbox{}{}&\diagbox{}{}\\
			\hline
		\end{tabular}
	}
\end{table}

Once the flat geometry is constructed, the real critical points $\left(\mathring{j}_h,\mathring{g}_{ve}, \mathring{\bold{z}}_{vf}\right)$ can be obtained by solving the critical equations Eq.(\ref{eom1}) and (\ref{eom2}). The solution of the critical point equations relates to the Lorentzian Regge geometry, as described in \cite{Han:2011re,Barrett:2009mw}. $\mathring{g}_{ve}$ relates to the Lorentzian transformation acting on each tetrahedron and glueing them together to form the $\Delta_3^2$ complex. In this model, we fix $g_{ve}$ to be constant $\Slc$ matrices for $v_1 e_5, v_2e_9, v_3 e_{12}, v_4e_{16},v_5e_{19}, v_6e_{21}$. The group elements $g_{ve}$ for the bulk tetrahedra $v_1 e_1, v_1e_2, v_2e_6,v_2e_7, v_3 e_3, v_3e_{10}, v_4e_{13}, v_5e_{17}, v_6e_{14}$ are fixed to be the upper triangular matrix. For the $\Delta^2_3$ triangulation, there are five internal faces $h(12k)$ with $k=3,4,5,6,7$. The areas of these internal faces are shown in Table \ref{a12k}. 
\begin{table}[h] \label{a12k}
	\centering\caption{Areas of internal faces $h$ in $\Delta_3^2$ complex.}
	\begin{tabular}{|c|c|c|c|c|}
		\hline	$\fa_{h(123)}$&$\fa_{h(124)}$&$\fa_{h(125)}$&$\fa_{h(126)}$&$\fa_{h(127)}$\\
		\hline
		0.971&0.333&1.55&1.78&1.93\\
		\hline
	\end{tabular}
\end{table}
The numerical results of the real critical point $(\mathring{g}_{ve},\,\mathring{\textbf{z}}_{vf})$ corresponding to the flat geometry are listed in Table \ref{tab:ga1}, \ref{tab:ga2}, \ref{tab:ga3}, \ref{tab:ga4}, \ref{tab:ga5} and \ref{tab:ga6}. 
\begin{table}[h]
	\centering\caption{The real critical point $(\mathring{g}_{ve},\, \mathring{\textbf{z}}_{vf})$ for the 4-simplex $v_1=(1,2,3,4,6)$}\label{tab:ga1}
	\scalebox{0.53}
	{
		\begin{tabular}{|c|c|c|c|}
			\hline
			\small{e}&$e_1$&$e_2$&$e_3$\\
			\hline
			$\mathring{g}_{v_1e}$ &$\left(\begin{matrix}
				0.96&0.42 + 0.04 i\\	0&1
			\end{matrix}\right)$&$\left(\begin{matrix}
				0.99& -0.05 - 0.15i\\	0&1
			\end{matrix}\right)$&$\left(\begin{matrix}
				0.77& -0.13 - 0.72i\\	0&1.3
			\end{matrix}\right)$\\
			\hline
			\small{e}&$e_4$&$e_5$&\\
			\hline
			$\mathring{g}_{v_1e}$&$\left(\begin{matrix}
				0& -1.0 i\\	-0.97i&0.34 + 0.12i
			\end{matrix}\right)$&$\left(\begin{matrix}
				0&-1.1 i\\	-0.91 i& 0.46 + 0.12i
			\end{matrix}\right)$&\\
			\hline
		\end{tabular}
		
		\begin{tabular}{|c|c|c|c|c|c|}
			\hline
			\diagbox{\small{$e$}}{$|\mathring{\textbf{z}}_{v_1f}\rangle$}{\small{$e'$}}&$e'_1$&$e'_2$&$e'_3$&$e'_4$&$e'_5$\\
			\hline
			$e_1$&\diagbox{}{}&\diagbox{}{}&(1,-0.94 + 0.69i)&\diagbox{}{}&(1,-0.82 + 0.45i)\\
			\hline
			$e_2$&(1,0.87 - 0.49i)&\diagbox{}{}&\diagbox{}{}&(1,-0.33 + 0.94i)&\diagbox{}{}\\
			\hline
			$e_3$&\diagbox{}{}&(1,-0.1 + 1.5i)&\diagbox{}{}&\diagbox{}{}&(1,2.5 + 6.0i)\\
			\hline
			$e_4$&(1,-0.92 + 0.40i)&\diagbox{}{}&(1,0.3 + 2.1i)&\diagbox{}{}&\diagbox{}{}\\
			\hline
                $e_5$&\diagbox{}{}&(1,-0.14 + 0.75i)&\diagbox{}{}&(1,0.2 - 1.4i)&\diagbox{}{}\\
			\hline
		\end{tabular}
		
	}
\end{table}

\begin{table}[h]
	\centering\caption{The real critical point $(\mathring{g}_{ve},\, \mathring{\textbf{z}}_{vf})$ for the 4-simplex $v_2=(1,2,3,5,6)$.}\label{tab:ga2}
	\scalebox{0.52}
	{
		\begin{tabular}{|c|c|c|c|}
			\hline
			\small{$e$}&$e_2$&$e_6$&$e_7$\\
			\hline
			$\mathring{g}_{v_2e}$ &$\left(\begin{matrix}
				0.99& -0.05 - 0.15 i\\	0.99&-0.05 - 0.15 i
			\end{matrix}\right)$&$\left(\begin{matrix}
				0.98& 0.32\\	0&1
			\end{matrix}\right)$&$\left(\begin{matrix}
				1.0& -0.031 + 0.044 i\\	0& 0.96
			\end{matrix}\right)$\\
			\hline
			\small{$e$}&$e_8$&$e_9$&\\
			\hline
			$\mathring{g}_{v_2 e}$ &$\left(\begin{matrix}
				0& -1.0 i\\	-1.0 i&0
			\end{matrix}\right)$&$\left(\begin{matrix}
				1.26&0.09-0.13i\\	0.09+0.13i&0.82
			\end{matrix}\right)$&\\
			\hline
		\end{tabular}
		
		\begin{tabular}{|c|c|c|c|c|c|}
			\hline
			\diagbox{\small{$e$}}{$|\mathring{\textbf{z}}_{v_2f}\rangle$}{\small{$e'$}}&$e'_2$&$e'_6$&$e'_7$&$e'_8$&$e'_9$\\
			\hline
			$e_2$&\diagbox{}{}&\diagbox{}{}&(1,-0.1 + 1.5 i)&\diagbox{}{}&(1,-0.14 + 0.75i)\\
			\hline
			$e_6$&(1,0.87 - 0.49i)&\diagbox{}{}&\diagbox{}{}&(1, 0.87 - 0.49i)&\diagbox{}{}\\
			\hline
			$e_7$&\diagbox{}{}&(1,-0.86 - 0.07i)&\diagbox{}{}&(1,1.8 + 2.6i)&\diagbox{}{}\\
			\hline
			$e_8$&(1,-0.33 + 0.94i)&\diagbox{}{}&\diagbox{}{}&\diagbox{}{}&(1,-1.8 - 2.6 i)\\
			\hline
                $e_9$&\diagbox{}{}&(1,-1.09 - 0.05i)&(1,4.9 + 7.0 i)&\diagbox{}{}&\diagbox{}{} \\
                \hline
		\end{tabular}
	}
\end{table}

\begin{table}[h]
	\centering\caption{The real critical point $(\mathring{g}_{ve},\, \mathring{\textbf{z}}_{vf})$ for the 4-simplex $v_3=(1,2,4,5,6)$.}\label{tab:ga3}
	\scalebox{0.56}
	{
		\begin{tabular}{|c|c|c|c|}
			\hline
			\small{$e$}&$e_3$&$e_7$&$e_{10}$\\
			\hline
			$\mathring{g}_{v_3e}$ &$\left(\begin{matrix}
				0.77& -0.13 - 0.72i\\	0& 1.3
			\end{matrix}\right)$&$\left(\begin{matrix}
				1.0& -0.031 + 0.044 i\\	0& 0.96
			\end{matrix}\right)$&$\left(\begin{matrix}
				0.96& 0.38\\	0&1
			\end{matrix}\right)$\\
			\hline
			\small{$e$}&$e_{11}$&$e_{12}$&\\
			\hline
			$\mathring{g}_{v_3 e}$ &$\left(\begin{matrix}
				0& -1.2i\\	-0.86i& -0.15 + 0.11i
			\end{matrix}\right)$&$\left(\begin{matrix}
				0& -1.8 i\\	-0.55 i& -0.16 + 0.12i
			\end{matrix}\right)$&\\
			\hline
		\end{tabular}
		
		\begin{tabular}{|c|c|c|c|c|c|}
			\hline
			\diagbox{\small{$e$}}{$|\mathring{z}_{v_3f}\rangle$}{\small{$e'$}}&$e'_{3}$&$e'_{7}$&$e'_{10}$&$e'_{11}$&$e'_{12}$\\
			\hline
			$e_{3}$&\diagbox{}{}&\diagbox{}{}&(1,-0.94 + 0.69i)&(1,0.3 + 2.1i)&\diagbox{}{}\\
			\hline
			$e_{7}$&(1,-0.1 + 1.5i)&\diagbox{}{}&\diagbox{}{}&\diagbox{}{}&(1, 4.9 + 7.0i)\\
			\hline
			$e_{10}$&\diagbox{}{}&(1,-0.86 - 0.07i)&\diagbox{}{}&\diagbox{}{}&(1,-0.45 - 0.08i)\\
			\hline
			$e_{11}$&\diagbox{}{}&(1,1.8 + 2.6i)&(1,-0.68 - 0.15i)&\diagbox{}{}&\diagbox{}{}\\
			\hline
			$e_{12}$&(1,2.5 + 6.0i)&\diagbox{}{}&\diagbox{}{}&(1,5.7 + 8.1 i)&\diagbox{}{}\\
			\hline
		\end{tabular}
	}
\end{table}   

\begin{table}[h]
	\centering\caption{The real critical point $(\mathring{g}_{ve},\, \mathring{\textbf{z}}_{vf})$ for the 4-simplex $v_4=(1,2,3,4,7)$.}\label{tab:ga4}
	\scalebox{0.52}
	{
		\begin{tabular}{|c|c|c|c|}
			\hline
			\small{$e$}&$e_1$&$e_{13}$&$e_{14}$\\
			\hline
			$\mathring{g}_{v_4e}$ &$\left(\begin{matrix}
				0.96& 0.42 + 0.04i\\	0& 1
			\end{matrix}\right)$&$\left(\begin{matrix}
				0.84& 0.82 + 0.19i\\	0& 1.2
			\end{matrix}\right)$&$\left(\begin{matrix}
				0.68& 1.3 + 0.9 i\\	0&1.5
			\end{matrix}\right)$\\
			\hline
			\small{$e$}&$e_{15}$&$e_{16}$&\\
			\hline
			$\mathring{g}_{v_4 e}$ &$\left(\begin{matrix}
				0& -1.3i\\	-0.79i & -0.34 - 0.92i
			\end{matrix}\right)$&$\left(\begin{matrix}
				0& -1.3 i\\	-0.77i& -0.49 - 1.01i
			\end{matrix}\right)$&\\
			\hline
		\end{tabular}
		
		\begin{tabular}{|c|c|c|c|c|c|}
			\hline
			\diagbox{\small{$e$}}{$|\mathring{z}_{v_4f}\rangle$}{\small{$e'$}}&$e'_{1}$&$e'_{13}$&$e'_{14}$&$e'_{15}$&$e'_{16}$\\
			\hline
			$e_{1}$&\diagbox{}{}&(1,0.87 - 0.49i)&\diagbox{}{}&(1,-0.92 + 0.40 i)&\diagbox{}{}\\
			\hline
			$e_{13}$&\diagbox{}{}&\diagbox{}{}&(1,-0.92 + 0.75i)&\diagbox{}{}&(1, -0.73 + 0.54i)\\
			\hline
			$e_{14}$&(1,-0.94 + 0.69i)&\diagbox{}{}&\diagbox{}{}&(1,-0.94 + 0.77i)&\diagbox{}{}\\
			\hline
			$e_{15}$&\diagbox{}{}&(1,-0.83 + 0.56i)&\diagbox{}{}&\diagbox{}{}&(1,-1.1 - 1.2i)\\
			\hline
			$e_{16}$&(1,-0.82 + 0.45i)&\diagbox{}{}&(1,-1.0 + 0.81i)&\diagbox{}{}&\diagbox{}{}\\
			\hline
		\end{tabular}
	}
\end{table} 

\begin{table}[H]
	\centering\caption{The real critical point $(\mathring{g}_{ve},\, \mathring{\textbf{z}}_{vf})$ for the 4-simplex $v_5=(1,2,3,5,7)$.}\label{tab:ga5}
	\scalebox{0.55}
	{
		\begin{tabular}{|c|c|c|c|}
			\hline
			\small{$e$}&$e_6$&$e_{13}$&$e_{17}$\\
			\hline
			$\mathring{g}_{v_5e}$ &$\left(\begin{matrix}
				0.98& 0.32\\	0& 1
			\end{matrix}\right)$&$\left(\begin{matrix}
				0.84& 0.82 + 0.19 i\\	0& 1.2
			\end{matrix}\right)$&$\left(\begin{matrix}
				0.84& 0.73 - 0.05 i\\	0&1.2
			\end{matrix}\right)$\\
			\hline
			\small{$e$}&$e_{18}$&$e_{19}$&\\
			\hline
			$\mathring{g}_{v_5 e}$ &$\left(\begin{matrix}
				0& -1.1i\\	-0.88 i& -0.72 i
			\end{matrix}\right)$&$\left(\begin{matrix}
				0& -1.2i\\	-0.86 i& 0.03 - 0.72i
			\end{matrix}\right)$&\\
			\hline
		\end{tabular}
		
		\begin{tabular}{|c|c|c|c|c|c|}
			\hline
			\diagbox{\small{$e$}}{$|\mathring{z}_{v_5f}\rangle$}{\small{$e'$}}&$e'_{6}$&$e'_{13}$&$e'_{17}$&$e'_{18}$&$e'_{19}$\\
			\hline
			$e_{6}$&\diagbox{}{}&\diagbox{}{}&(1,-0.86 - 0.07i)&\diagbox{}{}&(1,-1.09 - 0.05i)\\
			\hline
			$e_{13}$&(1,0.87 - 0.49i)&\diagbox{}{}&\diagbox{}{}&(1,-0.83 + 0.56i)&\diagbox{}{}\\
			\hline
			$e_{17}$&\diagbox{}{}&(1,-0.92 + 0.75i)&\diagbox{}{}&\diagbox{}{}&(1,1,-3.2 + 0.6i)\\
			\hline
			$e_{18}$&(1,-1)&\diagbox{}{}&(1,-1.9 + 2.2i)&\diagbox{}{}&\diagbox{}{}\\
			\hline
			$e_{19}$&\diagbox{}{}&(1,-0.73 + 0.54i)&\diagbox{}{}&(1,-1.8 - 0.8 i)&\diagbox{}{}\\
			\hline
		\end{tabular}
	}
\end{table} 

\begin{table}[h]
	\centering\caption{The real critical point $(\mathring{g}_{ve},\, \mathring{\textbf{z}}_{vf})$ for the 4-simplex $v_6=(1,2,4,5,7)$.}\label{tab:ga6}
	\scalebox{0.55}
	{
		\begin{tabular}{|c|c|c|c|}
			\hline
			\small{$e$}&$e_{10}$&$e_{14}$&$e_{17}$\\
			\hline
			$\mathring{g}_{v_6e}$ &$\left(\begin{matrix}
				0.96, 0.38 \\	0& 1
			\end{matrix}\right)$&$\left(\begin{matrix}
				0.68& 1.3 + 0.9 i\\	0& 1.5
			\end{matrix}\right)$&$\left(\begin{matrix}
				0.84& 0.73 - 0.05 i\\	0&1.2
			\end{matrix}\right)$\\
			\hline
			\small{$e$}&$e_{20}$&$e_{21}$&\\
			\hline
			$\mathring{g}_{v_6 e}$ &$\left(\begin{matrix}
				0& -1.1i\\	-0.93 i& 0.17-0.96 i
			\end{matrix}\right)$&$\left(\begin{matrix}
				0& -1.2i\\	-0.84 i& 0.4 - 2.3i
			\end{matrix}\right)$&\\
			\hline
		\end{tabular}
		
		\begin{tabular}{|c|c|c|c|c|c|}
			\hline
			\diagbox{\small{$e$}}{$|\mathring{z}_{v_6f}\rangle$}{\small{$e'$}}&$e'_{10}$&$e'_{14}$&$e'_{17}$&$e'_{20}$&$e'_{21}$\\
			\hline
			$e_{10}$&\diagbox{}{}&(1,-0.94 + 0.69i)&\diagbox{}{}&(1,-0.68 - 0.15i)&\diagbox{}{}\\
			\hline
			$e_{14}$&\diagbox{}{}&\diagbox{}{}&(1,-0.92 + 0.75i)&\diagbox{}{}&(1,-1+0.81i)\\
			\hline
			$e_{17}$&(1,-0.86 - 0.07i)&\diagbox{}{}&\diagbox{}{}&(1,-1.9+2.2i)&\diagbox{}{}\\
			\hline
			$e_{20}$&\diagbox{}{}&(1,-0.94 + 0.77i)&\diagbox{}{}&\diagbox{}{}&(1,-2.7 - 0.4i)\\
			\hline
			$e_{21}$&(1,-0.45 - 0.08i)&\diagbox{}{}&(1,-3.2+0.6i)&\diagbox{}{}&\diagbox{}{}\\
			\hline
		\end{tabular}
	}
\end{table}

All the boundary data $\mathring{r}=(\mathring{j}_{b},\mathring{\xi}_{eb})$ and the data of the real critical point $(\mathring{j}_h, \mathring{g}_{ve}, \mathring{\textbf{z}}_{vf})$ can be found in the Mathematica notebook in \cite{qudx.org}. 

\subsection{Boundary data and the pseudo critical points for the curved $\Delta_3^2$ complex} \label{curveddata}
The boundary data in Appendix \ref{flatdata} admits a flat geometry. To construct a curved geometry, we deform the segment length $l_{35}\rightarrow l_{35}+10^{-3}$ and keep the other boundary segment lengths unchanged. We list the boundary data for this curved geometry in Table  \ref{curveBDdata1}, \ref{curvedBDdata2}, \ref{curvedDdata3}, \ref{curvedBDdata4}, \ref{curvedBDdata5} and \ref{curvedBDdata6} as the internal segment length is $l_{12} = L_0+\delta L_c^{\text{Regge}}$. 
\begin{table}[h] 
	\centering \caption{Boundary data $(\fa_b,\xi_{eb})$ of the curved geometry for the 4-simplex $v_1=\{1,2,3,4,6\}$}\label{curveBDdata1}
	\scalebox{0.6}{
		\begin{tabular}{|c|c|c|c|c|c|} 
			\hline
			\diagbox{\small{$e$}}{$\xi_{eb}$}{\small{$e'$}}&$e'_1$&$e'_2$&$e'_3$&$e'_4$&$e'_5$\\
			\hline
			$e_1$&\diagbox{}{}&\diagbox{}{}&\diagbox{}{}&\diagbox{}{}&(-0.40 + 0.73i, -0.15 - 0.53i)\\
			\hline
			$e_2$&\diagbox{}{}&\diagbox{}{}&\diagbox{}{}&(-0.61 + 0.22i, - 0.76i)&\diagbox{}{}\\
			\hline
			$e_3$&\diagbox{}{}&\diagbox{}{}&\diagbox{}{}&\diagbox{}{}&(-0.079 - 0.033i, 0.04 - 1.0i)\\
			\hline
			$e_4$&(0.60, -0.66 - 0.46i)&\diagbox{}{}&(0.76, -0.04 - 0.65i)&\diagbox{}{}&\diagbox{}{}\\
			\hline
			$e_5$&\diagbox{}{}&(0.43, -0.18 - 0.88i)&\diagbox{}{}&(0.95, -0.03 + 0.31i)&\diagbox{}{}\\
			\hline
		\end{tabular}	
		
		\begin{tabular}{|c|c|c|c|c|c|}
			\hline
			\diagbox{\small{e}}{$\fa_b$}{\small{e'}}&$e'_1$&$e'_2$&$e'_3$&$e'_4$&$e'_5$\\
			\hline
			$e_1$&\diagbox{}{}&\diagbox{}{}&\diagbox{}{}&\diagbox{}{}&0.75\\
			\hline
			$e_2$&\diagbox{}{}&\diagbox{}{}&\diagbox{}{}&5&\diagbox{}{}\\
			\hline
			$e_3$&\diagbox{}{}&\diagbox{}{}&\diagbox{}{}&\diagbox{}{}&0.55\\
			\hline
			$e_4$&2&\diagbox{}{}&2&\diagbox{}{}&\diagbox{}{}\\
			\hline
                $e_5$&\diagbox{}{}&2.8&\diagbox{}{}&2.0&\diagbox{}{}\\
                \hline
		\end{tabular}	
	}
\end{table}

\begin{table}[h]
	\centering \caption{Boundary data $(\fa_b,\xi_{eb})$ of the curved geometry for the 4-simplex $v_2=\{1,2,3,5,6\}$}\label{curvedBDdata2}
	\scalebox{0.58}
	{
		\begin{tabular}{|c|c|c|c|c|c|} 
			\hline
			\diagbox{\small{$e$}}{$\xi_{eb}$}{\small{$e'$}}&$e'_2$&$e'_6$&$e'_7$&$e'_8$&$e'_9$\\
			\hline
			$e_2$&\diagbox{}{}&\diagbox{}{}&\diagbox{}{}&\diagbox{}{}&(-0.71 + 0.13i, 0.02 - 0.69i)\\
			\hline
			$e_6$&\diagbox{}{}&\diagbox{}{}&\diagbox{}{}&(0.81 i, -0.59i)&\diagbox{}{}\\
			\hline
			$e_7$&\diagbox{}{}&\diagbox{}{}&\diagbox{}{}&(-0.27 - 0.19i, -0.94i)&\diagbox{}{}\\
			\hline
			$e_8$&(0.71, -0.24 - 0.67 i)&\diagbox{}{}&\diagbox{}{}&\diagbox{}{}&(0.95, -0.17 + 0.25 i)\\
			\hline
			$e_9$&\diagbox{}{}&(0.74, -0.67 + 0.05i)&(1.0, 0.049 - 0.065i)&\diagbox{}{}&\diagbox{}{}\\
			\hline
		\end{tabular}
		
		\begin{tabular}{|c|c|c|c|c|c|}
			\hline
			\diagbox{\small{$e$}}{$\fa_b$}{\small{$e'$}}&$e'_2$&$e'_6$&$e'_7$&$e'_8$&$e'_9$\\
			\hline
			$e_2$&\diagbox{}{}&\diagbox{}{}&\diagbox{}{}&\diagbox{}{}&2.8\\
			\hline
			$e_6$&\diagbox{}{}&\diagbox{}{}&\diagbox{}{}&5&\diagbox{}{}\\
			\hline
			$e_7$&\diagbox{}{}&\diagbox{}{}&\diagbox{}{}&5&\diagbox{}{}\\
			\hline
			$e_8$&5&\diagbox{}{}&\diagbox{}{}&\diagbox{}{}&5\\
			\hline
                $e_9$&\diagbox{}{}&2.6&3.2&\diagbox{}{}&\diagbox{}{}\\
                \hline
		\end{tabular}
	}
\end{table}

\begin{table}[h]
	\centering\caption{Boundary data $(\fa_b,\xi_{eb})$ of curved geometry for the 4-simplex $v_3=\{1,2,4,5,6\}$}\label{curvedDdata3}
	\scalebox{0.57}
	{
		\begin{tabular}{|c|c|c|c|c|c|}
			\hline
			\diagbox{\small{$e$}}{$\xi_{eb}$}{\small{$e'$}}&$e'_3$&$e'_7$&$e'_{10}$&$e'_{11}$&$e'_{12}$\\
			\hline
			$e_3$&\diagbox{}{}&\diagbox{}{}&\diagbox{}{}&(-0.22 - 0.03 i, 0.07 - 0.97 i)&\diagbox{}{}\\
			\hline
			$e_7$&\diagbox{}{}&\diagbox{}{}&\diagbox{}{}&\diagbox{}{}&(-0.105 - 0.072i, -0.99i)\\
			\hline
			$e_{10}$&\diagbox{}{}&\diagbox{}{}&\diagbox{}{}&\diagbox{}{}&(0.18 + 0.98 i, 0.065 - 0.106 i)\\
			\hline
			$e_{11}$&\diagbox{}{}&(0.98, 0.12 - 0.18i)&(0.43, -0.87 + 0.25i)&\diagbox{}{}&\diagbox{}{}\\
			\hline
			$e_{12}$&(0.99, -0.01 - 0.17i)&\diagbox{}{}&\diagbox{}{}&(1.0, -0.018 + 0.025 i)&\diagbox{}{}\\
			\hline
		\end{tabular}
		
		\begin{tabular}{|c|c|c|c|c|c|}
			\hline
			\diagbox{\small{$e$}}{$\fa_b$}{\small{$e'$}}&$e'_3$&$e'_7$&$e'_{10}$&$e'_{11}$&$e'_{12}$\\
			\hline
			$e_3$&\diagbox{}{}&\diagbox{}{}&\diagbox{}{}&2.0&\diagbox{}{}\\
			\hline
			$e_7$&\diagbox{}{}&\diagbox{}{}&\diagbox{}{}&\diagbox{}{}&3.2\\
			\hline
			$e_{10}$&\diagbox{}{}&\diagbox{}{}&\diagbox{}{}&\diagbox{}{}&0.69\\
			\hline
			$e_{11}$&\diagbox{}{}&5&2&\diagbox{}{}&\diagbox{}{}\\
			\hline
			$e_{12}$&0.55&\diagbox{}{}&\diagbox{}{}&2&\diagbox{}{}\\
			\hline
		\end{tabular}
	}
\end{table}

\begin{table}[h]
	\centering\caption{Boundary data $(\fa_b,\xi_{eb})$ of curved geometry for the 4-simplex $v_4=\{1,2,3,4,7\}$}\label{curvedBDdata4}
	\scalebox{0.56}
	{
		\begin{tabular}{|c|c|c|c|c|c|}
			\hline
			\diagbox{\small{$e$}}{$\xi_{eb}$}{\small{$e'$}}&$e'_1$&$e'_{13}$&$e'_{14}$&$e'_{15}$&$e'_{16}$\\
			\hline
			$e_1$&\diagbox{}{}&\diagbox{}{}&\diagbox{}{}&(-0.33 + 0.75 i, -0.12 - 0.57 i)&\diagbox{}{}\\
			\hline
			$e_{13}$&\diagbox{}{}&\diagbox{}{}&\diagbox{}{}&\diagbox{}{}&(-0.52 + 0.71 i, -0.35 - 0.32 i)\\
			\hline
			$e_{14}$&\diagbox{}{}&\diagbox{}{}&\diagbox{}{}&(-0.58 + 0.71 i, -0.19 - 0.35 i)&\diagbox{}{}\\
			\hline
			$e_{15}$&\diagbox{}{}&(0.90, -0.14 - 0.41 i)&\diagbox{}{}&\diagbox{}{}&(0.63, 0.33 + 0.71 i)\\
			\hline
			$e_{16}$&(0.94, -0.25 - 0.22 i)&\diagbox{}{}&(0.94, 0.28 - 0.18i)&\diagbox{}{}&\diagbox{}{}\\
			\hline
		\end{tabular}
		
		\begin{tabular}{|c|c|c|c|c|c|}
			\hline
			\diagbox{\small{$e$}}{$\fa_b$}{\small{$e'$}}&$e'_1$&$e'_{13}$&$e'_{14}$&$e'_{15}$&$e'_{16}$\\
			\hline
			$e_1$&\diagbox{}{}&\diagbox{}{}&\diagbox{}{}&2&\diagbox{}{}\\
			\hline
			$e_{13}$&\diagbox{}{}&\diagbox{}{}&\diagbox{}{}&\diagbox{}{}&3.2\\
			\hline
			$e_{14}$&\diagbox{}{}&\diagbox{}{}& 2.1&\diagbox{}{}&\diagbox{}{}\\
			\hline
			$e_{15}$&\diagbox{}{}&5.6&\diagbox{}{}&\diagbox{}{}&2.3\\
			\hline
			$e_{16}$&0.75&\diagbox{}{}&0.5&\diagbox{}{}&\diagbox{}{}\\
			\hline
		\end{tabular}
	}
\end{table}

\begin{table}[H]
	\centering \caption{Boundary data $(\fa_b,\xi_{eb})$ of the curved geometry for the 4-simplex $v_5=\{1,2,3,5,7\}$}\label{curvedBDdata5}
	\scalebox{0.6}
	{
		\begin{tabular}{|c|c|c|c|c|c|} 
			\hline
			\diagbox{\small{$e$}}{$\xi_{eb}$}{\small{$e'$}}&$e'_{6}$&$e'_{13}$&$e'_{17}$&$e'_{18}$&$e'_{19}$\\
			\hline
			$e_6$&\diagbox{}{}&\diagbox{}{}&\diagbox{}{}&\diagbox{}{}&(0.04 + 0.77 i, 0.01 - 0.64 i)\\
			\hline
			$e_{13}$&\diagbox{}{}&\diagbox{}{}&\diagbox{}{}&(-0.48 + 0.71 i, -0.31 - 0.41 i)&\diagbox{}{}\\
			\hline
			$e_{17}$&\diagbox{}{}&\diagbox{}{}&\diagbox{}{}&(-0.19 + 0.17 i, -0.18 - 0.95 i)&(-0.05 + 0.25 i, -0.05 - 0.97 i)\\
			\hline
			$e_{18}$&(0.90, -0.43)&\diagbox{}{}&\diagbox{}{}&\diagbox{}{}&\diagbox{}{}\\
			\hline
			$e_{19}$&\diagbox{}{}&(0.71, -0.26 - 0.66 i)&\diagbox{}{}&(0.95, 0.19 + 0.24 i)&\diagbox{}{}\\
			\hline
		\end{tabular}
		
		\begin{tabular}{|c|c|c|c|c|c|}
			\hline
			\diagbox{\small{$e$}}{$\fa_b$}{\small{$e'$}}&$e'_{6}$&$e'_{13}$&$e'_{17}$&$e'_{18}$&$e'_{19}$\\
			\hline
			$e_6$&\diagbox{}{}&\diagbox{}{}&\diagbox{}{}&\diagbox{}{}&2.6\\
			\hline
			$e_{13}$&\diagbox{}{}&\diagbox{}{}&\diagbox{}{}&5.6&\diagbox{}{}\\
			\hline
			$e_{17}$&\diagbox{}{}&\diagbox{}{}&\diagbox{}{}&5.4&3.5\\
			\hline
			$e_{18}$&5&\diagbox{}{}&\diagbox{}{}&\diagbox{}{}&\diagbox{}{}\\
			\hline
                $e_{19}$&\diagbox{}{}&3.2&\diagbox{}{}&5.2&\diagbox{}{}\\
                \hline
		\end{tabular}
	}
\end{table}

\begin{table}[h]
	\centering\caption{Boundary data $(\fa_b,\xi_{eb})$ of the curved geometry for the 4-simplex $v_6=\{1,2,4,5,7\}$}\label{curvedBDdata6}
	\scalebox{0.55}
	{
		\begin{tabular}{|c|c|c|c|c|c|}
			\hline
			\diagbox{\small{$e$}}{$\xi_{eb}$}{\small{$e'$}}&$e'_{10}$&$e'_{14}$&$e'_{17}$&$e'_{20}$&$e'_{21}$\\
			\hline
			$e_{10}$&\diagbox{}{}&\diagbox{}{}&\diagbox{}{}&(0.20 + 0.91 i, 0.07 - 0.35 i)&\diagbox{}{}\\
			\hline
			$e_{14}$&\diagbox{}{}&\diagbox{}{}&\diagbox{}{}&\diagbox{}{}&(-0.55 + 0.68 i, -0.16 - 0.47 i)\\
			\hline
			$e_{17}$&\diagbox{}{}&\diagbox{}{}&\diagbox{}{}&\diagbox{}{}&\diagbox{}{}\\
			\hline
			$e_{20}$&\diagbox{}{}&(0.76, 0.22 - 0.61 i)&(0.74, 0.57 - 0.36 i)&\diagbox{}{}&(0.85, 0.52 - 0.1 i)\\
			\hline
			$e_{21}$&(0.95, -0.31 + 0.07 i)&\diagbox{}{}&(0.39, 0.89 - 0.23 i)&\diagbox{}{}&\diagbox{}{}\\
			\hline
		\end{tabular}
		
		\begin{tabular}{|c|c|c|c|c|c|}
			\hline
			\diagbox{\small{$e$}}{$\fa_b$}{\small{$e'$}}&$e'_{10}$&$e'_{14}$&$e'_{17}$&$e'_{20}$&$e'_{21}$\\
			\hline
			$e_{10}$&\diagbox{}{}&\diagbox{}{}&\diagbox{}{}&2&\diagbox{}{}\\
			\hline
			$e_{14}$&\diagbox{}{}&\diagbox{}{}&\diagbox{}{}&\diagbox{}{}&0.5\\
			\hline
			$e_{17}$&\diagbox{}{}&\diagbox{}{}&\diagbox{}{}&\diagbox{}{}&\diagbox{}{}\\
			\hline
			$e_{20}$&\diagbox{}{}&2.1&5.4&\diagbox{}{}&2.4\\
			\hline
			$e_{21}$&0.69&\diagbox{}{}&3.5&\diagbox{}{}&\diagbox{}{}\\
			\hline
		\end{tabular}
	}
\end{table}
The curved geometry does not have real critical point. However, we can find the pseudo-critical point $(j_{h}^0,g_{ve}^0,\bold{z}_{vf}^0)$, which is close to the real critical point inside the real integration domain. The pseudo-critical point satisfies the critical equation (\ref{eom1}) but violates critical equation (\ref{eom2}). The data for the pseudo-critical point is listed in Table \ref{curve:ga1}, \ref{curve:ga2}, \ref{curve:ga3}, \ref{curve:ga4}, \ref{curve:ga5} and \ref{curve:ga6}. 
\begin{table}[h]
	\centering\caption{The pseudo-critical point $(g^0_{ve},\, \textbf{z}^0_{vf})$ for the 4-simplex $v_1=(1,2,3,4,6)$}\label{curve:ga1}
	\scalebox{0.48}
	{
		\begin{tabular}{|c|c|c|c|}
			\hline
			\small{e}&$e_1$&$e_2$&$e_3$\\
			\hline
			$g^0_{v_1e}$ &$\left(\begin{matrix}
				0.96&0.40 + 0.02 i\\	0&1
			\end{matrix}\right)$&$\left(\begin{matrix}
				0.99& -0.06 - 0.16i\\	0&1
			\end{matrix}\right)$&$\left(\begin{matrix}
				0.78& -0.12 - 0.71i\\	-0.00024 - 0.00065i& 1.29 
			\end{matrix}\right)$\\
			\hline
			\small{e}&$e_4$&$e_5$&\\
			\hline
			$g^0_{v_1e}$&$\left(\begin{matrix}
				-0.0016 - 0.0001i& -1.0 i\\	-0.97i&0.34 + 0.12i
			\end{matrix}\right)$&$\left(\begin{matrix}
				0&-1.1 i\\	-0.91 i& 0.46 + 0.12i
			\end{matrix}\right)$&\\
			\hline
		\end{tabular}
		
		\begin{tabular}{|c|c|c|c|c|c|}
			\hline
			\diagbox{\small{$e$}}{$|\bold{z}^0_{v_1f}\rangle$}{\small{$e'$}}&$e'_1$&$e'_2$&$e'_3$&$e'_4$&$e'_5$\\
			\hline
			$e_1$&\diagbox{}{}&\diagbox{}{}&(1,-0.95 + 0.70i)&\diagbox{}{}&(1,-0.82 + 0.45i)\\
			\hline
			$e_2$&(1, 0.87 - 0.50i)&\diagbox{}{}&\diagbox{}{}&(1,-0.34 + 0.95i)&\diagbox{}{}\\
			\hline
			$e_3$&\diagbox{}{}&(1,-0.1 + 1.5i)&\diagbox{}{}&\diagbox{}{}&(1,2.5 + 6.0i)\\
			\hline
			$e_4$&(1,-0.92 + 0.40i)&\diagbox{}{}&(1,0.3 + 2.1i)&\diagbox{}{}&\diagbox{}{}\\
			\hline
                $e_5$&\diagbox{}{}&(1,-0.14 + 0.75i)&\diagbox{}{}&(1,0.2 - 1.4i)&\diagbox{}{}\\
			\hline
		\end{tabular}
			}
\end{table}

\begin{table}[h]
	\centering\caption{The pseudo-critical point $(g^0_{ve},\, \textbf{z}^0_{vf})$ for the 4-simplex $v_2=(1,2,3,5,6)$.}\label{curve:ga2}
	\scalebox{0.5}
	{
		\begin{tabular}{|c|c|c|c|}
			\hline
			\small{$e$}&$e_2$&$e_6$&$e_7$\\
			\hline
			$g^0_{v_2e}$ &$\left(\begin{matrix}
				0.99& -0.05 - 0.15 i\\	0.0024-0.0112i&1.01
			\end{matrix}\right)$&$\left(\begin{matrix}
				0.98& 0.30\\	0&1
			\end{matrix}\right)$&$\left(\begin{matrix}
				1.0& -0.029 + 0.048 i\\	0& 0.97
			\end{matrix}\right)$\\
			\hline
			\small{$e$}&$e_8$&$e_9$&\\
			\hline
			$g^0_{v_2 e}$ &$\left(\begin{matrix}
				0.0008+0.00056i& -1.0 i\\	-1.0 i&-0.0054-0.0011i
			\end{matrix}\right)$&$\left(\begin{matrix}
				0&-0.98i\\	-1.0i&-0.029+0.016i
			\end{matrix}\right)$&\\
			\hline
		\end{tabular}
		
		\begin{tabular}{|c|c|c|c|c|c|}
			\hline
			\diagbox{\small{$e$}}{$|\textbf{z}^0_{v_2f}\rangle$}{\small{$e'$}}&$e'_2$&$e'_6$&$e'_7$&$e'_8$&$e'_9$\\
			\hline
			$e_2$&\diagbox{}{}&\diagbox{}{}&(1,-0.1 + 1.5 i)&\diagbox{}{}&(1,-0.14 + 0.75i)\\
			\hline
			$e_6$&(1,0.87 - 0.48i)&\diagbox{}{}&\diagbox{}{}&(1, -1)&\diagbox{}{}\\
			\hline
			$e_7$&\diagbox{}{}&(1,-0.86 - 0.07i)&\diagbox{}{}&(1,1.8 + 2.6i)&\diagbox{}{}\\
			\hline
			$e_8$&(1,-0.33 + 0.94i)&\diagbox{}{}&\diagbox{}{}&\diagbox{}{}&(1,-1.8 - 2.6 i)\\
			\hline
                $e_9$&\diagbox{}{}&(1,-1.09 - 0.05i)&(1,4.7 + 6.9i)&\diagbox{}{}&\diagbox{}{} \\
                \hline
		\end{tabular}
	}
\end{table}

\begin{table}[h]
	\centering\caption{The real critical point $(g^0_{ve},\, \textbf{z}^0_{vf})$ for the 4-simplex $v_3=(1,2,4,5,6)$.}\label{curve:ga3}
	\scalebox{0.50}
	{
		\begin{tabular}{|c|c|c|c|}
			\hline
			\small{$e$}&$e_3$&$e_7$&$e_{10}$\\
			\hline
			$g^0_{v_3e}$ &$\left(\begin{matrix}
				0.78& -0.13 - 0.72i\\	0& 1.29
			\end{matrix}\right)$&$\left(\begin{matrix}
				1.04& -0.030 + 0.046 i\\	-0.0010+0.0018i& 0.96
			\end{matrix}\right)$&$\left(\begin{matrix}
				0.96& 0.38\\	0&1
			\end{matrix}\right)$\\
			\hline
			\small{$e$}&$e_{11}$&$e_{12}$&\\
			\hline
			$g^0_{v_3 e}$ &$\left(\begin{matrix}
				-0.00013-0.0001i& -1.2i\\	-0.85i& -0.15 + 0.11i
			\end{matrix}\right)$&$\left(\begin{matrix}
				0& -1.8 i\\	-0.55 i& -0.16 + 0.12i
			\end{matrix}\right)$&\\
			\hline
		\end{tabular}
		
		\begin{tabular}{|c|c|c|c|c|c|}
			\hline
			\diagbox{\small{$e$}}{$|\bold{z}^0_{v_3f}\rangle$}{\small{$e'$}}&$e'_{3}$&$e'_{7}$&$e'_{10}$&$e'_{11}$&$e'_{12}$\\
			\hline
			$e_{3}$&\diagbox{}{}&\diagbox{}{}&(1,-0.94 + 0.69i)&(1,0.3 + 2.1i)&\diagbox{}{}\\
			\hline
			$e_{7}$&(1,-0.1 + 1.5i)&\diagbox{}{}&\diagbox{}{}&\diagbox{}{}&(1, 4.9 + 7.0i)\\
			\hline
			$e_{10}$&\diagbox{}{}&(1,-0.86 - 0.07i)&\diagbox{}{}&\diagbox{}{}&(1,-0.45 - 0.08i)\\
			\hline
			$e_{11}$&\diagbox{}{}&(1,1.8 + 2.6i)&(1,-0.68 - 0.15i)&\diagbox{}{}&\diagbox{}{}\\
			\hline
			$e_{12}$&(1,2.5 + 6.0i)&\diagbox{}{}&\diagbox{}{}&(1,5.7 + 8.1 i)&\diagbox{}{}\\
			\hline
		\end{tabular}
	}
\end{table}   

\begin{table}[h]
	\centering\caption{The pseudo-critical point $(g^0_{ve},\, \textbf{z}^0_{vf})$ for the 4-simplex $v_4=(1,2,3,4,7)$.}\label{curve:ga4}
	\scalebox{0.48}
	{
		\begin{tabular}{|c|c|c|c|}
			\hline
			\small{$e$}&$e_1$&$e_{13}$&$e_{14}$\\
			\hline
			$g^0_{v_4e}$ &$\left(\begin{matrix}
				0.96& 0.42 + 0.04i\\	0.02-0.02i& 1.05
			\end{matrix}\right)$&$\left(\begin{matrix}
				0.84& 0.82 + 0.2i\\	0& 1.2
			\end{matrix}\right)$&$\left(\begin{matrix}
				0.68& 1.3 + 0.9 i\\	-0.0023+0.0038i&1.5+0.01i
			\end{matrix}\right)$\\
			\hline
			\small{$e$}&$e_{15}$&$e_{16}$&\\
			\hline
			$g^0_{v_4 e}$ &$\left(\begin{matrix}
				0.0032-0.0015i& -1.3i\\	-0.79i & -0.34 - 0.92i
			\end{matrix}\right)$&$\left(\begin{matrix}
				0& -1.3 i\\	-0.77i& -0.49 - 1.01i
			\end{matrix}\right)$&\\
			\hline
		\end{tabular}
		
		\begin{tabular}{|c|c|c|c|c|c|}
			\hline
			\diagbox{\small{$e$}}{$|\bold{z}^0_{v_4f}\rangle$}{\small{$e'$}}&$e'_{1}$&$e'_{13}$&$e'_{14}$&$e'_{15}$&$e'_{16}$\\
			\hline
			$e_{1}$&\diagbox{}{}&(1,0.88 - 0.46i)&\diagbox{}{}&(1,-0.91 + 0.40 i)&\diagbox{}{}\\
			\hline
			$e_{13}$&\diagbox{}{}&\diagbox{}{}&(1,-0.92 + 0.75i)&\diagbox{}{}&(1, -0.73 + 0.54i)\\
			\hline
			$e_{14}$&(1,-0.94 + 0.68i)&\diagbox{}{}&\diagbox{}{}&(1,-0.94 + 0.77i)&\diagbox{}{}\\
			\hline
			$e_{15}$&\diagbox{}{}&(1,-0.83 + 0.56i)&\diagbox{}{}&\diagbox{}{}&(1,-1.1 - 1.2i)\\
			\hline
			$e_{16}$&(1,-0.82 + 0.45i)&\diagbox{}{}&(1,-1.0 + 0.81i)&\diagbox{}{}&\diagbox{}{}\\
			\hline
		\end{tabular}
	}
\end{table} 

\begin{table}[H]
	\centering\caption{The pseudo-critical point $(g^0_{ve},\, \textbf{z}^0_{vf})$ for the 4-simplex $v_5=(1,2,3,5,7)$.}\label{curve:ga5}
	\scalebox{0.49}
	{
		\begin{tabular}{|c|c|c|c|}
			\hline
			\small{$e$}&$e_6$&$e_{13}$&$e_{17}$\\
			\hline
			$g^0_{v_5e}$ &$\left(\begin{matrix}
				0.98& 0.32\\	0.011+0.006i& 1.03
			\end{matrix}\right)$&$\left(\begin{matrix}
				0.84& 0.82 + 0.19 i\\	-0.0012+0.011i& 1.19
			\end{matrix}\right)$&$\left(\begin{matrix}
				0.84& 0.73 - 0.05 i\\	0&1.2
			\end{matrix}\right)$\\
			\hline
			\small{$e$}&$e_{18}$&$e_{19}$&\\
			\hline
			$g^0_{v_5 e}$ &$\left(\begin{matrix}
				-0.00066+0.00052& -1.1i\\	-0.88 i& -0.72 i
			\end{matrix}\right)$&$\left(\begin{matrix}
				0& -1.2i\\	-0.86 i& 0.03 - 0.72i
			\end{matrix}\right)$&\\
			\hline
		\end{tabular}
		
		\begin{tabular}{|c|c|c|c|c|c|}
			\hline
			\diagbox{\small{$e$}}{$|\bold{z}^0_{v_5f}\rangle$}{\small{$e'$}}&$e'_{6}$&$e'_{13}$&$e'_{17}$&$e'_{18}$&$e'_{19}$\\
			\hline
			$e_{6}$&\diagbox{}{}&\diagbox{}{}&(1,-0.86 - 0.07i)&\diagbox{}{}&(1,-1.09 - 0.06i)\\
			\hline
			$e_{13}$&(1,0.87 - 0.50i)&\diagbox{}{}&\diagbox{}{}&(1,-0.83 + 0.56i)&\diagbox{}{}\\
			\hline
			$e_{17}$&\diagbox{}{}&(1,-0.93 + 0.75i)&\diagbox{}{}&\diagbox{}{}&(1,1,-3.2 + 0.6i)\\
			\hline
			$e_{18}$&(1,-1)&\diagbox{}{}&(1,-2 + 2.2i)&\diagbox{}{}&\diagbox{}{}\\
			\hline
			$e_{19}$&\diagbox{}{}&(1,-0.73 + 0.54i)&\diagbox{}{}&(1,-1.8 - 0.8 i)&\diagbox{}{}\\
			\hline
		\end{tabular}
	}
\end{table} 

\begin{table}[h]
	\centering\caption{The pseudo-critical point $(g^0_{ve},\, \textbf{z}^0_{vf})$ for the 4-simplex $v_6=(1,2,4,5,7)$.}\label{curve:ga6}
	\scalebox{0.48}
	{
		\begin{tabular}{|c|c|c|c|}
			\hline
			\small{$e$}&$e_{10}$&$e_{14}$&$e_{17}$\\
			\hline
			$g^0_{v_6e}$ &$\left(\begin{matrix}
				0.96, 0.38 \\	0.00077+0.00070i& 1.05
			\end{matrix}\right)$&$\left(\begin{matrix}
				0.68& 1.3 + 0.9 i\\	0& 1.5
			\end{matrix}\right)$&$\left(\begin{matrix}
				0.83& 0.73 - 0.05 i\\	-0.0014-0.0019i&1.2
			\end{matrix}\right)$\\
			\hline
			\small{$e$}&$e_{20}$&$e_{21}$&\\
			\hline
			$g^0_{v_6 e}$ &$\left(\begin{matrix}
				-0.00019-0.00100i& -1.1i\\	-0.93 i& 0.17-0.96 i
			\end{matrix}\right)$&$\left(\begin{matrix}
				0& -1.2i\\	-0.84 i& 0.4 - 2.3i
			\end{matrix}\right)$&\\
			\hline
		\end{tabular}
		
		\begin{tabular}{|c|c|c|c|c|c|}
			\hline
			\diagbox{\small{$e$}}{$|\bold{z}^0_{v_6f}\rangle$}{\small{$e'$}}&$e'_{10}$&$e'_{14}$&$e'_{17}$&$e'_{20}$&$e'_{21}$\\
			\hline
			$e_{10}$&\diagbox{}{}&(1,-0.94 + 0.68i)&\diagbox{}{}&(1,-0.68 - 0.15i)&\diagbox{}{}\\
			\hline
			$e_{14}$&\diagbox{}{}&\diagbox{}{}&(1,-0.92 + 0.75i)&\diagbox{}{}&(1,-1+0.81i)\\
			\hline
			$e_{17}$&(1,-0.86 - 0.07i)&\diagbox{}{}&\diagbox{}{}&(1,-1.9+2.2i)&\diagbox{}{}\\
			\hline
			$e_{20}$&\diagbox{}{}&(1,-0.94 + 0.77i)&\diagbox{}{}&\diagbox{}{}&(1,-2.7 - 0.4i)\\
			\hline
			$e_{21}$&(1,-0.45 - 0.08i)&\diagbox{}{}&(1,-3.2+0.6i)&\diagbox{}{}&\diagbox{}{}\\
			\hline
		\end{tabular}
	}
\end{table}
The boundary data for the curved geometry and the corresponding pseudo-critical point can be found in Mathematica notebook \cite{qudx.org}.

\section{Regge Action} \label{AppendixA2}
Let's first recall the volume of the simplex. The volume formula for the Lorentzian 4-simplex $\sig$ is given by \cite{Tate:2011rm, Tate2011FixedtopologyLT}
\be 
\mathbb{V}_{\sigma}=\frac{(-1)^{4}}{2^{4}(4 !)^{2}}\det(\bold{C}_{\sigma}) \label{V4}
\ee 
where $\mathbb{V}_{\sigma}$ is the volume square and $\det(\bold{C}_{\sigma})$ is the Cayley–Menger determinant. The Cayley–Menger matrix $\bold{C}_{\sigma}$ is the $6\times6$ matrix  with entries $l^2_{i j} \text { for } i, j=0, \cdots, 4$, where $l_{ij}$ is the segment length. The Cayley–Menger matrix is augmented by an additional row and column with entries given by $({\bf C}_\sig)_{5,5}=0$ and $({\bf C}_\sig)_{i,5}=({\bf C}_\sig)_{5,j}=1$. That is 
\be 
\bold{C}_\sig=\left[\begin{array}{c|c}
	l^{2}_{ij} & 1_{i} \\
	\hline 1_{j} & 0
\end{array}\right]
\ee 
Similarly, the volume formula of the Euclidean tetrahedron is given by 
\be 
\mathbb{V}_{\tau}=\frac{(-1)^{3+1}}{2^{3}(3 !)^{2}}\det(\bold{C}_{\tau}) \label{V3}
\ee 
here, $\bf{C}_{\tau}$ is the Cayley–Menger matrix for the tetrahedron, which is a $5\times 5$ matrix defined similarly as the above. 

Given $\vec{a}$ and $\vec{b}$ as timelike normal vector of two tetrahedra $\tau_a,\t_b$ of the 4-simplex $\sigma$, the Lorentzian dihedral angles are \cite{Asante:2021phx,Dittrich:2021gww} 
\be 
\theta_t(\sigma)=\sgn{(\vec{a}\cdot\vec{b})}\cosh^{-1}\left(\sgn{(\vec{a}\cdot\vec{b})}\frac{\vec{a}\cdot\vec{b}}{|\vec{a}||\vec{b}|}\right), \quad \sgn{(\vec{a}\cdot\vec{b})}=\frac{\sqrt{(\vec{a}\cdot\vec{b})^2}}{\vec{a}\cdot\vec{b}}.
\ee 
In the 4-dimentional triangulation, the hinge of the angle is a triangle denoted by $t$. Given a triangle $t$, it is shared by $\tau_a$ and $\tau_b$, and $s_{\bar{t}}$ is the length square of the segment opposite to the triangle $t$ in $\sigma$. For example, in the 4-simplex $\sig=(12345)$, the tetrahedra $\tau_a=(1234)$ and $\tau_b=(1235)$ share the triangle $t=(123)$. Then $\bar{t}$ is the segment $(45)$. The dihedral angles w.r.t $t$ are given by \cite{Dittrich:2007wm} 
\be 
\theta_t(\sigma)=\frac{\sqrt{\left(\frac{1}{\mathbb{V}_t}\frac{\partial \mathbb{V}_\sigma}{\partial s_{\bar{t}}}\right)^2}}{\frac{1}{\mathbb{V}_t}\frac{\partial \mathbb{V}_\sigma}{\partial s_{\bar{t}}}}\cosh^{-1}\left(\frac{\sqrt{\left(\frac{1}{\mathbb{V}_t}\frac{\partial \mathbb{V}_\sigma}{\partial s_{\bar{t}}}\right)^2}}{\frac{1}{\mathbb{V}_t}\frac{\partial \mathbb{V}_\sigma}{\partial s_{\bar{t}}}}\frac{\frac{3^2\cdot 4^2}{\mathbb{V}_t}\frac{\partial \mathbb{V}_{\sigma}}{\partial s_{\bar{t}}}}{\sqrt{3^2\frac{\mathbb{V}_{\tau_a}}{\mathbb{V}_t}}\sqrt{3^2\frac{\mathbb{V}_{\tau_b}}{\mathbb{V}_t}}}\right)
\ee 
Here, $\mathbb{V}$ are volume square ($\mathbb{V}_t=\fa_t^2$ is the area square) and $s$ is length square. As we only consider the space-like triangles and tetrahedra, so all the volume square are positive. The above formula can be simplified as 
\be 
\theta_t(\sigma)=\frac{\sqrt{\left(\frac{1}{\mathbb{V}_t}\frac{\partial \mathbb{V}_\sigma}{\partial s_{\bar{t}}}\right)^2}}{\frac{1}{\mathbb{V}_t}\frac{\partial \mathbb{V}_\sigma}{\partial s_{\bar{t}}}}\cosh^{-1}\left(
\frac{4^2\sqrt{\left(\frac{1}{\mathbb{V}_t}\frac{\partial \mathbb{V}_\sigma}{\partial s_{\bar{t}}}\right)^2}}{\sqrt{\mathbb{V}_{\tau_a}}\sqrt{\mathbb{V}_{\tau_b}}}
\right).
\ee 
Here, the volume of 4-simplex, tetrahedra and areas of triangles can be computed by following Eq.(\ref{V4}) and Eq.(\ref{V3}). Given any simplicial complex $\ck$, Regge action can be defined as
\be 
S_{\operatorname{Regge}}=\sum_{\sigma\subset \ck} \sum_{t\subset\sigma} \fa_{t}\theta_t(\sigma),
\ee 
where $\fa_{t}$ are the areas of the triangles $t$ and $\theta_t$ is the dihedral angle of triangle $t$.

\end{appendix}

\bibliographystyle{jhep}
\bibliography{ref}
\end{document}